\documentclass{jfm}
\usepackage{graphicx,caption,subcaption}
\usepackage{epstopdf, epsfig}
\usepackage{bm}
\usepackage{amsmath}
\usepackage{hyperref}
\hypersetup{
	colorlinks=true,
	allcolors=blue}

\title{Inertial migration of a neutrally buoyant spheroid in plane Poiseuille flow}
\author{Prateek Anand\aff{1}, Ganesh Subramanian\aff{1}
	\corresp{\email{sganesh@jncasr.ac.in}}}
\affiliation{\aff{1}Engineering Mechanics Unit, Jawaharlal Nehru Centre for Advanced Scientific Research, Bengaluru-560064, India.}
\begin{document}
\maketitle
	\begin{abstract}
 We study the cross-stream inertial migration of a torque-free neutrally buoyant spheroid, of an arbitrary aspect ratio $\kappa$, in wall-bounded plane Poiseuille flow for small particle Reynolds numbers\,($Re_p\ll1$) and confinement ratios\,($\lambda\ll1$), with the channel Reynolds number, $Re_c = Re_p/\lambda^2$, assumed to be arbitrary; here, $\lambda=L/H$ where $L$ is the semi-major axis of the spheroid and $H$ denotes the separation between the channel walls. In the Stokes limit\,($Re_p =0)$ and for $\lambda \ll 1$, a spheroid rotates along any of an infinite number of Jeffery orbits parameterized by an orbit constant $C$, while translating with a time dependent speed along a given ambient streamline. Weak inertial effects stabilize either the spinning\,($C=0$) or the tumbling orbit\,($C=\infty$), or both, depending on $\kappa$. The separation of the Jeffery-rotation and orbital drift time scales, from that associated with cross-stream migration, implies that the latter occurs due to a Jeffery-averaged lift velocity. Although the magnitude of this averaged lift velocity depends on $\kappa$ and $C$, the shape of the lift profiles are identical to those for a sphere, regardless of $Re_c$. In particular, the equilibrium positions for a spheroid remain identical to the classical Segre-Silberberg ones for a sphere, starting off at a distance of about $0.6(H/2)$ from the channel centerline for small $Re_c$, and migrating wallward with increasing $Re_c$. For spheroids with $\kappa \sim O(1)$, the Jeffery-averaged analysis is valid for $Re_p\ll1$; for extreme aspect ratio spheroids, the regime of validity becomes more restrictive being given by $Re_p\,\kappa/\ln \kappa \ll 1$ and $Re_p/\kappa^2 \ll 1$ for $\kappa \rightarrow \infty$\,(slender fibers) and $\kappa \rightarrow 0$\,(flat disks), respectively.
	\end{abstract}


\section{Introduction} \label{sec:intro}
In an experimental study of pressure-driven pipe flow of a suspension of neutrally buoyant spheres, \cite{segresilberberg1962a,segresilberberg1962b} reported the migration of particles to an intermediate annular location. This was termed the \textit{tubular pinch} effect to indicate that the initially uniform distribution of particles over the pipe cross-section is `pinched' to a narrow annulus with increasing downstream distance. The authors performed experiments for $0.03\leq d_p/D\leq0.15$ and for $Re$ upto 700, where $d_p$ and $D$ are the particle and pipe diameters, respectively, and $Re$ is the Reynolds number based on $D$ and the mean velocity. The aforementioned pinch occurred at $0.6\times$pipe radius for the smaller $Re$'s, and shifted towards the walls with increasing $Re$. In any unidirectional shearing flow, including pressure-driven flow in the pipe and channel geometries, reversibility of the Stokes equations prohibits cross-stream migration of a neutrally buoyant sphere, and the pinch effect above is a consequence of migration driven by inertial lift forces.

Since the original experiments of Segre and Silberberg, inertial migration in varying flow configurations has been investigated in a number of theoretical (\cite{saffman1965}, \cite{coxbrenner1968}, \cite{holeal1974}, \cite{vasseur1976},  \cite{cox1977}, \cite{schonberghinch1989}, \cite{asmolov1999}, \cite{matas2009}), numerical (\cite{ladd2006}, \cite{shao2008inertial}, \cite{seki2017}, \cite{nakayama2019}, \cite{glowinski2021}) and experimental (\cite{repetti1964}, \cite{jeffrey1965}, \cite{karnis1966}, \cite{tachibana1973}, \cite{aoki1979}, \cite{matas2004}, \cite{masaeli2012}, \cite{nakayama2019}) studies. From the rheological perspective, the pinch effect in the original pipe-flow experiments is undesirable, since it manifests as an apparent non-Newtonian behaviour for a dilute suspension of spheres. For instance, in a capillary viscometer, depending on the suspension flow rate, the residence time of the particles may or may not be sufficient for inertial migration to be complete; incomplete migration leads to a non-linear dependence of the inferred viscosity on the shear rate\citep{segre1963}. On the other hand, inertial migration has recently been used as a tool in microfluidics to separate particles based on size \citep{dicarlo2007}. Due to the robust fault-tolerant physical effects employed, and high rates of operation\,(the latter being a natural consequence of being in the inertial regime), inertial microfluidic systems are expected to have a broad range of applications in continuous bioparticle separation, cell and particle manipulation, and filtration systems\citep{dicarlo2009}.

A first theoretical explanation of the phenomenon was given by \cite{holeal1974} who determined the lift force on a sphere in plane Poiseuille flow for $Re_p, Re_c\ll1$, where $Re_p$ and $Re_c$ are the particle and channel Reynolds numbers, respectively. A more accurate calculation was performed in \cite{vasseur1976} using the framework developed earlier in \cite{coxbrenner1968}, wherein the lift velocity was expressed in terms of a volume integral involving the Green's function for creeping flow in the presence of a pair of plane boundaries\,(the channel walls). The resulting lift force profile had a pair of zero crossings, symmetrically located on either side of the channel centerline, and that corresponded to stable equilibrium locations. These were interpreted as the analog of the intermediate annulus observed in the experiments, implicitly pointing to the similarity of the physics governing inertial migration in the channel and pipe geometries. Later, \cite{schonberghinch1989} calculated the lift velocity of a sphere in plane Poiseuille flow for $Re_p\ll1$ and for $Re_c$ upto $150$, finding the equilibrium locations to move towards the respective walls with increasing $Re_c$, consistent with the original observations. \cite{asmolov1999} confirmed these  findings, and further extended the calculation to $Re_c=3000$\footnote{This upper bound was regarded as reasonable based on the threshold ($Re_c\sim11544$) for the Tolmein-Schlichting instability. It is now known, however, that the actual transition of plane Poiseuille flow to turbulence has a subcritical character, occurring at much lower $Re_c$'s of $O(2000)$.}.

There are numerous instances in microfluidic and other settings where the particles of interest are anisotropic, for example, cancer cells \citep{suresh2007}, blood cells \citep{irimia2005} or polymeric microstructures \citep{chung2008}. For anisotropic particles, the inertial lift force, and any equilibria that arise as a result of this force vanishing at specific locations, are expected to depend on particle shape. Motivated by this, \cite{masaeli2012} conducted experiments on spheroids of aspect ratio $1\leq\kappa\leq5$, suspended in pressure-driven flow through a rectangular duct, for $Re_c$ upto $80$\,($Re_c$ defined based on the smaller cross-sectional dimension). The study confirmed the existence of shape-sensitive equilibria, with large-aspect-ratio spheroids migrating to locations near the channel centerline and those with order unity aspect ratios migrating towards the duct walls, over a range of cross-sectional aspect ratios. 

Motivated by the experiments of \cite{masaeli2012}, we take a first step towards analyzing the inertial migration of a freely rotating neutrally buoyant spheroid in plane Poiseuille flow. Specifically, we calculate the leading order time-averaged lift velocity for $Re_p\ll 1$ within the framework of a point-particle approximation; $Re_c$, while much larger than $Re_p$, is otherwise arbitrary. $\S$\ref{sec:Ch3formulation} below presents the governing equations and boundary conditions in the context of the problem formulation. Next, in $\S$\ref{sec:Ch3Recsmall}, we examine the small-$Re_c$ limit where the time-averaged lift velocity is determined semi-analytically using a generalized reciprocal theorem formulation originally used by \cite{holeal1974}, and that is derived in $\S$\ref{sec:Ch3GRT}. Scaling arguments given in $\S$\ref{sec:Ch3Scaling} show that the dominant contribution to the lift velocity in this limit comes from scales of order the channel width $H$ which is much smaller than the inertial screening length of $O(HRe_c^{-1/2})$. Inertia therefore has a regular character, with the lift velocity being $O(Re_p)$, and its calculation requiring knowledge of only the Stokesian disturbance fields in the confined domain. These Stokesian fields are calculated in $\S$\ref{sec:Ch3boundedstresslet}, and are then used to obtain the time-averaged lift velocity for spheroids in $\S$\ref{sec:avglift_Calcn}. This is followed by a presentation of the results in $\S$\ref{sec:smallRecresults}. In $\S$\ref{sec:Ch3Reclarge}, the time-averaged lift velocity is calculated numerically for $Re_c\gtrsim O(1)$, with inertia now acting as a singular perturbation. Following \cite{schonberghinch1989}, this calculation involves a partial Fourier transform of the linearized Navier-Stokes equations, leading to coupled ODEs\,(in the transverse coordinate) for the transformed pressure and normal velocity fields, and their numerical solution using a shooting method. A brief outline of the aforementioned calculation procedure is given in $\S$\ref{sec:finiteRecformulation}, with the results obtained discussed in $\S$\ref{sec:finiteRecResults}. In both $\S$\ref{sec:Ch3Recsmall} and $\S$\ref{sec:Ch3Reclarge}, a time averaging is necessary on account of the separation between spheroid rotation and inertial migration time scales, and at leading order in $Re_p$, is based on the Jeffery angular velocity. This, together with the fact that the dominant contributions to the lift velocity come from scales much larger than the particle size for all $Re_c$, lead to the equilibrium locations being independent of the spheroid aspect ratio, and identical to those for spheres; the magnitude of the lift force does depend on aspect ratio. In $\S$\ref{sec:Ch3conclusion}, we summarize the main results, and briefly discuss their implications for shape-sorting in microfluidic settings.

\section{Problem Formulation} \label{sec:Ch3formulation}
\begin{figure}
	\centering
    \begin{subfigure}[b]{0.59\textwidth}
		\includegraphics[width=\textwidth]{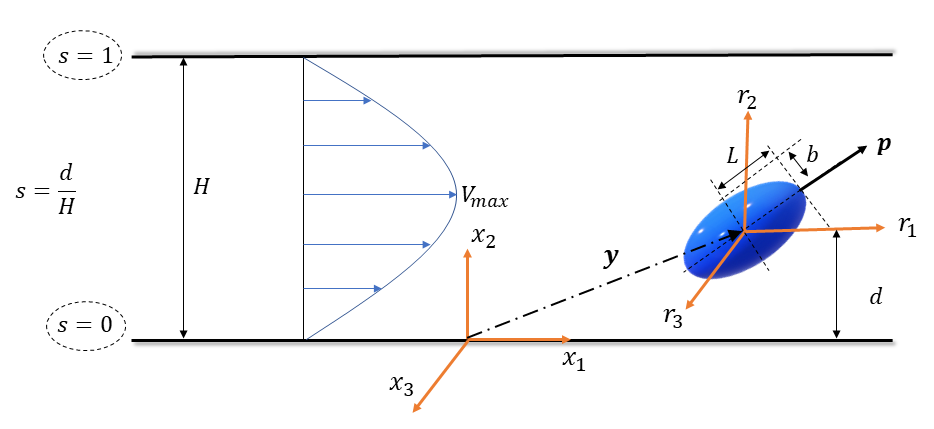}
        \caption{}
    \end{subfigure}
    \hfill
    \begin{subfigure}[b]{0.39\textwidth}
		\includegraphics[width=\textwidth]{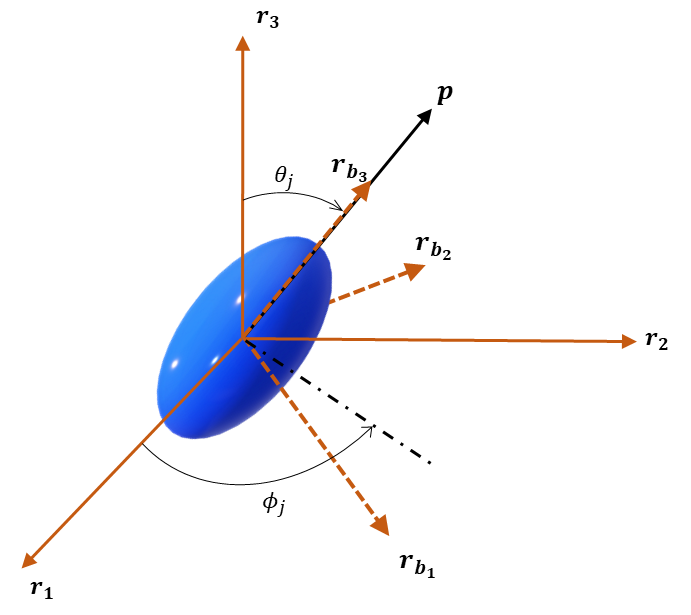}
        \caption{}
    \end{subfigure}
	\caption{(a) A neutrally buoyant spheroid with symmetry axis $\bm{p}$ in plane Poiseuille flow. The position of the spheroid relative to the lab frame ($x_1,x_2,x_3$) is denoted by $\bm{y}$; $(r_1,r_2,r_3)$ represents the Cartesian frame with origin at the spheroid center, and translating with it. (b) shows the body-fixed coordinate system aligned with $\bm{p}$, along with the polar\,($\theta_j$) and azimuthal\,($\phi_j$) angles that define the spheroid orientation.}
	\label{fig:Ch3channelGeometry}
\end{figure}
Figure \ref{fig:Ch3channelGeometry}a shows a neutrally buoyant spheroid of aspect ratio $\kappa=L/b$\,($L$ and $b$ are the semi-major and minor axes) freely suspended in a wall-bounded plane Poiseuille flow at a distance $d$ from the lower wall; $\kappa<1$ and $>1$ for oblate and prolate spheroids, respectively.
The non-dimensional equations governing the velocity field $\bm{u}$ and pressure field $p$ are given by:
\begin{subequations}
	\begin{align}
	\nabla^2 \bm{u}-\bm{\nabla}p &=Re_p\,\left(\frac{\partial\bm{u}}{\partial t}+\bm{u\cdot\nabla u}\right),\\
	\bm{\nabla}\cdot\bm{u}&=0 ,
	\end{align} \label{eq:Ch3NS1}
\end{subequations}
where $\bm{u}$ satisfies the following boundary conditions:
\begin{subequations}
	\begin{align}
	\bm{u}&=\bm{\Omega_p}\wedge \bm{r} \text{ for } \bm{r}\in S_p, \label{spheroidsurf_BC}\\
	\bm{u}&\rightarrow \bm{u}^\infty \text{ for } r_1,r_3\rightarrow \infty\,(r_2\hspace*{0.05in}\text{fixed}), \label{ambient_infBC}\\
	\bm{u}&=-\bm{U}_p \text{ at } r_2=-s\lambda^{-1}, (1-s)\lambda^{-1}.
	\end{align} \label{eq:Ch3BC1}
\end{subequations}
In (\ref{eq:Ch3NS1}a-\ref{eq:Ch3BC1}c), all variables are non-dimensionalized using $L$ and the velocity scale $V_c=V_\text{max}L/H$, this being the order of the velocity change across the ends of the spheroid. 
The particle Reynolds number $Re_p=V_c L/\nu = V_\text{max}L^2/H\nu$, $\nu$ being the kinematic viscosity of the fluid. $\bm{\Omega}_p$ in (\ref{spheroidsurf_BC}) is the angular velocity of the spheroid whose surface is denoted by $S_p$. The coordinate system chosen in writing the above equations translates with the spheroid velocity $\bm{U}_p$, with its origin at the spheroid center. Thus, $\bm{u}^\infty$ in (\ref{ambient_infBC}), the ambient plane Poiseuille flow in this frame, is given by,
\begin{align}
\bm{u}^\infty=(\alpha+\beta r_2 +\gamma {r_2}^2)\bm{1}_1-\bm{U}_p,
\label{eq:Ch3uamb}
\end{align}
where $\alpha \bm{1}_1 - \bm{U}_p$, with $\alpha=4 \lambda^{-1} s(1-s)$, is the ambient slip velocity at the spheroid center, $\beta=4(1-2s)$  is the local shear rate that varies linearly across the channel, and $\gamma=-4\lambda$ is the constant curvature of the plane Poiseuille flow; $s = d/H$ here  being the (non-dimensional)\,spheroid location. $\lambda=L/H$ is the confinement ratio assumed small, so the channel Reynolds number  $Re_c=Re_p/\lambda^2 \gg Re_p$. As will be argued below, at leading order in $Re_p$ and $\lambda$, $\bm{U}_p$ is along the flow direction.

We now define the disturbance fields $\bm{u}'=\bm{u}-\bm{u}^\infty$, $p'=p-p^\infty$ which satisfy the following governing equations:,
\begin{subequations}
	\begin{align}
	\bm{\nabla} \cdot \bm{\sigma}' = \nabla^2 \bm{u}'-\bm{\nabla} p' &=Re_p\left(\frac{\partial\bm{u}'}{\partial t}+\bm{u}'\cdot\bm{\nabla u}'+\bm{u}'\cdot\bm{\nabla u}^\infty+\bm{u}^\infty\cdot\bm{\nabla u}'\right),\\
	\bm{\nabla}\cdot \bm{u}'&=0,
	\end{align} \label{eq:Ch3NS2}
\end{subequations}
where $\bm{u}'$ satisfies:
\begin{subequations}
	\begin{align}	\bm{u}'&=\bm{U}_p+\bm{\Omega}_p\wedge \bm{r}-(\alpha+\beta r_2 +\gamma r_2^2)\bm{1}_1 \text{ for } \bm{r}\in S_p,\\
	\bm{u}'&\rightarrow 0 \text{ for } r_1,r_3\rightarrow \infty\,(r_2\hspace*{0.05in}\text{fixed}),\\
	\bm{u}'&=0 \text{ at } r_2=-s\lambda^{-1}, (1-s)\lambda^{-1}.
	\end{align} \label{eq:Ch3BC2} 
\end{subequations}
Unlike a sphere, one needs to include the unsteady terms in the inertial acceleration on account of spheroid rotation even in the Stokes limit\,($Re_p = 0$). We analyze the inertial migration problem defined above in the limit $\lambda, Re_p\ll1$ with $Re_c=Re_p/\lambda^2$ arbitrary. It is also assumed that $s,\,(1-s)\gg\lambda$, implying that the analysis is restricted to the spheroid being at distances from either wall that are much larger than $O(L)$.

Before embarking on a detailed analysis for weak inertia, it is worth summarizing the nature of spheroid motion in plane Poiseuille flow in the Stokes limit. Since the local linear flow approximation for the plane Poiseuille profile is simple shear flow, the neutrally buoyant spheroid must rotate along Jeffery orbits\citep{jeffery}, this rotation being characterized by the polar ($\theta_j$) and azimuthal ($\phi_j$) angles of the spheroid symmetry axis\,(see Figure \ref{fig:Ch3channelGeometry}b) being functions of time; these equations are given later in $\S$\ref{sec:Ch3Recsmall}\,(see \ref{eq:Ch3Spheroidangles}a,b). Jeffery rotation is best described in $(C,\tau$) coordinates, where $C\in[0,\infty)$ is the orbit constant, and $\tau$ is the orbit phase that changes at a constant albeit $\kappa$-dependent rate from $0$ to $2\pi$ over a single period\citep{leal1971,navaneeth2016}; the two limiting orbits correspond to the spinning ($C=0$) and tumbling ($C=\infty$) modes. The Jeffery period is independent of $C$, being given by $T_\text{jeff}=2\pi(\beta V_\text{max}/H)^{-1} (\kappa+\kappa^{-1})$, where $\beta$ accounts for the linearly varying shear rate. The curvature of the plane Poiseuille profile only affects the spheroid translation velocity that is now a function of its orientation, and thence, of time. A neutrally buoyant spheroid in an unbounded plane Poiseuille flow continues to move along a given ambient streamline, with a speed that ranges from a maximum, corresponding to the flow-aligned orientation\,($\phi_j=0$ and $\frac{\pi}{2}$ for $\kappa >1$ and $<1$), to a minimum when the spheroid is aligned orthogonal to the flow direction\,($\phi_j=\frac{\pi}{2}$ and $0$ for $\kappa >1$ and $<1$)\citep{chwang1975}. For very large\,(small) $\kappa$, the spheroid spends an increasing fraction of a Jeffery period in the flow\,(gradient-vorticity plane)-aligned orientation, and the translational motion thereby acquires an increasingly jerky character owing to the spheroid abruptly slowing down during the brief periods of mis-alignment. 

In presence of walls, the leading order correction to the motion above arises from interaction of the spheroid with time dependent image stresslets induced by each wall, resulting in an $O(\lambda^2)$ lateral velocity component even in the Stokes limit, with an additional $O(\lambda^3)$ correction to the Jeffery angular velocity. Note that this image-stresslet interaction does not lead to transverse motion for a sphere, as may be seen from the stresslet orientation\,(along the local extensional axis) and the associated purely radial velocity field; as already mentioned in $\S$\ref{sec:intro}, this must be so, independent of $\lambda$, due to reversibility constraints. Thus, for $\lambda\ll 1$, the center of mass of a neutrally buoyant spheroid in wall-bounded plane Poiseuille flow exhibits a small-amplitude oscillatory motion about an ambient streamline, the amplitude being $O(\lambda^2)$. In other words, unlike a sphere, Stokesian reversibility does not preclude an instantaneous lift force for an anisotropic particle. Evidence for such oscillation-cum-tumbling spheroid trajectories, for finite $\lambda$, is available from earlier computations, with there being a transition from tumbling\,(rotation) to angular oscillations beyond a threshold $\lambda$ close to unity, corresponding to sufficiently narrow channels,\citep{sugihara1993,sugihara1996,staben2003,staben2006}. However, Stokesian reversibility still forbids a net migration of the spheroid in the transverse direction, and to allow for such a motion, one needs inertia. It is this net cross-stream migration, for small $Re_p$ and $\lambda$, that is analyzed in the following sections. We obtain the time-averaged motion of a spheroid in this limit, and since the wall and inertia-induced modification of the  primary translational and rotational motion are weak, the time average corresponds to an average over a Jeffery period. 

It is well-known that the Stokes equations do not provide a uniformly valid leading order approximation for small but finite $Re_p$, and that in general one requires a matched asymptotic expansion approach \citep{proudman_pearson_1957} involving an inner expansion in the neighbourhood of the particle, and an outer expansion at distances of order an inertial screening length, to calculate inertial corrections. The screening length for the present problem is $L Re_p^{-1/2}$, or equivalently, $HRe_c^{-1/2}$. For small $Re_c$, the screening length is larger than $H$, so the channel walls lie in the inner Stokesian region where fluid inertia may be treated as a regular perturbation. The inertial migration problem for a sphere in this limit was investigated by \cite{holeal1974}, and in a series of papers by Brenner, Cox and collaborators (\cite{coxbrenner1968}, \cite{vasseur1976}, \cite{cox1977}). We examine the analogous problem for a spheroid in this limit in $\S$\ref{sec:Ch3Recsmall} below. When $Re_c\gtrsim O(1)$, the inertial screening length is of $O(H)$ or smaller, and the solution procedure involves the outer expansion, the leading order term of which satisfies the linearized Navier-Stokes equations. This limit was first examined for a sphere in \cite{schonberghinch1989}, and  we examine the same for a spheroid in $\S$\ref{sec:Ch3Reclarge}.

\section{The inertial lift velocity for $Re_c\ll1$} \label{sec:Ch3Recsmall}

Although inertial effects are a regular perturbation for small $Re_c$, following \cite{holeal1974}, we don't calculate this perturbation explicitly, and instead use the generalized reciprocal theorem that relates the velocity and stress fields of the problem of interest with the velocity and stress fields of a simpler test problem whose solution is known. The problem of interest ($\bm{u}',\bm{\sigma}'$) corresponds to the motion of a torque-free neutrally buoyant spheroid in a wall-bounded plane Poiseuille flow, taking into account the inertial acceleration of the suspending fluid. Since the quantity of interest is the inertial lift velocity, the test problem ($\bm{u}^t,\bm{\sigma}^t$) is taken to be a torque-free spheroid, in a quiescent ambient between parallel walls, acted on by an arbitrarily oriented unit force; the test spheroid has the same instantaneous orientation as the one in the actual problem. The governing equations and boundary conditions for the actual problem have already been given in (\ref{eq:Ch3NS2}a,b) and (\ref{eq:Ch3BC2}a-c). Those for the test problem are:
\begin{subequations}
\begin{align}
\bm{\nabla}\cdot\bm{\sigma}^t&=\nabla^2 \bm{u}^t-\bm{\nabla} p^t = 0 ,\\
\bm{\nabla}\cdot \bm{u}^t&=0,
\end{align} \label{eq:Ch3NStest}
\end{subequations}
with the boundary conditions:
\begin{subequations}
	\begin{align} 
	\bm{u}^t&=\bm{U}_p^t+\bm{\Omega}_p^t\wedge\bm{r} \text{ for } \bm{r}\in S_p,\\
	\bm{u}^t &\rightarrow 0 \text{ for } r_1,r_3 \rightarrow \infty\,(r_2\hspace*{0.05in}\text{fixed}),\\
	\bm{u}^t&=0 \text{ at } r_2=-s\lambda^{-1}, (1-s)\lambda^{-1},
	\end{align} \label{eq:Ch3BCtest}
\end{subequations}
where $\bm{U}_p^t$ and $\bm{\Omega}_p^t$ are the spheroid translational and angular velocities.

\subsection{The Generalized Reciprocal theorem} \label{sec:Ch3GRT}
To derive the generalized reciprocal theorem identity, we contract  (\ref{eq:Ch3NS2}a) with $\bm{u}^t$ and (\ref{eq:Ch3NStest}a) with $\bm{u}'$, and subtract the resulting expressions\citep{holeal1974} to obtain:
\begin{align}
(\bm{\nabla}\cdot\bm{\sigma}')\cdot \bm{u}^t-(\bm{\nabla}\cdot\bm{\sigma}^t) \cdot \bm{u}'&=Re_p\,\,\bm{u}^t\cdot\bm{f}, \label{eq:Ch3RT2}
\end{align}
where $\bm{f}=\left(\frac{\partial\bm{u}'}{\partial t}+\bm{u}'\cdot\bm{\nabla u}'+\bm{u}'\cdot\bm{\nabla u}^\infty+\bm{u}^\infty\cdot\bm{\nabla u}'\right)$. Integrating \eqref{eq:Ch3RT2} over the fluid volume\,($V^F$) between the channel walls:
\begin{align}
\int_{V^F} \bm{\nabla}\cdot(\bm{\sigma}'\cdot \bm{u}^t-\bm{\sigma}^t\cdot \bm{u}')dV + \int_{V^F} (\bm{\sigma}^t\cdot\bm{\nabla u}'-\bm{\sigma}'\cdot \bm{\nabla u}^t)dV &=Re_p\int_{V^F}\bm{u}^t\cdot\bm{f}\,dV. \label{eq:Ch3RT3}
\end{align}
Using (\ref{eq:Ch3NS2}b) and (\ref{eq:Ch3NStest}b), along with the definitions of $\bm{\sigma}'$ and $\bm{\sigma}^t$, the second integral on the LHS in \eqref{eq:Ch3RT3} can be shown to be identically zero. Applying the divergence theorem to the first integral yields,
\begin{align}
-\int_{S_p+S_w+S_\infty} (\bm{\sigma}'\cdot\bm{u}^t-\bm{\sigma}^t\cdot \bm{u}')\cdot \bm{n}\,dS &=Re_p\int_{V^F}\bm{u}^t\cdot\bm{f}\,dV. \label{eq:Ch3RT4}
\end{align}
where $\bm{n}$ is the unit normal at all bounding surfaces pointing into the fluid domain. The set of bounding surfaces include the particle surface ($S_p$), the channel walls ($S_w$) and the surface at infinity ($S_\infty$); the latter can be thought of as the curved surface of a cylinder of radius $R$, with its axis along the gradient direction, in the limit $R\to\infty$. For $R\gg H$, the disturbance fields in the actual and test problems are exponentially small (see \eqref{eq:Ch3BoundedStokesletfield} and \eqref{eq:Ch3stressletForm}), and therefore, the integral over $S_\infty$ in \eqref{eq:Ch3RT4} is vanishingly small. The integral $\int_{S_w} (\bm{\sigma}'\cdot \bm{u}^t-\bm{\sigma}^t\cdot \bm{u}')\cdot \bm{n}\,dS\,\,$ can be shown to be identically zero on using the no-slip conditions (\ref{eq:Ch3BC2}c) and (\ref{eq:Ch3BCtest}c) on the channel walls. Thus, \eqref{eq:Ch3RT4} reduces to:
\begin{align}
-\int_{S_p} (\bm{\sigma}'\cdot \bm{u}^t-\bm{\sigma}^t\cdot \bm{u}')\cdot \bm{n}\,dS &=Re_p\int_{V^F}\bm{u}^t\cdot\bm{f}\,dV. \label{eq:Ch3RT5}
\end{align}
Applying the no-slip conditions (\ref{eq:Ch3BC2}a) and (\ref{eq:Ch3BCtest}a)  to \eqref{eq:Ch3RT5},
\begin{align}
&-\bm{U}_p^t\cdot\int_{S_p} \bm{\sigma}'\cdot \bm{n}\,dS -\bm{\Omega}_p^t\cdot\int_{S_p}\bm{r}\wedge(\bm{\sigma}'\cdot\bm{n})\,dS + \bm{\Omega}_p\cdot\int_{S_p} \bm{r}\wedge(\bm{\sigma}^t\cdot\bm{n})\,dS\nonumber\\ 
&+\bm{U}_p\cdot\int_{S_p} \bm{\sigma}^t\cdot \bm{n}\,dS-\int_{S_p} (\bm{\sigma}^t\cdot \bm{n})\cdot(\alpha+\beta r_2+\gamma r_2^2)\bm{1}_1\,dS=Re_p\int_{V^F}\bm{u}^t\cdot\bm{f} dV. \label{eq:Ch3RT6}
\end{align}
The hydrodynamic force and torque experienced by the particle may be written as $\int_{S_p} \bm{\sigma}'\cdot \bm{n}\,dS=Re_p\,\, d\bm{U}_p/dt$ and $\int_{S_p} \bm{r}\wedge (\bm{\sigma}'\cdot\bm{n})\,dS=Re_p\,\, d(\bm{I}_p\cdot\bm{\Omega}_p)/dt$, $\bm{I}_p$ being the spheroid moment of inertia tensor. Next, to $O(Re_p)$, one may replace $\bm{u}'$ by $\bm{u}_s$ in the volume integral in \eqref{eq:Ch3RT6}, $\bm{u}_s$ being the corresponding Stokesian approximation. The validity of such a replacement requires the resulting volume integral to be convergent, this being related to inertia acting as a regular perturbation. Detailed arguments in this regard are given in the next subsection. Further, noting that $\int_{S_p} \bm{r}\wedge(\bm{\sigma}^t\cdot \bm{n})\,dS = 0$ owing to the spheroid being torque-free in the test problem, one obtains:
\begin{align}
&-Re_p\,\,\bm{U}_p^t\cdot \frac{d\bm{U}_p}{dt} -Re_p\,\,\bm{\Omega}_p^t\cdot\frac{d(\bm{I}_p\cdot\bm{\Omega}_p)}{dt}+\bm{U}_p\cdot\int_{S_p} \bm{\sigma}^t\cdot \bm{n}\,dS\nonumber\\
&-\int_{S_p} (\bm{\sigma}^t\cdot \bm{n})\cdot(\alpha+\beta r_2+\gamma r_2^2)\bm{1}_1\,dS=Re_p\int_{V^F}\bm{u}^t\cdot\bm{f}_s\,dV. \label{eq:Ch3RT7}
\end{align}
where $\bm{f}_s$ denotes the  approximation of $\bm{f}$ based on replacing $\bm{u}'$ by $\bm{u}_s$, being given by:
\begin{align}
\bm{f}_s=\left(\frac{\partial\bm{u}_s}{\partial t}+\bm{u}_s\cdot\bm{\nabla u}_s+\bm{u}_s\cdot\bm{\nabla u}^\infty+\bm{u}^\infty\cdot\bm{\nabla u}_s\right).
\end{align}
 One may now use the small-$Re_p$ expansions $\bm{U}_p=\bm{U}_{p0}+Re_p \bm{U}_{p1}+O(Re_p^2)$ and $\bm{\Omega}_p=\bm{\Omega}_{p0}+Re_p \bm{\Omega}_{p1}+O(Re_p^2)$, of the spheroid translational and angular velocities, in \eqref{eq:Ch3RT7}, and obtain the following relations at successive orders in $Re_p$:
\begin{itemize} 
	\item $O(1)$:
\begin{align}
	\bm{U}_{p0}\cdot\int_{S_p} \bm{\sigma}^t\cdot \bm{n}\,dS=\int_{S_p} (\bm{\sigma}^t\cdot \bm{n})\cdot(\alpha+\beta r_2+\gamma r_2^2)\bm{1}_1\,dS.
	\label{eq:Ch3RT8}
\end{align}
	\item $O(Re_p)$:
\begin{align}
-\,\,\bm{U}_p^t\cdot \frac{d\bm{U}_{p0}}{dt}-\,\,\bm{\Omega}_p^t\cdot\frac{d(\bm{I}_p\cdot\bm{\Omega}_{p0})}{dt}+\bm{U}_{p1}\cdot\int_{S_p} \bm{\sigma}^t\cdot \bm{n}\,dS=\int_{V^F}\bm{u}^t\cdot\bm{f}_s\, dV .\label{eq:Ch3RT9}
\end{align}
\end{itemize}

\underline{\paragraph{The $O(1)$ problem:}}\hspace{0pt}\\\\
For the $O(1)$ problem defined by \eqref{eq:Ch3RT8}, we choose $\int_{S_p} \bm{\sigma}^t\bm{\cdot n}\,dS=\bm{1}_1$, corresponding to the test spheroid translating due to a unit flow-aligned force. In the absence of boundaries, the induced velocity field is known in terms of vector spheroidal harmonics\citep{navaneeth2015}, and the surface force density to be used on the RHS is given by\citep{kushch2003}:
\begin{align}
\bm{\sigma}^t\bm{\cdot n} =& -p^t \bm{1}_\xi+2\left(\frac{(\xi^2-1)^{1/2}}{\xi_0 (\xi^2-\eta^2)^{1/2}}\frac{\partial \bm{u}^t}{\partial \xi}+\frac{1}{2}\bm{1}_\xi\wedge\bm{\nabla}\wedge\bm{u}^t\right), \label{fdensity_transSpheroid}
\end{align}
where $\xi>\xi_0>1$ is the coordinate characterizing a family of confocal spheroids with $\xi_0$ representing the spheroid surface, $\bm{1}_\xi$ is the unit normal to the spheroid surface and $p^t$ is the pressure field. Evaluating the surface integral in (\ref{eq:Ch3RT8}), using \eqref{fdensity_transSpheroid}, one obtains:  
\begin{equation}
\bm{U}_{p0}= \left[ \alpha+\frac{\gamma}{3\kappa^2}[\cos^2\phi_j+(\cos^2\theta_j+\kappa^2\sin^2\theta_j)\sin^2\phi_j] \right]\bm{1}_1, \label{eq:Ch3UpFaxen} \end{equation}
which is Faxen's law for the translation of a force-free spheroid truncated to second order, the term proportional to $\gamma$ arising from the curvature of the ambient flow; higher order terms in the expansion are zero for an ambient quadratic flow\citep{brennerbook}. \eqref{eq:Ch3UpFaxen} pertains to spheroid of a given orientation, and a time trajectory requires knowing $\theta_j$ and $\phi_j$ as functions of time. For this purpose, one may use Faxen's law relating the torque and angular velocity of a spheroid. This is again an infinite expansion in the general case, but with only the leading order term, proportional to the ambient velocity gradient, surviving for a quadratic flow. Thus, spheroid rotation remains identical to that in an ambient linear flow, and application of the torque-free constraint yields:
\begin{align}
&\bm{\Omega}_{p0}=-\frac{\beta(\kappa^2-1)}{4(\kappa^2+1)}\cos\phi_j\sin 2\theta_j\bm{1}_1+\frac{\beta(\kappa^2-1)}{4(\kappa^2+1)}\sin\phi_j\sin 2\theta_j\bm{1}_2\nonumber\\
&+\frac{\beta}{2}\big[-1+\frac{(\kappa^2-1)}{(\kappa^2+1)}\cos 2\phi_j\sin^2 \theta_j\big]\bm{1}_3.
\label{eq:Ch3OmegapFaxen}
\end{align}
The equations governing Jeffery orbits arise as components of (\ref{eq:Ch3OmegapFaxen}) in the body-aligned coordinate system, and are given below in (\ref{eq:Ch3JefferyAngularvelocity}ab). With $\theta_j$ and $\phi_j$ defined in this manner, $\bm{U}_{p0}$ as defined in (\ref{eq:Ch3UpFaxen}) exhibits the time dependence described earlier in $\S$\ref{sec:Ch3formulation}. Substituting $\bm{U}_{p0}$ in \eqref{eq:Ch3uamb}, the ambient flow in the  reference frame chosen for the reciprocal theorem is given by:
\begin{align}
\bm{u}^\infty=\left[ (\beta r_2 +\gamma r_2^2)-\frac{\gamma}{3\kappa^2}[\cos^2\phi_j+(\cos^2\theta_j+\kappa^2\sin^2\theta_j)\sin^2\phi_j] \right] \bm{1}_1.
\label{eq:VambFaxen}
\end{align}
Note that the $O(\lambda^2)$ center-of-mass oscillations mentioned in $\S$\ref{sec:Ch3formulation} can only arise from $\bm{U}_{p0}$ having a component along the gradient direction, and this requires an expression for $\bm{\sigma}^t \cdot \bm{n}$ that includes the influence of the plane boundaries.

\underline{\paragraph{The $O(Re_p)$ problem:}}\hspace{0pt}\\\\
At $O(Re_p)$, one chooses the test force as $\int_{S_p} \bm{\sigma}^t\bm{\cdot n}\,dS=-\bm{1}_2$, which leads to:
\begin{align}
V_p=Re_p\,\bm{U}_{p1}\cdot\bm{1}_2 &=-Re_p\int_{V^F}\bm{u}^{t2}\cdot\left(\frac{\partial\bm{u}_s}{\partial t}+\bm{u}_s\cdot\bm{\nabla u}_s+\bm{u}_s\cdot\bm{\nabla u}^\infty+\bm{u}^\infty\cdot\bm{\nabla u}_s\right) dV\nonumber\\
&-Re_p\,\,\bm{U}_p^{t2}\cdot \frac{d\bm{U}_{p0}}{dt}-Re_p\,\,\bm{\Omega}_p^{t2}\cdot\frac{d(\bm{I}_p\cdot\bm{\Omega}_{p0})}{dt}
\label{eq:Ch3RTliftvel}
\end{align}
for the instantaneous lift velocity of a neutrally buoyant spheroid in plane Poiseuille flow, which  is a function of the changing spheroid orientation via $\bm{u}^{t2}$ and $\bm{u}_s$, both of them depend on $\bm{p}$. The additional superscript `$2$' for the test problem quantities indicate the choice of the test force orientation above.

As already mentioned, for small confinement ratios, a spheroid in wall-bounded plane Poiseuille flow rotates along Jeffery orbits \citep{jeffery} in the Stokes limit. For small but finite $Re_p$, there is an additional $O(Re_p)$ drift across orbits which stabilizes the tumbling mode\,($C=\infty$) for prolate spheroids, the spinning mode\,($C=0$) for oblate spheroids with $0.14<\kappa<1$ and, depending on the initial orientation, either the spinning or tumbling mode for oblate spheroids with $\kappa<0.14$ \citep{einarsson2015,navaneeth2016}. The time scale for rotation along a Jeffery orbit is $T_\text{jeff}\sim O(H/V_\text{max})$ for $\kappa\sim O(1)$. The inertia-driven orbital drift above occurs on longer time scales $t_\text{drift} \sim O(Re_p^{-1}H/V_\text{max})$. Scaling arguments in $\S$\ref{sec:Ch3Scaling} below show that the volume integral in (\ref{eq:Ch3RTliftvel}) is the dominant contribution to the cross-stream migration, being $O(1)$ for $\lambda \ll 1$, so the time scale for inertial migration is $t_\text{lift} \sim O(Re_p^{-1}\lambda^{-1}H/V_\text{max})$. Since $t_\text{lift}/t_\text{drift}\sim \lambda^{-1}\gg1$, for purposes of the migration calculation, one may assume that the spheroid has settled into its stable Jeffery orbit. Moreover, since $t_\text{lift}/T_\text{jeff}\sim Re_p^{-1}\lambda^{-1} \gg1$, the leading order migration is the result of an orientation-averaged lift velocity, the average being over the orientations sampled in the stabilized Jeffery orbit. As indicated below, this average may be approximated based on the Jeffery angular velocity for small $Re_p$. Superposed on this orientation-averaged migration trajectory would be small-amplitude oscillations of $O(Re_p\lambda)$, corresponding to the fluctuations in the instantaneous lift velocity arising from the rapidly changing spheroid orientation, as may be established formally using the method of multiple scales.

Owing to the aforementioned time scale separation in the limit $\lambda, Re_p \ll 1$, it is of interest to determine the Jeffery-averaged rather than the instantaneous lift velocity. Averaging both sides of (\ref{eq:Ch3RTliftvel}) over a Jeffery period, one obtains:
\begin{align}
\langle V_p\rangle=-Re_p\int_{V^F}\left\langle\bm{u}^{t2}\cdot\left(\frac{\partial\bm{u}_s}{\partial t}+\bm{u}_s\cdot\bm{\nabla u}_s+\bm{u}_s\cdot\bm{\nabla u}^\infty+\bm{u}^\infty\cdot\bm{\nabla}\bm{u}_s\right)\right\rangle\,\, dV,
\label{eq:VpTimeAvgd1}
\end{align}
where the averaging operation is defined as $\langle.\rangle = \frac{1}{T_\text{jeff}}\int_0^{T_\text{jeff}} (.)\,dt$, and pertains to the inertially stabilized Jeffery orbit\,(that is, $C$ is either $0$ or $\infty$) for times longer than $O(Re_p^{-1}H/V_\text{max})$. 

 In writing down (\ref{eq:VpTimeAvgd1}), we have neglected the terms on the RHS of (\ref{eq:Ch3RTliftvel}) involving the translational and angular accelerations in the actual problem in the Stokesian limit. These depend on $\lambda$ at leading order, and are therefore asymptotically smaller than the Jeffery-averaged volume integral that, as mentioned above, is independent of $\lambda$ for $\lambda \ll 1$. For a spinning spheroid both acceleration terms are trivially zero because, independent of $\lambda$, a spinning spheroid\,(like a sphere) translates with a constant speed along an ambient streamline, while rotating at a uniform rate about the ambient vorticity direction. For a spheroid rotating in any other Jeffery orbit, both terms do lead to non-zero Jeffery-averaged contributions. The first of these terms is non-zero because of a correlation between the time-periodic variations of $\frac{d\bm{U}_{p0}}{dt}$ along the flow direction, and the analogous variation of the flow-directed component of $\bm{U}_p^{t2}$, the latter arising due to the periodically varying test spheroid orientation for a fixed force\,(along $-\bm{1}_2$). The acceleration in the actual problem is only $O(\lambda^2)$, however, as may be seen from the Faxen's translation law given earlier. The smallness of the second term is because $\bm{\Omega}_p^{t2}$, at leading order, is driven by the velocity gradient associated with the equal and oppositely directed image-Stokeslet, and is again $O(\lambda^2)$; the angular acceleration in the actual problem, $\frac{d(\bm{I}_p\cdot\bm{\Omega}_{p0})}{dt}$, is $O(1)$. Thus, both acceleration terms in (\ref{eq:Ch3RTliftvel}) are $O(\lambda^2)$ smaller than the volume integral.
 
The averaging operation in (\ref{eq:VpTimeAvgd1}) corresponds to a fixed $C$, and is naturally evaluated out in $(C,\tau)$ coordinates\citep{leal1971,navaneeth2016}. Here, $C=\frac{\tan\theta_j(\kappa^2\sin^2\phi_j+\cos^2\phi_j)^{1/2}}{\kappa}$ and $\tau=\tan^{-1}\left[\frac{1}{\kappa\tan\phi_j}\right]$, where $\theta_j$ and  $\phi_j$ are the polar and azimuthal angles characterizing the spheroid orientation. The latter is denoted by the unit vector $\bm{p}$ in Figure \ref{fig:Ch3channelGeometry}b with $\bm{p}=\sin\theta_j\cos\phi_j\bm{1}_1+\sin\theta_j\sin\phi_j\bm{1}_2+\cos\theta_j\bm{1}_3$. The equations:
\begin{subequations}
\begin{align}
\frac{d\phi_j}{dt} =&\beta \left(-\frac{1}{2}+\frac{\kappa^2-1}{2 (\kappa^2+1)}\cos 2\phi_j\right),\\
\frac{d\theta_j}{dt} =&\beta\frac{(\kappa^2-1)}{4(\kappa^2+1)}\sin 2\theta_j\sin 2\phi_j,
\end{align}
\label{eq:Ch3JefferyAngularvelocity}
\end{subequations}
describe rotation along Jeffery orbits, and are obtained as the individual Cartesian components of the Faxen angular velocity relation, \eqref{eq:Ch3OmegapFaxen}, given above\,($\bm{\Omega}_{p0} \cdot \bm{1}_{r_{b_2}} = \frac{d\theta_j}{dt}$, $\bm{\Omega}_{p0} \cdot \bm{1}_{r_{b_1}} = -\sin \theta_j\frac{d\phi_j}{dt}$). In terms of $C$ and $\tau$, (\ref{eq:Ch3JefferyAngularvelocity}a) and (\ref{eq:Ch3JefferyAngularvelocity}b) take the form $\frac{dC}{dt}=0$ and $\frac{d\tau}{dt}=\frac{\beta}{\kappa+\kappa^{-1}}$. The former must be the case by definition, while the latter may be used to transform the time-averaged integral in (\ref{eq:VpTimeAvgd1}) into a $\tau$-averaged one. Inverting the definitions above, one may write $\theta_j$ and $\phi_j$ as:
\begin{subequations}
\begin{align}
\phi_j&=\sin^{-1}\Big[\frac{\cos\tau}{(\kappa^2-(\kappa^2-1) \cos^2\tau)^{1/2}}\Big],\\
\theta_j&=\cos^{-1}\Big[\frac{1}{(1+C^2\kappa^2-C^2(\kappa^2-1)\cos^2\tau)^{1/2}}\Big],
\end{align}  \label{eq:Ch3Spheroidangles}
\end{subequations}
which are used below in evaluating $\langle V_p \rangle$.

For $Re_p$ fixed, the Jeffery-averaged description of inertial migration, given by  (\ref{eq:VpTimeAvgd1}), breaks down for sufficiently small or large $\kappa$. This is because the leading order Jeffery angular velocity becomes small in these limits, being $O(\kappa^{-2})$ for fibers\,($\kappa \gg 1$) close to flow-alignment, and $O(\kappa^2)$ for flat disks\,($\kappa \ll 1$) close to alignment with the gradient-vorticity plane. As a result, the $O(Re_p)$ inertial correction becomes comparable in magnitude to the Jeffery contribution, leading to a slow down and eventual arrest of rotation, first shown by \cite{subkoch2005} for the case of a slender fiber. Herein, we will nevertheless confine ourselves to analyzing the Jeffery-averaged approximation, only noting that it remains valid provided $Re_p\,\kappa/\ln \kappa \ll 1\,(Re_p/\kappa^2 \ll 1)$ for $\kappa \gg 1\,(\kappa \ll 1)$, an increasingly restrictive assumption for extreme-aspect-ratio particles\citep{subkoch2005,navaneeth2017}. The consequence of an inertia-induced slow down at leading order, for the said particles, will be analyzed in a later communication.

A final point worth mentioning is that, for thin oblate spheroids with $\kappa < 0.14$, the aforementioned inertial drift time scale of $O(Re_p^{-1}H/V_\text{max})$ only applies in the absence of stochastic orientation fluctuations. In presence of such fluctuations, either of a thermal origin or otherwise\,(for instance, due to pair-hydrodynamic interactions or weak turbulence; see \cite{subramanian2022}), there is a barrier-hopping time associated with the eventual equilibration between the numbers of spinning and tumbling spheroids in a manner independent of the initial orientation distribution \citep{navaneeth2016,navaneethRapids,marath2018}. This time scale increases exponentially with decreasing amplitude of the fluctuations, becoming much longer than the nominal drift time scale, and likely  comparable to the time scale for inertial migration. Under these conditions, the migration dynamics will have a probabilistic rather than deterministic character, being described by a kinetic equation for the probability density that is a function of both $C$ and the transverse channel coordinate\,($s$) - we briefly revisit this issue when calculating the lift velocity for arbitrary $C$ later in this section.

\subsection{Scaling analysis and the Point-particle formulation} \label{sec:Ch3Scaling}

The dominant contribution to the volume integral in \eqref{eq:VpTimeAvgd1}, for $\lambda\ll1$, can arise from either scales of $O(L)$ or those of $O(H)$, the inertial screening length being irrelevant in the small-$Re_c$ limit. In order to assess the relative importance of these contributions, we consider the intermediate asymptotic interval $1\ll r\ll\lambda^{-1}$ ($r$ is measured in units of $L$), in which case $\bm{u}^{t2}\sim 1/r$, $\bm{u}_s\sim \beta/r^2+\gamma/r^3$, corresponding to the Stokeslet scaling for the test velocity field, and the stresslet-cum-force-quadrupole scaling for the velocity field in the actual problem. Using these $r$-scalings along with $\bm{u}^\infty\sim \beta r + \gamma r^2$ for the ambient flow, and separating the estimates for the linear and non-linear parts of the integrand in \eqref{eq:VpTimeAvgd1}, one obtains:
\begin{itemize}
	\item $\bm{u}^{t2}\cdot\left(\dfrac{\partial\bm{u}_s}{\partial t}+ \bm{u}_s\cdot\bm{\nabla u}^\infty+\,\,\bm{u}^\infty\cdot\bm{\nabla u}_s\right)\sim \dfrac{\beta^2}{r^3}+\dfrac{\beta\gamma}{r^2}+\dfrac{\gamma\beta}{r^4}+\dfrac{\gamma^2}{r^3}\,\,\,\,$ (linear in $\bm{u}_s$),
	\item $\bm{u}^{t2}\cdot(\bm{u}_s\cdot\bm{\nabla} \bm{u}_s)\sim \dfrac{\beta^2}{r^6}+\dfrac{\beta\gamma}{r^7}+\dfrac{\gamma^2}{r^8}\,\,\,\,\,\,\,$ (non-linear in $\bm{u}_s$).
\end{itemize}
Note that, over the range of scales under consideration, wall effects only contribute at a smaller order in $\lambda$, and hence, use of the unbounded domain estimates above. Next, using $dV\sim O(r^2 dr)$, one obtains the following estimates for the contributions of the different terms to the lift velocity integral:
\begin{subequations}
\begin{align}
&V_p^\text{linear}\sim Re_p\left(\beta^2\ln r+\beta\gamma r+\dfrac{\gamma\beta}{r}+\gamma^2\ln r\right), \\
&V_p^\text{non-linear}\sim Re_p\left(\frac{\beta^2}{r^3}+\dfrac{\beta\gamma}{r^4}+\dfrac{\gamma^2}{r^5}\right).
\end{align}  \label{eq:Ch3Vpscalings}
\end{subequations}
The algebraically growing contribution in (\ref{eq:Ch3Vpscalings}a) will be dominated by scales of $O(H)$ ($r\sim\lambda^{-1}$: the outer region), while contributions in (\ref{eq:Ch3Vpscalings}a,b) that decay with $r$ will be dominated by length scales of $O(L)$ ($r\sim O(1)$: the inner region). The algebraic growth with $r$ will be cut off for $r\gtrsim\lambda^{-1}$ due to the more rapid decay of the disturbance velocity fields induced by wall-induced screening\,(recall that this more rapid decay was used in neglecting the integral over $S_\infty$ in (\ref{eq:Ch3RT4})). Likewise, the apparent divergence for $r\to0$, for the algebraically decaying terms, will be cut off at $r \sim O(1)$ by the finite size of the spheroid. The terms proportional to $\ln r$ in (\ref{eq:Ch3Vpscalings}a) imply the dominance of the matching interval $1 \ll r \ll \lambda$, resulting in a leading order contribution proportional to $\ln \lambda^{-1}$, with logarithmically smaller 
contributions arising from the inner and outer regions. Use of these cutoffs leads to the estimates:
\begin{subequations}
	\begin{align}
	&V_p^\text{linear}\sim Re_p\left[\beta^2(1+\ln\lambda^{-1})+\beta\gamma\lambda^{-1}+\gamma\beta+\gamma^2(1+\ln\lambda^{-1})\right], \label{eq:Ch3Vpscalings2}\\
	&V_p^\text{non-linear}\sim Re_p\left(\beta^2+\beta\gamma+\gamma^2\right),
	\label{eq:Ch3VpscalingsNL2}
	\end{align}
\end{subequations}
for the contributions from terms linear and nonlinear in $\bm{u}_s$. Using the definitions of $\beta$ and $\gamma$ given below \eqref{eq:Ch3uamb}, and reverting to the dimensional form, imply a leading order lift velocity of the form $V_\text{max}^2 L^3/(\nu H^2)[\ln(H/a)+O(1)]$. Here, the contribution to the $O(1)$ term within brackets arises from the $\beta\gamma\lambda^{-1}$ term in \eqref{eq:Ch3Vpscalings2} pertaining to the outer region, and from the $\beta^2$ terms in both \eqref{eq:Ch3Vpscalings2} and \eqref{eq:Ch3VpscalingsNL2} pertaining, respectively, to the outer and inner regions; the logarithmically larger lift contribution arises from the $\beta^2\ln \lambda^{-1}$ term in \eqref{eq:Ch3Vpscalings2} that pertains to the matching region. 
However, the matching-region $\beta^2\ln \lambda^{-1}$ contribution and the inner-region $\beta^2$-contribution must both vanish by symmetry because they cannot involve boundaries, and therefore relate to the time-averaged lift on a neutrally buoyant spheroid in an unbounded simple shear flow. Since such a spheroid cannot exhibit a net lateral drift, only the outer-region $\beta^2$-contribution survives. This and the outer-region $\beta\gamma$-contribution both yield a dimensional lift velocity of $V_\text{max}^2L^3/(\nu H^2)$; in dimensionless terms, the Jeffery-averaged volume integral in \eqref{eq:VpTimeAvgd1} is $O(1)$, with the scaled lift velocity being $O(Re_p)$. More detailed arguments along these lines given in \cite{anandfinitesize2022} show that the next order contribution to the volume integral for a neutrally buoyant sphere arises from the inner region, involves an additional factor of $\lambda$, but leads to a  qualitative alteration of the lift velocity profiles for large $Re_c$. Owing to the acceleration terms in \eqref{eq:Ch3RTliftvel} being $O(\lambda^2)$, the aforementioned inner-region contribution will also be relevant to a neutrally buoyant spheroid.

Owing to the dominant contribution arising from length scales of $O(H)$ implied by the above arguments, the spheroids in the actual and test problems can be replaced by the corresponding point singularities. Thus, $\bm{u}_s$ and $\bm{u}^{t2}$ may be approximated as being induced by a time dependent stresslet\,($\bm{u}_\text{str}$) and a Stokeslet due to a point force directed along the negative gradient direction\,($\bm{u}_\text{St}$), respectively. The Jeffery-averaged lift velocity given by (\ref{eq:VpTimeAvgd1}), at leading order in $\lambda$, may therefore be written as:
\begin{align}
\langle V_p\rangle &=-Re_p\int_{V^F+V^P}\Big\langle\bm{u}_\text{St}\cdot\Big(\dfrac{\partial\bm{u}_\text{str}}{\partial t}+ \langle\bm{u}_\text{str}\rangle \cdot\nabla\bm{u}^\infty+\bm{u}^\infty\cdot\bm{\nabla}\langle \bm{u}_\text{str}\rangle\Big)\Big\rangle\,d\bm{r}. \label{eq:Ch3VpOuterTimeAvgFull}
\end{align}
where, on account of the subdominant nature of the length scales of $O(L)$, the domain of integration has now been extended to include the particle volume $V^P$. Thus, \eqref{eq:Ch3VpOuterTimeAvgFull} is the leading order point particle approximation for the lift velocity for $Re_c\ll1$. 

  In \eqref{eq:Ch3VpOuterTimeAvgFull}, $\bm{u}_\text{St}$ is time independent since the unit force points in a fixed direction, despite the changing orientation of the test spheroid. Further,  $\langle\partial\bm{u}_\text{str}/\partial t\rangle=0$ for rotation along Jeffery orbits. Also noting that the ambient flow is steady in the chosen non-rotating reference frame, \eqref{eq:Ch3VpOuterTimeAvgFull} reduces to:
\begin{align}
\langle V_p\rangle &=-Re_p\int_{V^F+V^P}\bm{u}_\text{St}\cdot(\langle\bm{u}_\text{str}\rangle\cdot\bm{\nabla u}^\infty+\bm{u}^\infty\cdot\bm{\nabla}\langle \bm{u}_\text{str}\rangle)\,d\bm{r}. \label{eq:Ch3VpOuterTimeAvg}
\end{align}
Using $r_2 \sim O(\lambda^{-1})$ in \eqref{eq:VambFaxen} on account of the outer-region dominance, the Faxen's correction to spheroid translation turns out to be $O(\lambda^2)$ smaller than the terms linear and quadratic in $r_2$, and one may therefore approximate the ambient flow   in \eqref{eq:Ch3VpOuterTimeAvg} as $\bm{u}^\infty\approx(\beta r_2 +\gamma r_2^2)\bm{1}_1$. 

With the spheroid volume neglected, the volume integral in \eqref{eq:Ch3VpOuterTimeAvg} is most easily evaluated by Fourier transforming the flow\,($r_1$) and vorticity\,($r_3$) coordinates, the partial Fourier transform being defined as:
\begin{align}
\hat{f}(k_1,r_2,k_3) =\int_{-\infty}^\infty \int_{-\infty}^\infty dr_1 dr_3\,\,e^{\iota(k_1 r_1+k_3 r_3)}\,\,f(r_1,r_2,r_3).
\label{eq:Ch3FTdefn}
\end{align}
Applying the convolution theorem along these coordinates \citep{arfkenweber}, and substituting the above approximate form of $\bm{u}^\infty$ in \eqref{eq:Ch3VpOuterTimeAvg}, one obtains:
\begin{align}
\langle V_p\rangle =-\frac{Re_p}{4\pi^2}\int_{-s\lambda^{-1}}^{(1-s)\lambda^{-1}} dr_2\int d\bm{k}_\perp\,\,\hat{\bm{u}}_\text{St}(-\bm{k}_\perp,r_2;y_2)\cdot
\big[&\langle\hat{\bm{u}}_\text{str}\rangle(\bm{k}_\perp,r_2;y_2)\cdot\bm{1}_2(\beta+2\gamma r_2)\bm{1}_1\nonumber\\
-\iota k_1(\beta r_2 +\gamma r_2^2)& \langle\hat{\bm{u}}_\text{str}\rangle(\bm{k}_\perp,r_2;y_2)\big], \label{eq:Ch3VpOuterTimeAvgFT}
\end{align}
in terms of the partial Fourier transforms of the Stokeslet and stresslet velocity fields; here, $\bm{k}_\perp \equiv (k_1,k_3)$ and $y_2 =s\lambda^{-1}$ is the transverse distance of the stresslet from the lower wall. The expressions for $\hat{\bm{u}}_\text{St}$ is derived in Appendix A, while that for $\langle\hat{\bm{u}}_\text{str}\rangle$ is developed in the next subsection.

\subsection{\texorpdfstring{Solution for $(\langle\bm{u}_\text{str}\rangle$, $\langle p_\text{str}\rangle)$}{Solution for the velocity field due to bounded Stresslet}}\label{sec:Ch3boundedstresslet}

The Jeffery-averaged disturbance fields ($\langle\bm{u}_\text{str}\rangle$, $\langle p_\text{str}\rangle$), that appear in the point-particle approximation of the inertial lift velocity given by (\ref{eq:Ch3VpOuterTimeAvg}), satisfy:
    \begin{subequations} \label{eq:Ch3StressletEqn}
    	\begin{align}
    	\nabla^2 \langle\bm{u}_\text{str}\rangle-\bm{\nabla}\langle p_\text{str}\rangle&=\beta\langle\bm{S}(\bm{p})\rangle\cdot \bm{\nabla}\delta(\bm{r}),\\
    	\bm{\nabla}\cdot\langle\bm{u}_\text{str}\rangle&=0,
    	\end{align}  \label{stresslet:eqns}
    \end{subequations}
    with the boundary conditions: 
    \begin{subequations} \label{eq:Ch3StressletBC}
    	\begin{align}
    	\langle\bm{u}_\text{str}\rangle&=0 \text{ at } r_2=-s\lambda^{-1}, (1-s)\lambda^{-1},\\    	\langle\bm{u}_\text{str}\rangle&\rightarrow 0 \text{ for } r_1,r_3\rightarrow\infty\,(r_2\hspace*{0.05in}\text{fixed}),
    	\end{align}
    \end{subequations}
where the stresslet singularity on the RHS of (\ref{stresslet:eqns}a) approximates the torque-free neutrally buoyant spheroid, in an unbounded simple shear flow, on scales much larger than $O(L)$. The stresslet coefficient $\bm{S}$ is a function of the spheroid orientation $\bm{p}$, being given by\citep{navaneeth2017}:
\begin{align}
\bm{S}(\bm{p})&= A_1 \frac{3}{2}(\bm{E}:\bm{pp})\left(\bm{pp}-\frac{\bm{I}}{3}\right) + A_2 ((\bm{I}-\bm{pp})\cdot\bm{E}\cdot\bm{pp}+\bm{pp}\cdot\bm{E}\cdot(\bm{I}-\bm{pp}))\nonumber\\& +A_3\left((\bm{I}-\bm{pp})\cdot\bm{E}\cdot(\bm{I}-\bm{pp})+(\bm{I}-\bm{pp})\frac{\bm{E}:\bm{pp}}{2}\right),
\label{eq:Ch3DipoleStrengthProlate}
\end{align}
where $E_{ij}=\frac{\beta}{2}(\delta_{i1}\delta_{j2}+\delta_{i2}\delta_{j1})$ is the rate of strain tensor associated with the local simple shear. The simple shear flow may be resolved into an axisymmetric extension aligned with $\bm{p}$, longitudinal planar extensions in a pair of orthogonal planes containing $\bm{p}$, and a pair of transverse planar extensions in the plane perpendicular to $\bm{p}$\citep{subkoch2006,navaneeth2016}. The $A_i$'s in \eqref{eq:Ch3DipoleStrengthProlate} are the $\kappa$-dependent stresslet amplitudes corresponding to the aforesaid component flows, there being only three of these\,(in contrast to the five component flows) owing to the axisymmetry of the spheroidal geometry\citep{kimkarrila,navaneeth2017}. For $\kappa>1$, these amplitudes are given by:
\begin{align}
A_1&=-\frac{16\pi(\kappa^2-1)^{5/2}}{9\kappa^3[-3(\kappa^2-1)^{1/2}\kappa+2\kappa^2 \cosh^{-1}(\kappa)+\cosh ^{-1}(\kappa)]},\\
A_2&=-\frac{16\pi(\kappa^2-1)^3}{3\kappa^2(\kappa^2+1)(\kappa^4+\kappa^2-3(\kappa^2-1)^{1/2}\kappa\cosh^{-1}(\kappa)-2)},\\
A_3&=-\frac{32\pi(\kappa^2-1)^3}{3\kappa^3(2\kappa^5-7\kappa^3+3(\kappa^2-1)^{1/2}\cosh^{-1}(\kappa)+5\kappa)}.
\end{align}
The corresponding expressions for an oblate spheroid ($\kappa<1$), may be obtained by first substituting $\kappa=\xi_0/(\xi_0^2-1)^{1/2}$ in terms of the coordinate $\xi_0\,(>1)$ labeling the surface of the spheroid, and then using the transformation $d\to-\iota d, \xi_0\to\iota(\xi_0^2-1)^{1/2}$ in the dimensional form of the stresslet obtained from multiplying $\bm{S}$ above by $\mu L^3 V_\text{max}/H$ with $L= d\xi_0$\citep{navaneeth2017}. Note that $\bm{p}$, and thence $\bm{S}$, is a function of time owing to rotation along Jeffery orbits. In the limit of a sphere, $A_1 = A_2 = A_3 = -\frac{20\pi}{3}$, and $\bm{S}$ reduces to $-20\pi\bm{E}/3$, independent of time, corresponding to the stresslet induced by a freely rotating sphere that leads to the well known Einstein coefficient in the suspension viscosity.

While one may, in principle, solve for ($\langle\bm{u}_\text{str}\rangle$, $\langle p_\text{str}\rangle$) in a manner similar to that for the bounded-domain Stokeslet fields\,(see Appendix \ref{App:A}) by defining the disturbance fields as the sum of an unbounded-domain contribution, and a contribution that accounts for the confinement induced by channel walls, one may also obtain them directly as a gradient of the bounded-domain Stokeslet fields, derived in Appendix \ref{App:A}; the gradient is with respect to the location $\bm{y}$ of the Stokeslet\,\citep{brady2010}. Therefore,
\begin{align}
\langle\bm{u}_\text{str}\rangle=\beta\langle\bm{S}\rangle:\frac{\partial\bm{J}}{\partial\bm{y}}.
\label{eq:Ch3stressletForm}
\end{align}
where the bounded-domain Stokeslet field corresponding to a point force $\bm{F}$ is given by $\bm{J}\cdot \bm{F}$; the Fourier transform\,($\hat{\bm{J}}$) of the tensor $\bm{J}$ is defined in Appendix \ref{App:A}\,(see text below \eqref{eq:Ch3BoundedStokesletfield}). Since it is $\langle \hat{\bm{u}}_\text{str} \rangle$ that appears in the final expression for the point-particle approximation of the 
lift velocity, \eqref{eq:Ch3VpOuterTimeAvgFT}, we Fourier transform \eqref{eq:Ch3stressletForm} to obtain:
\begin{align}
\langle\hat{u}_{\text{str},i}\rangle=\beta \langle S_{jm}\rangle\hat{N}_{ijm},
\label{eq:Ch3stressletFT}
\end{align}
where the third-order tensor $\hat{N}_{ijm}$ is defined as:
\begin{align}
\hat{N}_{ijm}=\iota\hat{J}_{im}(k_1 \delta_{j1}+k_3 \delta_{j3})+\dfrac{\partial\hat{J}_{im}}{\partial y_2}\delta_{j2}.
\end{align}
In \eqref{eq:Ch3stressletFT}, $\langle\bm{S}\rangle=\frac{1}{2\pi}\int_0^{2\pi} \bm{S}\,d\tau$ denotes the time-averaged stresslet corresponding to a neutrally buoyant spheroid rotating along a given Jeffery orbit in an unbounded simple shear flow\,(with unit velocity gradient). The expression for $\langle \bm{S}\rangle$ is given below in the next subsection.


\subsection{The time-averaged lift velocity $\langle V_p \rangle$} \label{sec:avglift_Calcn}

Using \eqref{eq:Ch3stressletFT} for the stresslet velocity field in \eqref{eq:Ch3VpOuterTimeAvgFT}, the Jeffery-averaged lift velocity takes the form:
\begin{align}
\langle V_p\rangle =-\frac{Re_p}{4\pi^2}\int_{-s\lambda^{-1}}^{(1-s)\lambda^{-1}} dr_2\int d\bm{k}_\perp\,\, &\hat{\bm{u}}_\text{St}(-\bm{k}_\perp,r_2;y_2)\cdot
\big[\beta\bm{1}_2\cdot\hat{\bm{N}}:\langle\bm{S}\rangle(\bm{k}_\perp,r_2;y_2)(\beta+2\gamma r_2)\bm{1}_1\nonumber\\
&-\iota k_1(\beta r_2 +\gamma r_2^2) \beta\hat{\bm{N}}:\langle\bm{S}\rangle(\bm{k}_\perp,r_2;y_2)\big]. \label{eq:Ch3VpTimeAvg1}
\end{align}
Note that the averaging operation\,($\langle.\rangle$) in \eqref{eq:Ch3VpTimeAvg1} is independent of the shear rate since it involves the scaled Jeffery angular velocity which is only a function of $\kappa$ and $\phi_j$ for the inertially stabilized orbits. Thus, $\langle \bm{S}\rangle$ must be linear in $\bm{E}$, with a prefactor that is a function of $C$ and $\kappa$. For simple shear flow, this implies $\langle S_{12}\rangle=\langle S_{21}\rangle$ are the only non-zero components of the Jeffery-averaged stresslet. The following expressions for these components may be obtained by starting from (\ref{eq:Ch3DipoleStrengthProlate}), with $\bm{p}$ expressed in terms of $C$ and $\tau$ using (\ref{eq:Ch3Spheroidangles}ab), followed by an integration over $\tau$:
\begin{align}
\langle S_{12}\rangle&=\frac{1}{4(\kappa^2-1)^2[(C^2+1)(C^2\kappa^2+1)]^{1/2}}\Bigg\{3 A_1 \kappa^2\bigg(2+C^2(\kappa^2+1)\nonumber\\
&-2 [(C^2+1) (C^2 \kappa^2+1)]^{1/2}\bigg)+2 A_2(\kappa^2+1)\bigg(\kappa^2 \Big([(C^2+1) (C^2 \kappa^2+1)]^{1/2}-2 C^2-1\Big)\nonumber\\
&+[(C^2+1) (C^2 \kappa^2+1)]^{1/2}-1\bigg)+A_3 \bigg(\kappa^4(C^2+2)+\kappa^2 (-2 [(C^2+1) (C^2 \kappa^2+1)]^{1/2}\nonumber\\
&+C^2-2)+2\bigg)\Bigg\}, \label{eq:Ch3TimeAvgDipoleStrength}
\end{align}
for prolate spheroids, and
\begin{align}
\langle S_{12}\rangle &= \frac{1}{4(\kappa^2-1)^2(C^2+1)(C^2\kappa^2+1)}\Bigg\{3 A_1 \kappa^2\bigg(-2C^4\kappa^2 +C^2(\kappa^2+1)(-2\nonumber\\
&+[(C^2+1)(C^2\kappa^2+1)]^{1/2})+2 (-1+[(C^2+1)(C^2\kappa^2+1)]^{1/2})\bigg)+2 A_2(\kappa^2+1)\nonumber\\
&\bigg(C^4 (\kappa^2+\kappa^4)-(1+\kappa^2)(-1+[(C^2+1)(C^2\kappa^2+1)]^{1/2})+C^2(1+\kappa^2(2+\kappa^2\nonumber\\
&-2[(C^2+1)(C^2\kappa^2+1)]^{1/2}))\bigg)+A_3 \bigg(2[(C^2+1)(C^2\kappa^2+1)]^{1/2}+\kappa^2(-2-2C^4\kappa^2\nonumber\\
&+2(\kappa^2-1)[(C^2+1)(C^2\kappa^2+1)]^{1/2})+C^2(1+\kappa^2)(-2+[(C^2+1)(C^2\kappa^2+1)]^{1/2})\bigg)\Bigg\},\label{eq:Ch3TimeAvgDipoleStrength2}
\end{align}
for oblate spheroids. For both \eqref{eq:Ch3TimeAvgDipoleStrength} and \eqref{eq:Ch3TimeAvgDipoleStrength2}$, \lim_{\kappa \rightarrow 1} \langle S_{12}\rangle =-10\pi/3$ independent of $C$, which yields the Einstein coefficient\,($5/2$) in the $O(\phi)$ contribution to the suspension viscosity, $\phi$ being the sphere volume fraction; the $C$-independence arises owing to the sphere orientation being a degenerate degree of freedom. At the other extreme, one has $\lim_{\kappa \to \infty} \langle S_{12}\rangle|_{C=\infty}=-2\pi/(3\kappa\ln\kappa)$ in \eqref{eq:Ch3TimeAvgDipoleStrength}. Here, the factor $1/\ln\kappa$ arises from viscous slender body theory\citep{subkoch2005}, while the additional $\kappa^{-1}$-factor reflects the probability of occurrence of non-aligned orientations that contribute dominantly to the viscosity of a dilute non-interacting suspension of slender fibers\,\citep{leal1971,navaneeth2016}.

For times much longer than $O(Re_p^{-1}H/V_\text{max})$, only the inertially stabilized orbits are relevant to cross-stream migration. As mentioned earlier, for prolate spheroids, this is the tumbling orbit\,($C=\infty$), in which case, $\langle S_{12}\rangle$ in (\ref{eq:Ch3TimeAvgDipoleStrength}) reduces to:
\begin{align}
\langle S_{12}\rangle|_{C=\infty} &= \frac{ (3 A_1+A_3)\kappa+2 A_2 (\kappa ^2+1)}{4 (\kappa +1)^2}. \label{tumbling_prolatestresslet}
\end{align}
For oblate spheroids, one can have either the spinning\,($C=0$) or tumbling\,($C=\infty$) orbits depending on $\kappa$, and accordingly, $\langle S_{12}\rangle$ in (\ref{eq:Ch3TimeAvgDipoleStrength2}) reduces to:
\begin{align}
\langle S_{12}\rangle|_{C=\infty} &= \frac{ (3 A_1+A_3)\kappa+2 A_2 (\kappa ^2+1)}{4 (\kappa +1)^2},\label{tumbling_oblatestresslet}\\
\langle S_{12}\rangle|_{C=0} &= \frac{A_3}{2}. \label{spinning_oblatestresslet}
\end{align}
\eqref{tumbling_prolatestresslet}-\eqref{spinning_oblatestresslet} will be used to determine $\langle V_p \rangle$ below. In light of the above, one may write (\ref{eq:Ch3stressletFT}) in the form:
\begin{align}
\langle\hat{u}_{\text{str},i}\rangle=\beta \langle S_{12}\rangle\,(\hat{N}_{i12}+\hat{N}_{i21}),
\label{eq:Ch3TimeAveragedActualVelocity}
\end{align}
with the Jeffery-averaged lift velocity given by: 
\begin{align}
\langle V_p\rangle =&-\frac{Re_p\langle S_{12}\rangle} {4\pi^2} \int_{-s\lambda^{-1}}^{(1-s)\lambda^{-1}} dr_2\int d\bm{k}_\perp\,\, \hat{u}_{\text{St},i} (-\bm{k}_\perp,r_2;y_2) \nonumber\\
&\big[\beta(\beta+2\gamma r_2)(\hat{N}_{212}+\hat{N}_{221}) (\bm{k}_\perp,r_2;y_2)\delta_{i1}-\iota \beta k_1(\beta r_2 +\gamma r_2^2) (\hat{N}_{i12}+\hat{N}_{i21}) (\bm{k}_\perp,r_2;y_2)\big]. \label{eq:Ch3VpTimeAvg2}
\end{align}
Since the dominant contributions to the integral in (\ref{eq:Ch3VpTimeAvg2}) come from scales of $O(H)$\,(or, equivalently, $k_\perp \sim O(H^{-1})$), it is natural to transform  to rescaled variables $\bm{k}_\perp=\bm{k}_\perp''\lambda, r_2=r_2''/\lambda$. The original disturbance fields may be written in terms of the rescaled ones as: $\hat{\bm{u}}_\text{St} =\hat{\bm{u}}_\text{St}^{''} /\lambda$, $\hat{\bm{N}}=\hat{\bm{N}}''$, and further, noting that $y_2''= s$ and $d\bm{k}_\perp\,dr_2\equiv \lambda d\bm{k}_\perp''dr_2''$, one obtains,
\begin{align}
\langle V_p\rangle =&-\frac{Re_p\langle S_{12}\rangle} {4\pi^2} \int_{-s}^{1-s} dr_2''\int d\bm{k}''_\perp\,\, \hat{u}_{\text{St},i}^{''} (-\bm{k}''_\perp,r_2'';s) \nonumber\\
&\big[\beta(\beta+2\gamma'' r_2'')(\hat{N}_{212}''+\hat{N}_{221}'') (\bm{k}''_\perp,r_2'';s)\delta_{i1}-\iota \beta k_1''(\beta r_2'' +\gamma'' r_2''^2) (\hat{N}_{i12}''+\hat{N}_{i21}'') (\bm{k}''_\perp,r_2'';s)\big], 
\label{eq:Ch3VpTimeAvg3}
\end{align}
where $\gamma''=\gamma\lambda^{-1}=-4$, and $\langle S_{12} \rangle$ is given by \eqref{tumbling_prolatestresslet}, or \eqref{tumbling_oblatestresslet} and \eqref{spinning_oblatestresslet}, depending on $\kappa$. The essential consequence of the rescaling above is to show that the volume integral in \eqref{eq:Ch3VpTimeAvg3} is independent of $\lambda$. In fact, the integral is only a function of $s$\,(the particle location in the channel; see Figure \ref{fig:Ch3channelGeometry}), with the additional dependence on $\kappa$ entirely contained in $\langle S_{12} \rangle$. Accounting for the scaling $V_\text{max}\lambda$ used for $V_p$, \eqref{eq:Ch3VpTimeAvg3} shows that the point-particle framework leads to a lift velocity of $O(V_\text{max}\lambda Re_p)$, which was used earlier in section \ref{sec:Ch3GRT} to estimate the inertial migration time scale\,($t_\text{lift}$). 

A consequence of the integral in \eqref{eq:Ch3VpTimeAvg3} being independent of $\kappa$ is that, within the Jeffery-averaged framework used, the inertial lift velocity of a sphere, and the time-averaged lift velocity of a spheroid, only differ by a multiplicative function of $C$ and $\kappa$ in general, and only by a function of $\kappa$ if one assumes the spheroid to have settled onto the inertially stabilized Jeffery orbit 
- this multiplicative function is the ratio of the Jeffery-averaged spheroid stresslet to the sphere stresslet, being equal to $-\frac{3\langle S_{12} \rangle}{10\pi}$. An immediate consequence of this proportionality relation is that the shapes of the lift velocity profiles for the two cases must be identical, with the zero-crossings in particular\,(that correspond to the Segre-Silberberg equilibria for  a sphere) being identical. In other words, a change in $\kappa$ only affects the magnitude of the inertial lift, the equilibrium positions being unaffected. In light of the obvious importance of the above conclusion for shape-sorting applications, it is worth documenting the reasons that leads to the simple proportionality relation:
\begin{enumerate}
	\item The dominant contribution from the linearized inertial terms on scales of $O(H)$. The latter outer-region dominance led to the point-particle approximation, and thence, to the time independence of the test velocity field in the reciprocal theorem volume integral. 
        \item The asymptotic smallness, for small confinement ratios, of the additional contributions arising from the translational and angular accelerations associated with the motion of the neutrally buoyant spheroid in the Stokes limit, 
	\item The approximation of the time averaging based on the $Re_p-$independent Jeffery angular velocity, which leads to the time-averaged spheroid stresslet being proportional to $\bm{E}$\,(just as the sphere stresslet).
\end{enumerate}

The Jeffery-averaged lift velocity in (\ref{eq:Ch3VpTimeAvg3}) may be written in the compact form:
\begin{align}
\langle V_p\rangle=Re_p\langle S_{12}\rangle(\kappa)\Big(\beta^2 F(s)+\beta\gamma''G(s)\Big),
\label{eq:Ch3VpExpression}
\end{align}
where the functions $F(s)$ and $G(s)$ are given by:
\begin{align}
F(s)=&-\frac{1} {4\pi^2} \int_{-s}^{1-s} dr_2''\int d\bm{k}''_\perp\,\, \hat{u}_{\text{St},i}^{''} (-\bm{k}''_\perp,r_2'';s) \nonumber\\
&\big[(\hat{N}_{212}''+\hat{N}_{221}'') (\bm{k}''_\perp,r_2'';s)\delta_{i1}-\iota k_1'' r_2'' (\hat{N}_{i12}''+\hat{N}_{i21}'') (\bm{k}''_\perp,r_2'';s)\big], \label{eq:Ch3FsPrimitive}\\
G(s)=&-\frac{1} {4\pi^2} \int_{-s}^{1-s} dr_2''\int d\bm{k}''_\perp\,\, \hat{u}_{\text{St},i}^{''} (-\bm{k}''_\perp,r_2'';s) \nonumber\\
&\big[2 r_2''(\hat{N}_{212}''+\hat{N}_{221}'') (\bm{k}''_\perp,r_2'';s)\delta_{i1}-\iota k_1''r_2''^2 (\hat{N}_{i12}''+\hat{N}_{i21}'') (\bm{k}''_\perp,r_2'';s)\big], \label{eq:Ch3GsPrimitive}
\end{align}
and satisfy $F(s)=-F(1-s)$ and $G(s)=G(1-s)$, consistent with the antisymmetry of the lift velocity profile about the channel centerline.

In  (\ref{eq:Ch3VpExpression}), the $\beta^2$ contribution characterizes the effect of the asymmetrically located walls on the disturbance stresslet, and leads to migration away from the walls. In contrast, the $\beta\gamma''$ contribution, which characterizes the interaction of the stresslet with the ambient profile curvature, causes migration away from the channel centerline. The Segre-Silberberg equilibria emerge from a balance between these two opposing effects. Substituting $\beta=4(1-2s)$ and $\gamma''=-4$ in \eqref{eq:Ch3VpExpression}, the final expression for the time-averaged lift velocity for a neutrally buoyant spheroid of an arbitrary aspect ratio $\kappa$, for $Re_c\ll1$, is given by:
\begin{align}
\langle V_p\rangle=Re_p\langle S_{12}\rangle(\kappa)\big[16(1-2s)^2 F(s)-16(1-2s)G(s)\big].
\label{eq:Ch3VpFinalExpressionsmallRec}
\end{align}

The Fourier integrals in \eqref{eq:Ch3FsPrimitive} and \eqref{eq:Ch3GsPrimitive} can be calculated using plane polar coordinates, $k_1''=k_\perp''\cos\phi$, $k_3''=k_\perp''\sin\phi$, with $d\bm{k}_\perp''\equiv k_\perp''dk_\perp''d\phi$; here $k_\perp''\in[0,\infty)$ and $\phi\in[0,2\pi]$. The integrals over the transverse coordinate $r_2''$ and the azimuthal angle $\phi$ may be done analytically, reducing \eqref{eq:Ch3FsPrimitive} and \eqref{eq:Ch3GsPrimitive} to the following one-dimensional integrals:
\begin{align}
F(s)&=\int_0^\infty dk_\perp'' \dfrac{k_\perp''\,\,e^{-k_\perp'' (27 s+16)} I(k_\perp'',s)}{48 \pi \left(e^{2 k_\perp''}-1\right) \left[-2 e^{2 k_\perp''} \left(2 k_\perp''^2+1\right)+e^{4 k_\perp''}+1\right]^2}\label{eq:Ch3Fs},\\ 
G(s)&=\int_0^\infty dk_\perp'' \dfrac{e^{-k_\perp'' (27 s+16)} J(k_\perp'',s)}{192 \pi\, k_\perp''^2 \left(e^{2 k_\perp''}-1\right) \left[-2 e^{2 k_\perp''} \left(2 k_\perp''^2+1\right)+e^{4 k_\perp''}+1\right]^2}\label{eq:Ch3Gs},
\end{align}
with $I(k_\perp'',s)$ and $J(k_\perp'',s)$ defined in Appendix \ref{App:B}. The $k_\perp''$-integrals above are evaluated numerically for various $s$, using Gauss-Legendre quadrature, after replacing the infinite interval with a finite one - $(0,K_\text{max})$. Numerical convergence, especially near the channel walls, depends sensitively on $K_\text{max}$. As the spheroid approaches either wall, the contribution to the $k_\perp''$-integrals comes from progressively smaller scales\,(compared to $H$) or, equivalently, from larger and larger $k_\perp''$. In fact, the relevant scale changes from $H$ ($k_\perp''\sim O(1)$) to the distance from the wall - either $s$ (lower wall) or $1-s$ (upper wall). To analyze the lift close to the lower wall, for example, one defines a rescaled wavenumber $k_\perp''=k_w/s$; next, expanding the integrands in \eqref{eq:Ch3Fs} and \eqref{eq:Ch3Gs} for $s\to0$ with $k_w$ fixed, the near-wall lift velocity takes the form:
\begin{align}
\lim_{s\to 0}\,\,\langle V_p\rangle=\langle V_p\rangle^\text{wall}= -\frac{Re_p\langle S_{12}\rangle(\kappa)}{3\pi} \int_0^\infty dk_w\,\, e^{-2 k_w} k_w (3 k_w^2-2 k_w+3).
\label{eq:Ch3VpnearWall}
\end{align}
The limiting value near the upper wall is equal in magnitude, but of an opposite sign, as must be the case by symmetry. The integral in \eqref{eq:Ch3VpnearWall} can be evaluated analytically, and yields 
\begin{align}
\langle V_p\rangle^\text{wall}= \pm \dfrac{11 Re_p\langle S_{12}\rangle(\kappa)}{24\pi},
\label{eq:Ch3VpnearWall2}
\end{align}
with the upper and lower signs pertaining to the corresponding channel wall.
For a sphere, (\ref{eq:Ch3VpnearWall2}) reduces to $55 Re_p/36$, a value originally given by \cite{vasseur1976}, albeit without an accompanying explanation.

 The near-wall value above is proportional to $\beta^2$, implying that the inertial lift in this limit is generated due to interactions on scales of order the spheroid-wall separation, and the sign points to a repulsion. The curvature-induced $\beta\gamma''$ contribution is asymptotically small in this limit, implying that the near-wall lift at leading order may be determined independently from considering a sphere or a spheroid in the vicinity of a single plane wall subject to a linear shearing flow. Another important point is that the near-wall lift velocity approaches a finite value, rather than zero, as one approaches the channel walls. The latter must be the case on account of the eventual diverging lubrication resistance associated with the thin intervening fluid layer between the particle and the wall, and the discrepancy is on account of the regime of validity of the present calculation. As stated in the problem definition in  $\S$\ref{sec:Ch3formulation}, the near-wall limit above pertains to spheroid-wall separations that, although much smaller than $O(H)$ are nevertheless much larger than $L$. We comment further on the nature of the lift force profile for small separations in $\S$\ref{sec:Ch3Reclarge}.

Based on the discussion above, it is clear that the choice of any finite $K_\text{max}$, however large, only leads to converged lift velocities down to a certain nonzero $s$. In the results shown in the next subsection, we have chosen $K_\text{max}=10^4$ to ensure accuracy down to $s=0.001$ without the aid of a large-$k_\perp''$ asymptote; this choice also enables a close approach to the near-wall limiting value obtained above. With this choice, $200$ quadrature points are sufficient to obtain converged results for arbitrary $\kappa$.

\subsection{Results and Discussion} \label{sec:smallRecresults}

\begin{figure}
	\centering
	\includegraphics[width=\textwidth]{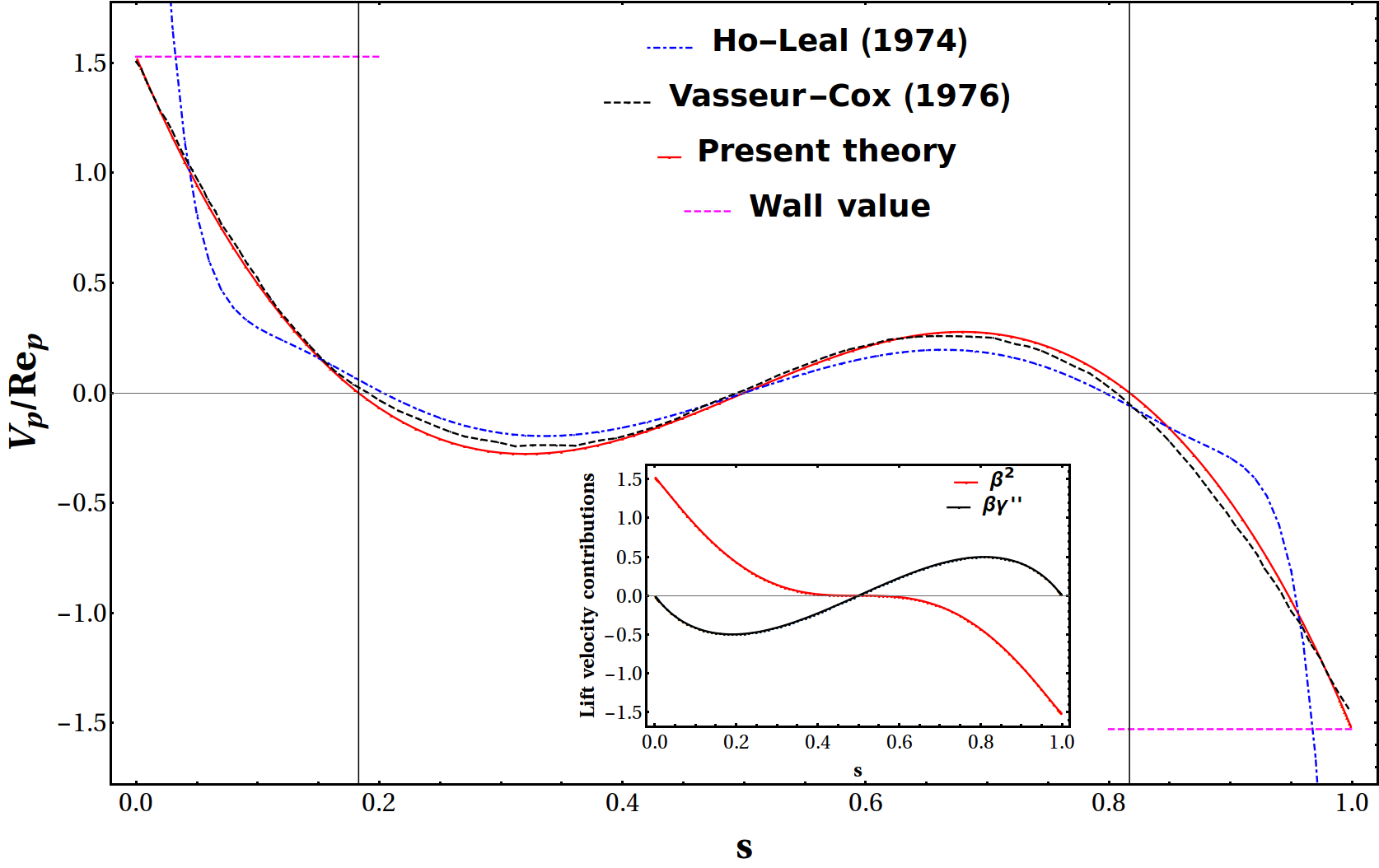}
	\caption{Comparison of the small-$Re_c$ lift profile for a sphere obtained from (\ref{eq:Ch3GsPrimitive})  with profiles extracted from \cite{holeal1974} and \cite{vasseur1976}. The vertical lines mark the Segre-Silberberg equilibria\,($s\approx0.182$ and $0.818$) on either side of the centerline; the horizontal dotted lines denote the near-wall limiting values given by \eqref{eq:Ch3VpnearWall2}. The inset shows the profiles of the component $\beta^2$ and $\beta\gamma''$ contributions.}
	\label{fig:Ch3smallRecComparison}
\end{figure}

In Figure \ref{fig:Ch3smallRecComparison}, we compare the lift velocity profile for a sphere, obtained by taking $\langle S_{12} \rangle = -\frac{10\pi}{3}$  in \eqref{eq:Ch3VpFinalExpressionsmallRec}, with profiles plotted using the data from figure 2 of \cite{holeal1974} and figure 8 of \cite{vasseur1976}; the data was obtained by digitizing the latter figure. There is a clear mismatch between our profile and the Ho-Leal one, an aspect that we will comment on in the conclusions. The result of \cite{vasseur1976} shows good agreement with the present calculation throughout the channel. The Segre-Silberberg equilibria, corresponding to the zeroes of $\langle V_p\rangle$, are located at $s\approx 0.182$ and $0.818$; these are at a (dimensional)\,distance of $0.636\times H/2$ from the centerline, which agrees well with the intermediate annulus at $\sim0.6\times$pipe radius in the original experiments\citep{segresilberberg1962a}. Note that the lift velocity profiles in both our and the \cite{vasseur1976} analysis asymptote to the wall values $\pm 55/36$\,(the horizontal dashed magenta lines in Figure \ref{fig:Ch3smallRecComparison}) given in the previous subsection. As already argued therein, $H$ is no longer the relevant scale\,(at leading order) close to the wall. This implies that $Re_c$ should also not be relevant, and the near-wall lift values should therefore be independent of $Re_c$. This is validated in the next section where the lift velocity profiles are seen to approach (\ref{eq:Ch3VpnearWall2}) for $s \to 0,1$ even for $Re_c\gtrsim O(1)$, although this approach occurs in a shrinking neighborhood of the wall with increasing  $Re_c$. The inset in Figure \ref{fig:Ch3smallRecComparison} plots the $\beta^2$ and $\beta\gamma''$ contributions to the inertial lift, confirming that the latter curvature-induced contribution becomes vanishingly small at the walls.

\begin{figure}
	\centering
    \begin{subfigure}[b]{0.49\textwidth}
		\includegraphics[width=\textwidth]{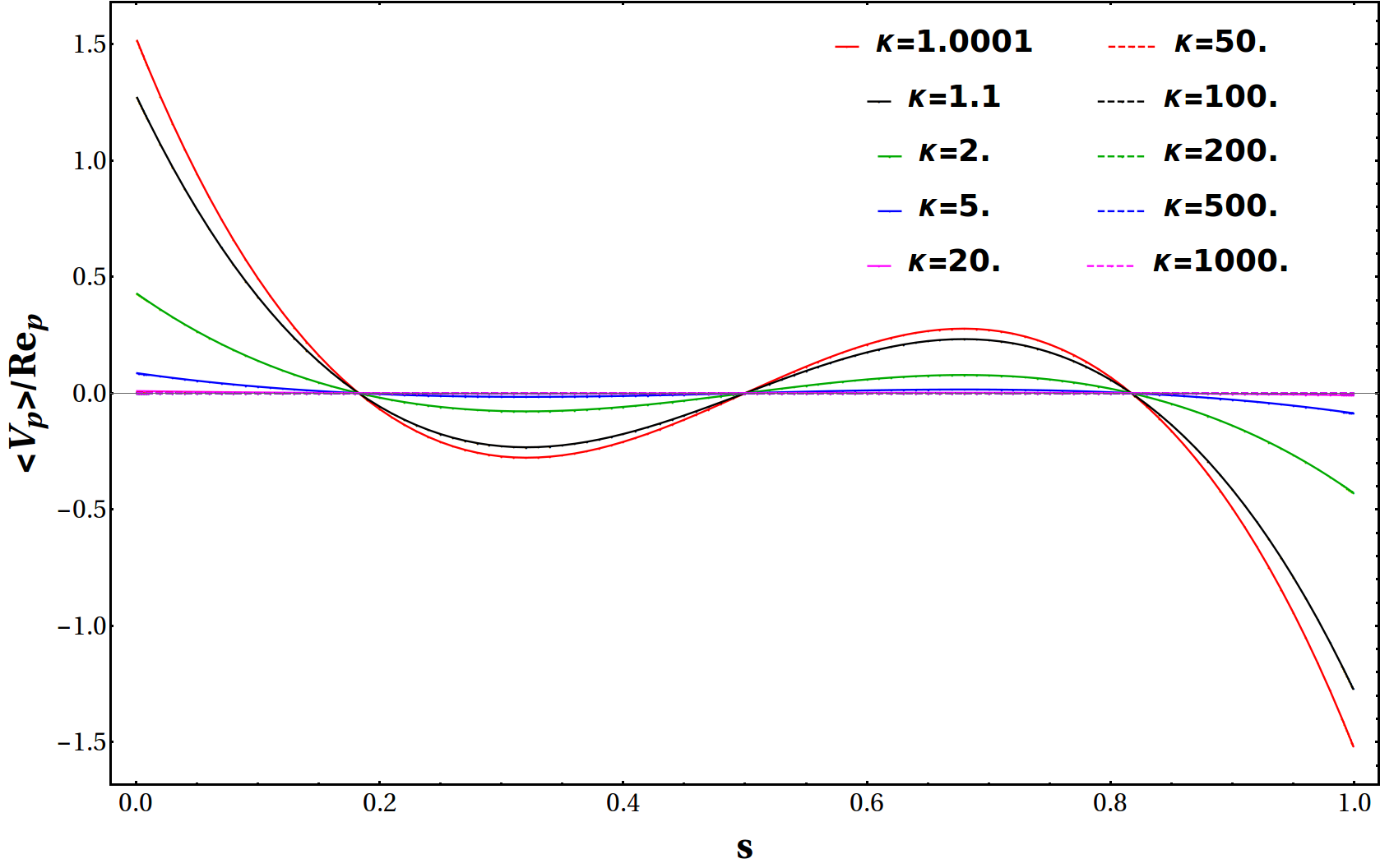}
        \caption{}
    \end{subfigure}
    \hfill
    \begin{subfigure}[b]{0.49\textwidth}
		\includegraphics[width=\textwidth]{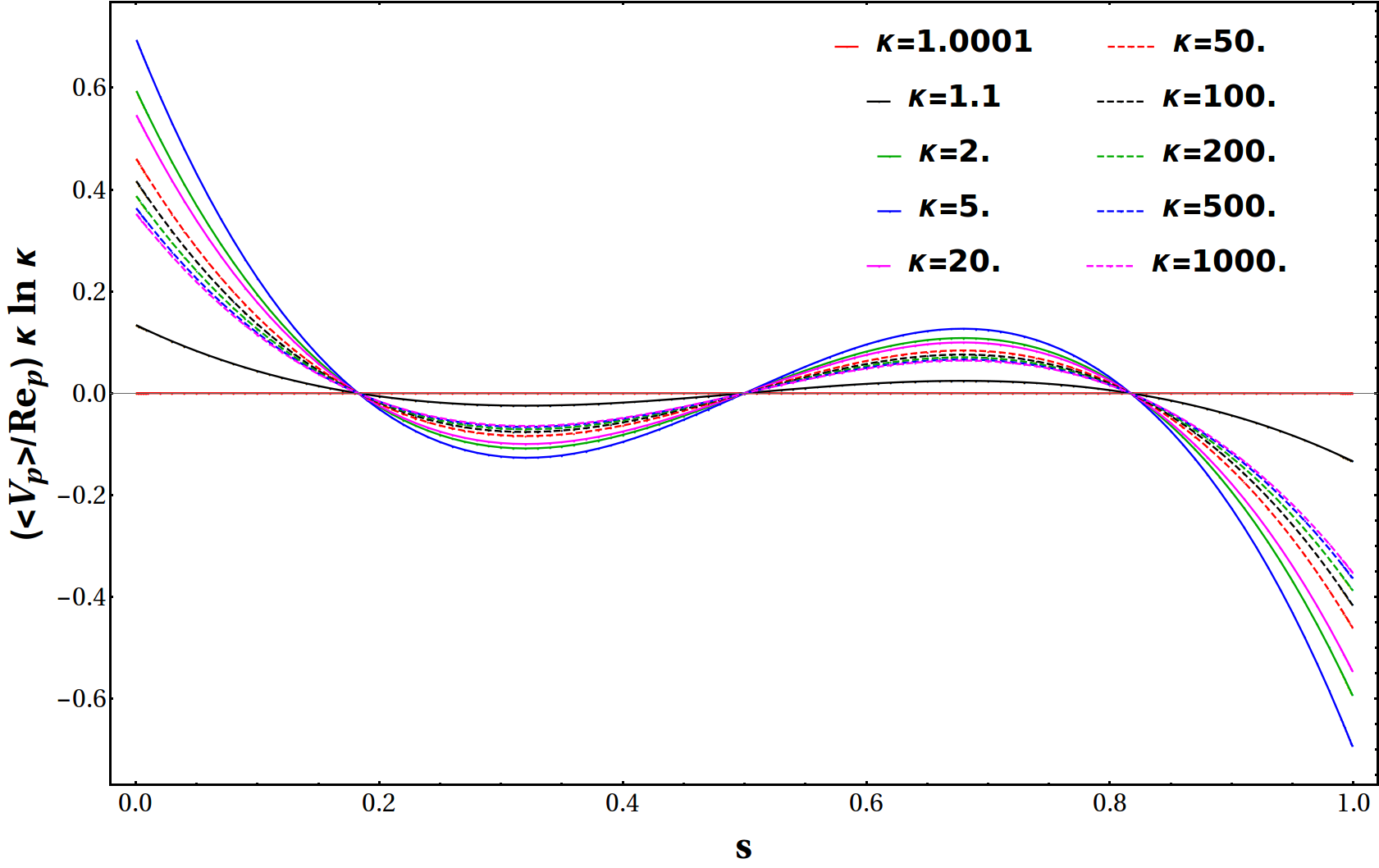}
        \caption{}
    \end{subfigure}
	\caption{(a) The small-$Re_c$-lift profiles for tumbling prolate spheroids of various aspect ratios; (b) Lift profiles in (a) re-scaled using the infinite-$\kappa$ $\langle S_{12}\rangle$-scaling.}
	\label{fig:Ch3VpTumblingvskappa}
\end{figure}

\begin{figure}
	\centering
    \begin{subfigure}[b]{0.49\textwidth}
		\includegraphics[width=\textwidth]{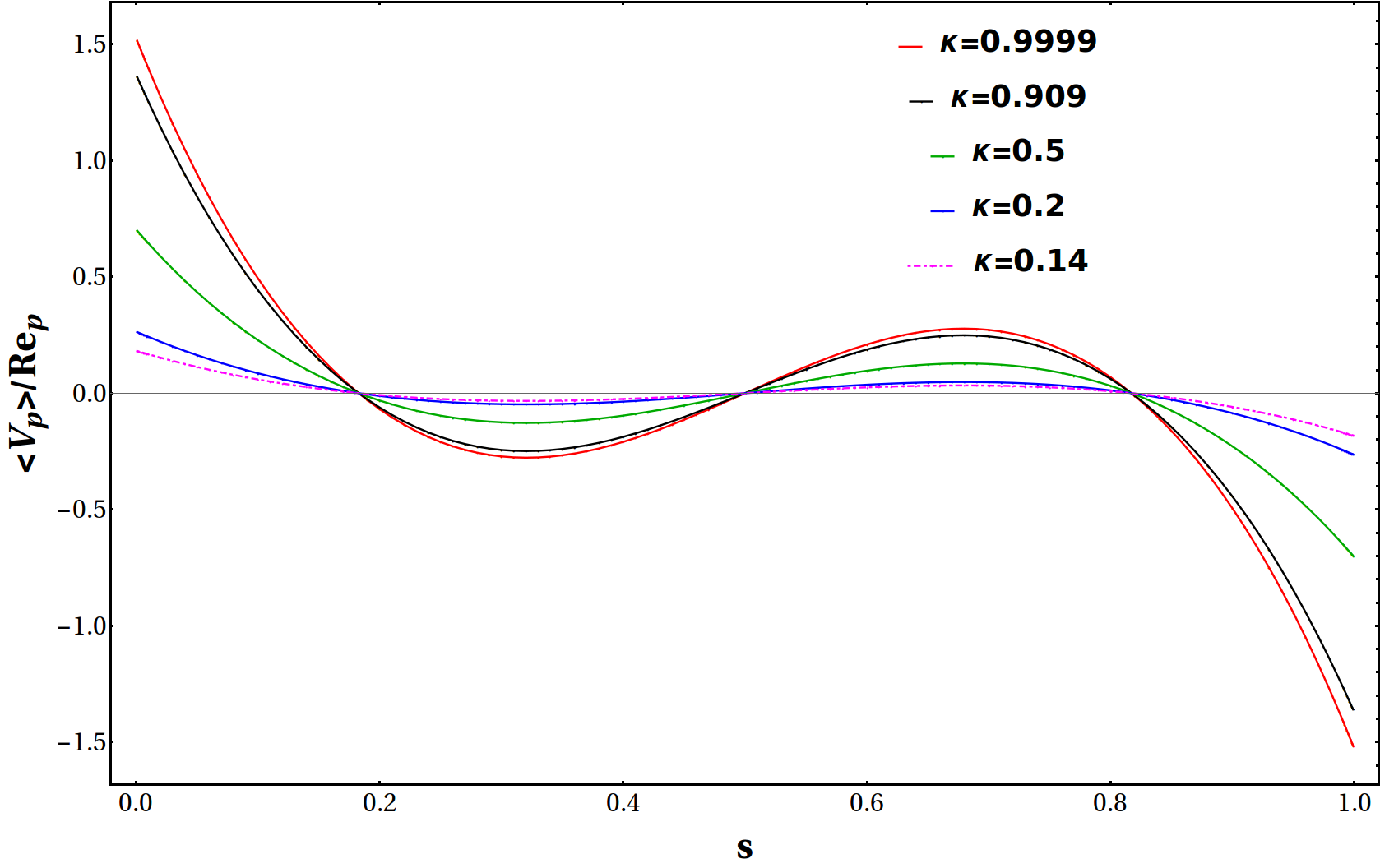}
        \caption{}
    \end{subfigure}
    \hfill
    \begin{subfigure}[b]{0.49\textwidth}
		\includegraphics[width=\textwidth]{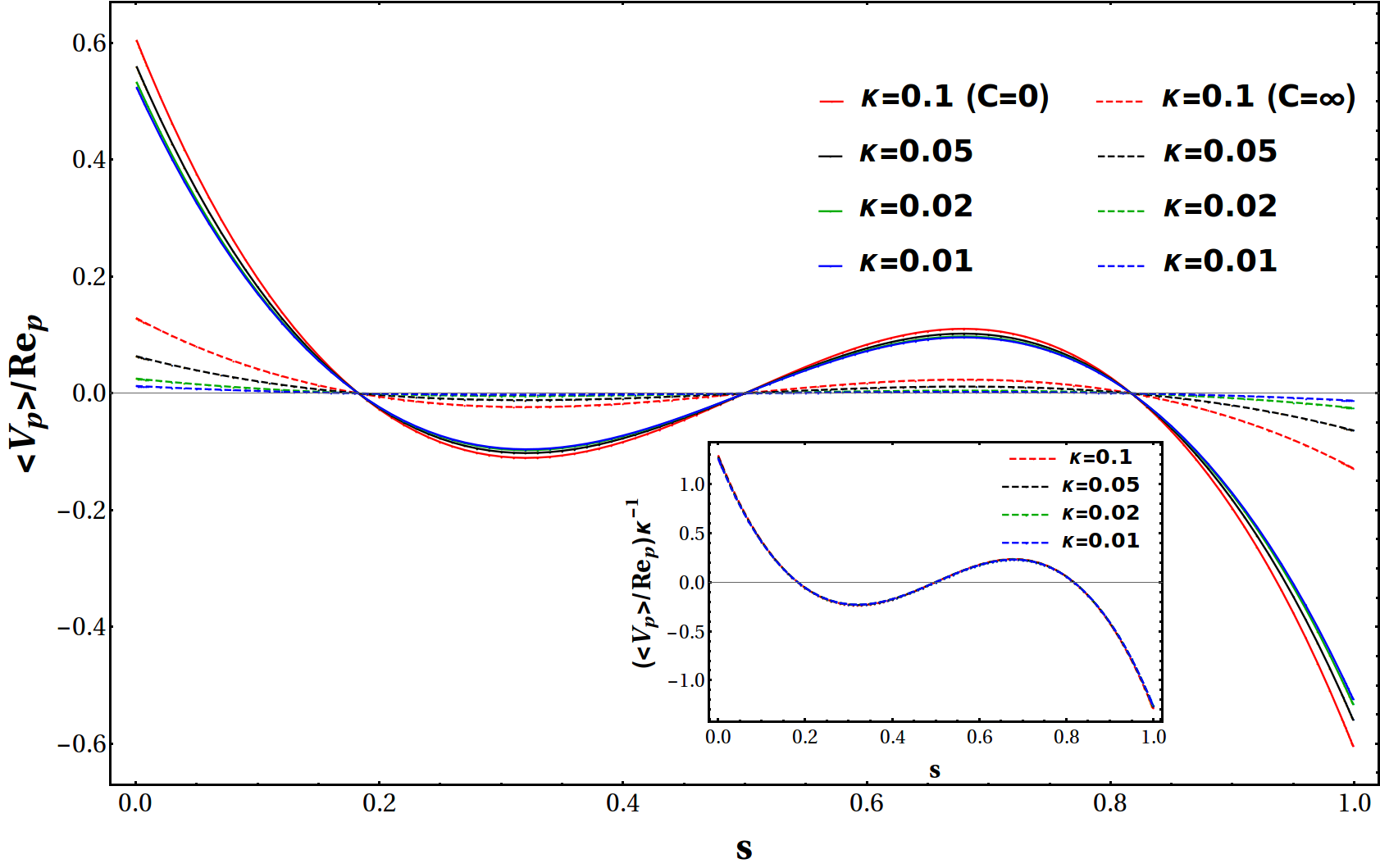}
        \caption{}
    \end{subfigure}
	\caption{The small-$Re_c$ lift profiles for (a) spinning oblate spheroids of $0.14\leq\kappa<1$, and b) spinning/tumbling oblate spheroids of $\kappa<0.14$, for $Re_c\ll1$. The inset shows the collapsed profiles on re-scaling with the $\kappa\to0$ limit of $\langle S_{12}\rangle$.}
	\label{fig:Ch3VpOblatevskappa}
\end{figure}

As mentioned after \eqref{eq:Ch3VpTimeAvg3}, while the magnitude of the Jeffery-averaged lift velocity is sensitive to the spheroid aspect ratio, the equilibrium locations remain the same as those for a sphere. This is seen from Figures \ref{fig:Ch3VpTumblingvskappa} and \ref{fig:Ch3VpOblatevskappa}, which show that the inertial lift profiles for both tumbling prolate spheroids and spinning/tumbling oblate spheroids of different $\kappa$ have the same zero crossings. In Figure \ref{fig:Ch3VpTumblingvskappa}a, the inertial lift for a tumbling prolate spheroid at a fixed $s$ is seen to decrease with increasing $\kappa$, consistent with the decreasing magnitude of the disturbance velocity field. To better examine the  large-$\kappa$ limit, in Figure \ref{fig:Ch3VpTumblingvskappa}b we plot the lift profiles normalized by the large-$\kappa$ scaling of $\langle S_{12} \rangle$ found earlier. The scaled profiles approach a $\kappa-$independent limiting form for $\kappa\to\infty$, although the approach is non-monotonic - the scaled profile with the maximum amplitude corresponds to $\kappa = 5$ - reflecting the non-monotonic approach of the scaled stresslet, $\kappa \ln \kappa\langle S_{12} \rangle$, to its infinite-$\kappa$ limit of $-\frac{2\pi}{3}$. In Figure \ref{fig:Ch3VpOblatevskappa}a, the inertial lift velocity of a spinning oblate spheroid is seen to decrease in magnitude as $\kappa$ decreases from $1$ to $0.14$, again consistent with a decrease in the magnitude of the disturbance velocity field. At the latter $\kappa$, a reversal of the inertia-induced orbital drift leads to both tumbling and spinning modes being stabilized, with the respective basins of attraction being demarcated by a pair of unstable limit cycles on the unit sphere\,\citep{navaneeth2016,einarsson2015}. In Figure \ref{fig:Ch3VpOblatevskappa}b, we therefore plot the inertial lift profiles of both spinning and tumbling oblate spheroids for $0 < \kappa < 0.14$. The lift profiles for the spinning spheroids approach a finite limiting form for $\kappa \rightarrow 0$, while those for tumbling spheroids are much smaller in magnitude, indicative of the much weaker disturbance field in this case. The latter arises from the much smaller value of the induced stresslet during the prolonged gradient-aligned phase of a thin oblate spheroid in the tumbling mode. The inertial lift for a tumbling oblate spheroid, in fact, goes to zero on account of $\langle S_{12} \rangle$ being $O(\kappa)$ for $\kappa\to 0$, this scaling reflective of the preponderance of orientations with $\bm{p}$ close to the gradient-vorticity plane, as mentioned earlier; the inset in Figure \ref{fig:Ch3VpOblatevskappa}b shows that the tumbling-oblate-spheroid lift profiles, when scaled by $\kappa$, approach a finite limiting form for $\kappa \to 0$.

The invariance of the equilibrium locations associated with the lift profiles in Figures \ref{fig:Ch3VpTumblingvskappa} and \ref{fig:Ch3VpOblatevskappa}, with changing $\kappa$, implies that inertia-induced shape-sorting of spheroidal particles is precluded in sufficiently long channels. All spheroids regardless of $\kappa$ will end up migrating to the same pair of Segre-Silberberg locations in such channels. However, one may achieve separation in channels short enough for the residence time of $O(L_c/V_\text{max})$ to be comparable or smaller than $t_\text{lift}$; this requires $L_c \lesssim HRe_p^{-1}\lambda^{-1}$, $L_c$ being the channel length. Shape-sorting of spheroids would occur in these channels owing to differential rates of migration: for instance, prolate spheroids with $\kappa\sim O(1)$ will migrate to the equilibrium locations relatively rapidly, while those with larger $\kappa$ remain close to their initial positions due to weaker lift forces. A similar scenario would prevail for oblate spheroids provided $\kappa \gtrsim 0.14$. For $\kappa \lesssim 0.14$, the disparity in magnitudes of the lift velocities for spinning and tumbling oblate spheroids would lead instead to sorting of spheroids  with the same $\kappa$, but in different\,(tumbling vis-a-vis spinning) orientation modes. 

\begin{figure}
	\centering
    \begin{subfigure}[b]{0.49\textwidth}
		\includegraphics[width=\textwidth]{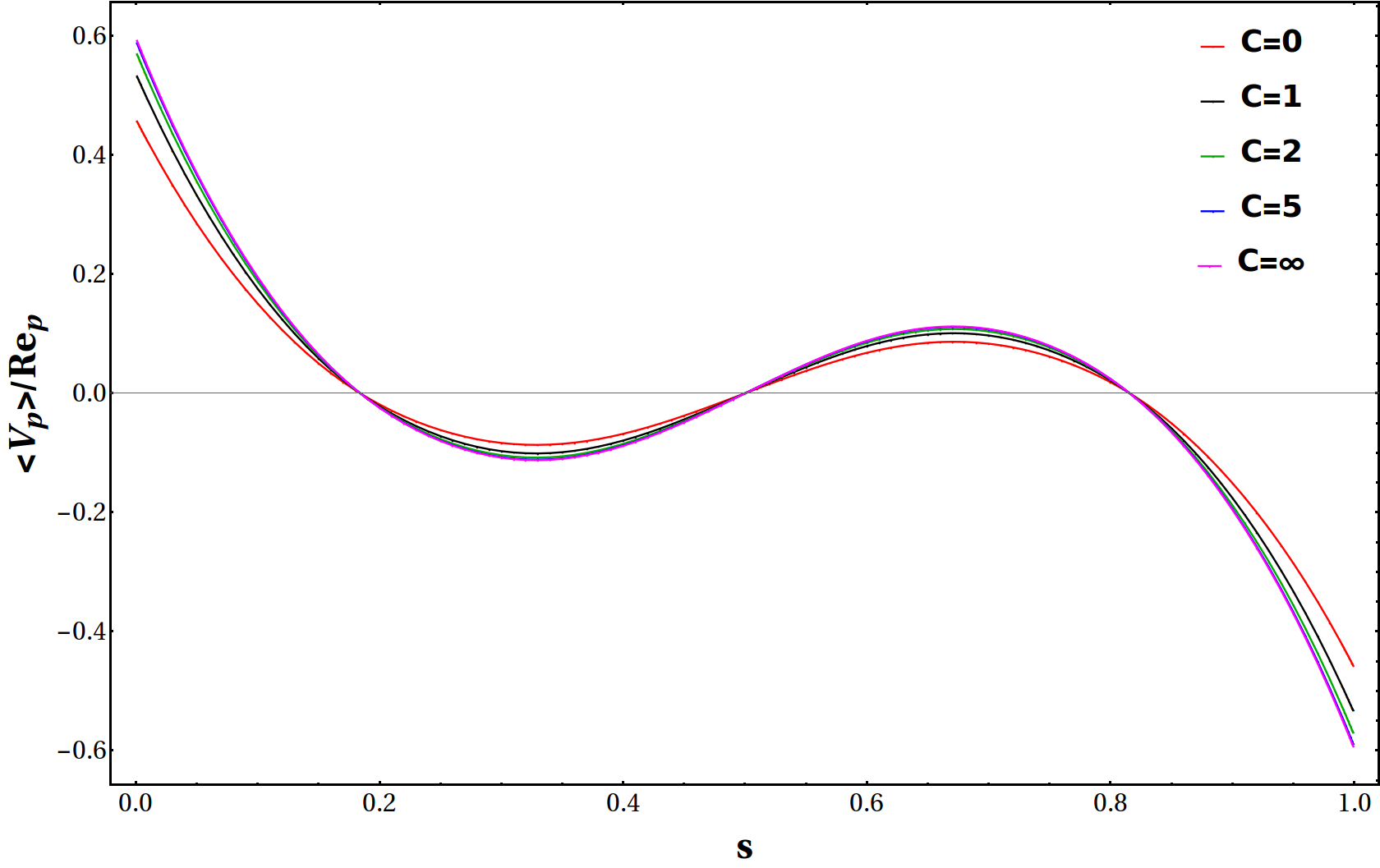}
        \caption{$\kappa=2$}
    \end{subfigure}
    \hfill
    \begin{subfigure}[b]{0.49\textwidth}
		\includegraphics[width=\textwidth]{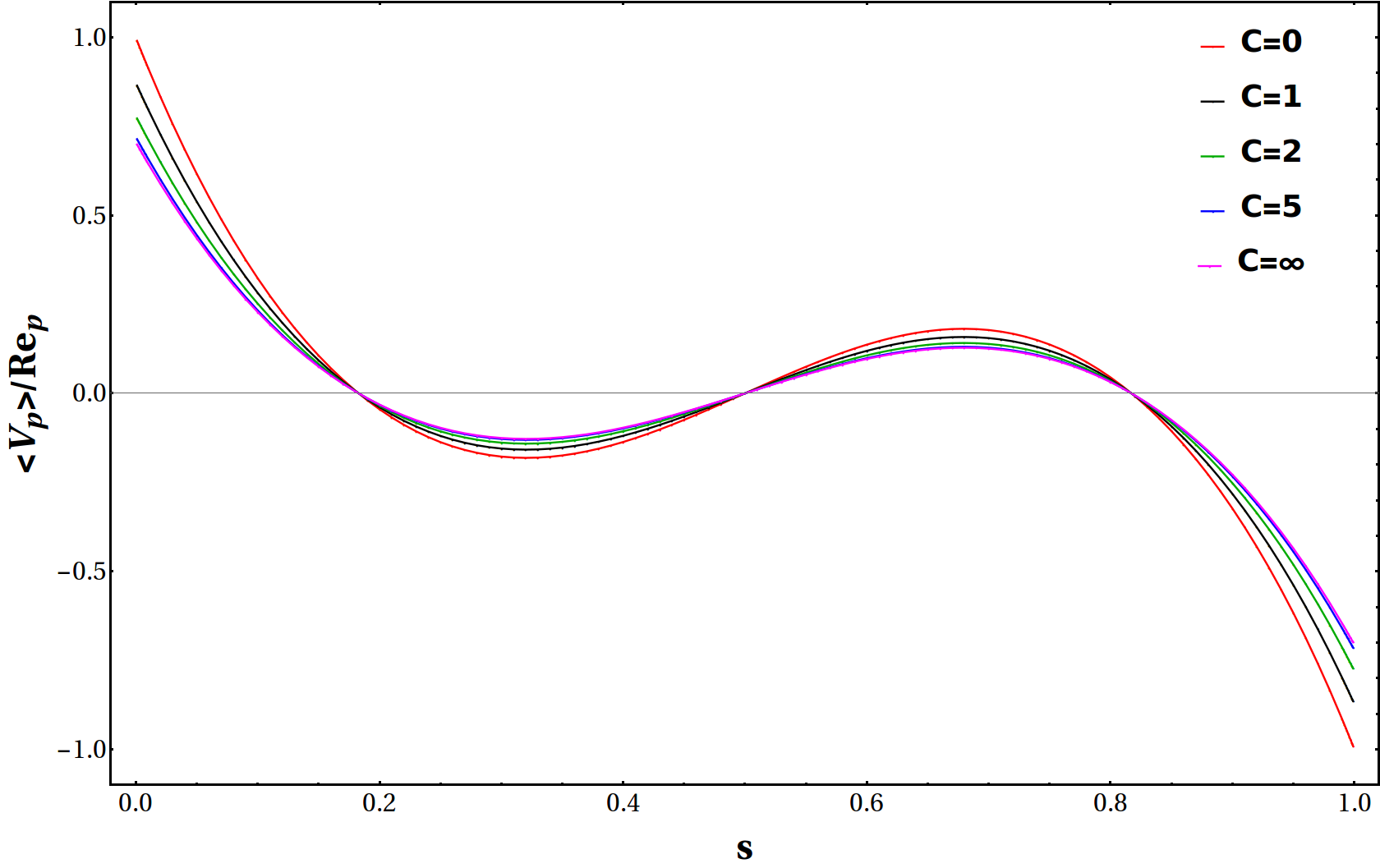}
        \caption{$\kappa=0.5$}
    \end{subfigure}
	\caption{Lift velocity profiles for prolate and oblate spheroids, of the indicated aspect ratios, rotating in different Jeffery orbits; $C=0$ and $\infty$ correspond to the spinning and tumbling modes.}
	\label{fig:Ch3VpvsCorbit}
\end{figure}

For even shorter channels with $L_c \lesssim O(H Re_p^{-1})$,  the residence time is no longer enough for a spheroid to be able to migrate to its stable Jeffery orbit.  For channel residence times of $O(L_c/V_\text{max}) \lesssim O(H/V_\text{max}Re_p^{-1}) \gg H/V_\text{max}$, the cross-stream migration is still determined, at leading order, by a Jeffery-averaged lift velocity, but one pertaining to orbits other than the tumbling or spinning mode. Thus, $\langle V_p \rangle$ is now also a function of $C$, being given by \eqref{eq:Ch3VpExpression} with $\langle S_{12} \rangle$ given by its full form involving both $C$ and $\kappa$; see \eqref{eq:Ch3TimeAvgDipoleStrength} and \eqref{eq:Ch3TimeAvgDipoleStrength2}. The $C-$dependent lift velocity profiles for $\kappa=2$ and $\kappa=0.5$ are shown in Figures \ref{fig:Ch3VpvsCorbit}a and b, respectively. Since, for a fixed $\kappa$, the amplitude of the disturbance field is the largest for a tumbling prolate spheroid and a spinning oblate spheroid, the largest amplitude profiles in Figures \ref{fig:Ch3VpvsCorbit}a and b correspond to $C = \infty$ and $C = 0$, respectively. Determining single-spheroid trajectories now requires solving the following system of coupled ODEs in $s$ and $C$:
\begin{align}
\frac{ds}{dt_s} =& \lambda\,\,\langle S_{12}\rangle(C,\kappa)\big[16(1-2s)^2 F(s)-16(1-2s)G(s)\big],\, \label{eqn:sevol} \\
\frac{dC}{dt_s} =&  \frac{\beta\,C}{2\pi} \left(\sum_{n=1}^{6} I_n(C,\kappa) F_n^f(\kappa)+\sum_{n=1}^{4} J_n(C,\kappa) G_n^f(\kappa)\right), \label{eqn:Cevol}
\end{align}
where $t_s = Re_pt$ is a slow time variable, and the functions $F_n^f,\,G_n^f,\,I_n$ and $J_n$ have been defined in \cite{navaneeth2016} and \cite{marath2018}. 

At higher volume fractions, the trajectory of the spheroid will acquire a stochastic component on account of occasional pair-interactions that may be modeled for via a scattering kernel involving pre- and post-interaction orbit constants\,\citep{navaneethRapids}. In this case, the evolution of an initial distribution of spheroids, on time scales long compared to $T_\text{jeff}$, is described by a probability density $P(s,C)$ which satisfies a kinetic equation of the form:
\begin{align}
\frac{\partial P}{\partial t_s} + \frac{\partial}{\partial s}(\dot{s}P) + \frac{\partial}{\partial C}(\dot{C}P) =\,nL^3\displaystyle\int dC' \displaystyle\int d\bm{r}_\perp (s-s')\displaystyle\int d\hat{C}d\hat{C}'[&\hat{P}\hat{P}'{\mathcal K}(\hat{C},\hat{C}'|C,C';s)\nonumber\\
&- P P'], \label{pairinteractions_migration}
\end{align}
where $P' \equiv P(s,C')$, $\hat{P} \equiv P(s,\hat{C})$, $\hat{P} \equiv P(s,\hat{C}')$; $\dot{s}$ and $\dot{C}$ being given by \eqref{eqn:sevol} and \eqref{eqn:Cevol}, respectively. The RHS of (\ref{pairinteractions_migration}) denotes a Boltzmann-type kernel involving the pre-\,($[\hat{C},\hat{C}']$) and post-($[C,C']$) interaction orbit-constant pairs.

At the end of $\S$\ref{sec:avglift_Calcn}, we had highlighted the finite value of the inertial lift velocity attained even as the particle approaches either wall. This runs counter to one's expectation of the lift velocity vanishing in this limit due to the diverging resistance associated with the thin lubricating  layer of fluid between the particle and the wall. As mentioned therein, the discrepancy arises due to the present analysis only being valid for $s,1-s\gg\lambda$, a restriction that comes from treating the spheroid as a point\,(stresslet) singularity. Our analysis has to be supplemented by one that accounts for the finite size of the spheroid and is therefore valid for $s,1-s \sim O(\lambda)$, which in turn would connect to the lubrication regime corresponding to $s-\lambda,(1-s)+\lambda\ll\lambda$. Such a connection is possible for the case of a sphere based on results available in the literature. The inertial lift on a sphere, in presence of a single plane boundary subject to a linear shearing flow, has been evaluated numerically for $s \sim O(\lambda)$ by \cite{cherukat1994}, with the aid of an integral expression obtained using bispherical coordinates. The authors found the lift force to always have a repulsive character\,(that is, to be directed away from the wall). The numerical value asymptoted to the near-wall limit\,($\approx 55\pi Re_p/6$) of the two-wall\,(channel) problem mentioned above for $s\gg\lambda$, while asymptoting to a different finite value\,($9.22 Re_p$) when the sphere touches the wall\,($s=\lambda$). The latter value was shown to agree with the inertial lift force acting on a stationary non-rotating sphere, in contact with a plane boundary, evaluated by \cite{acrivos1985}. These authors examined a non-rotating sphere in light of results obtained earlier\,\citep{goldman1967b} which showed that the angular velocity of a torque-free sphere must decrease to zero logarithmically in the limit of a vanishing sphere-wall separation on account of the lubrication resistance associated with the relative tangential motion in the narrow gap; the finiteness of the force implies that the dominant contributions to the lift arise from the fluid domain outside the thin gap. When combined with the known $O(s-\lambda)^{-1}$ divergence of the translational resistance for normal approach towards the wall, one concludes that the inertial lift velocity for a sphere must start from $55 Re_p/36$ for $\lambda \ll s \ll 1$, and eventually approach zero linearly in the limit $s-\lambda \ll \lambda$. In the reciprocal theorem formulation used here, the approach to zero would appear via the divergence of the test problem resistance coefficient. To our knowledge, analogous results for a spheroid when $s, 1-s \sim O(\lambda)$ are not available, and will involve more effort. This is both because one does not have a spheroidal analog of the bispherical coordinate system used in \cite{cherukat1994}, and because the orientation dynamics change from Jeffery rotation to pole-vaulting with approach towards either wall. 


\section{The inertial lift velocity for $Re_c\gtrsim O(1)$} \label{sec:Ch3Reclarge}

Herein, we calculate the inertial lift profiles for $Re_c \gtrsim O(1)$ using a numerical shooting method employed originally by \cite{schonberghinch1989}, and later for higher $Re_c$'s by \cite{asmolov1999}, both for the case of a sphere. For $Re_c\gtrsim O(1)$, the inertial screening length\,($H Re_c^{-1/2}$) is of order the channel width or smaller, and one must solve the governing equations (\ref{eq:Ch3NS2}a,b) with the boundary conditions (\ref{eq:Ch3BC2}a-c) using a matched asymptotic expansions approach. The inner region is characterized by scales of $O(L)$, and the outer region by scales of $O(HRe_c^{-1/2})$, with the channel width $H$ also entering via the wall boundary conditions. Although we use a shooting method below to evaluate the lift velocity, it is worth noting that the reciprocal theorem formulation in section \ref{sec:Ch3GRT} remains valid for any $Re_c$, provided one uses the finite-$Re_c$ disturbance velocity field\,($\bm{u}'$) in the volume integral. Thus, the scaling arguments given in section \ref{sec:Ch3Scaling} may be extended to finite $Re_c$, and in doing so, one finds that the dominant contribution to the inertial lift continues to come from scales asymptotically larger than $O(L)$. For $Re_c\ll1$ in the previous section, this outer-region dominance meant solving the Stokes equations in a domain confined by plane channel walls with the spheroid approximated as a stresslet singularity. In the present section, this means solving the linearized Navier-Stokes equations at leading order, driven by the same stresslet singularity\,(since $Re_p$ is assumed small). Note that $Re_c$, interpreted as the square of the ratio of the two outer-region length scales\,($H$ and $HRe_c^{-\frac{1}{2}}$) appears as a parameter in the governing equations below, and the inertial lift velocity is therefore now a function of $Re_c$. 

\subsection{The finite-$Re_c$ formulation}\label{sec:finiteRecformulation}

The Stokesian disturbance field due to a freely suspended particle in an ambient linear flow only decays algebraically, as $1/r^2$, at large distances, a fact already used in the scaling arguments in section \ref{sec:Ch3Scaling}. The algebraic decay implies that distinct outer-region expansions for both the velocity and pressure fields become necessary on scales of $O(HRe_c^{-1/2})$, with the leading terms satisfying the linearized Navier-Stokes and continuity equations. To write down these equations, one transforms to outer coordinates using $\bm{r}=Re_p^{-1/2}\bm{R}$, which corresponds to using $HRe_c^{-\frac{1}{2}}$\,(rather than $L$ as in section \ref{sec:Ch3formulation}) as the relevant length scale. The Stokesian rates of decay in the inner region suggest the scalings $\bm{u}'=Re_p\bm{U}$ and $p'=Re_p^{3/2} P$ for the leading order terms in the outer expansions, where $\bm{U}$ and $P$ satisfy:  
\begin{subequations} 
	\begin{align} 
	\frac{\partial^2 U_i}{\partial R_m^2}-\frac{\partial P}{\partial R_i}-\frac{\partial U_i}{\partial t}-U_2(\beta+2\gamma''R_2 Re_c^{-1/2})\delta_{i1}&-(\beta R_2+\gamma''R_2^2 Re_c^{-1/2})\frac{\partial U_i}{\partial R_1}=\beta S_{im}\frac{\partial\delta(\bm{R})}{\partial R_m},\\
	\bm{\nabla}\cdot \bm{U}&=0.
	\end{align} \label{eq:Ch3NSHinchOuterEqn}
\end{subequations}
 The Faxen correction contributes at a higher order in $\lambda$ and therefore, $\bm{u}^\infty\approx Re_p^{-1/2}(\beta R_2+\gamma''\,Re_c^{-1/2} R_2^2) \delta_{i1}$ has been used in (\ref{eq:Ch3NSHinchOuterEqn}a).
Equations (\ref{eq:Ch3NSHinchOuterEqn}a,b) must be supplemented by the following conditions: 
\begin{subequations} 
	\begin{align} 
	U_i &\sim \frac{3\beta R_i R_j S_{jm}R_m}{4\pi R^5} \frac{}{} \text{ for } \bm{R}\rightarrow 0, \\
	\bm{U}&=0 \text{ at } R_2=-s\,Re_c^{1/2}, (1-s)\,Re_c^{1/2},
	\end{align} \label{eq:Ch3NSHinchOuterBC}
\end{subequations}
where (\ref{eq:Ch3NSHinchOuterBC}b) denotes the no-slip conditions on the channel walls, while (\ref{eq:Ch3NSHinchOuterBC}a) is the requirement of matching to the stresslet velocity field that arises as the far-field form of the inner-region Stokesian field. In (\ref{eq:Ch3NSHinchOuterEqn}a) and (\ref{eq:Ch3NSHinchOuterBC}a),  $\bm{S}$ is the tensorial amplitude defined in \eqref{eq:Ch3DipoleStrengthProlate}.

As explained in $\S$\ref{sec:Ch3Scaling}, the separation between the migration and drift time scales implies one need only solve for the Jeffery-averaged lift velocity, and towards this end, we average (\ref{eq:Ch3NSHinchOuterEqn}a,b) and (\ref{eq:Ch3NSHinchOuterBC}a,b) over a single period of the stable Jeffery orbit. One obtains:
\begin{subequations}  \label{eq:Ch3NSHinchOuterTimeAvg}
	\begin{align} 
	\frac{\partial^2\langle U_i\rangle}{\partial R_m^2} -\frac{\partial\langle P\rangle}{\partial R_i}-\langle U_2\rangle(\beta+2\gamma''R_2 Re_c^{-1/2})\delta_{i1}&-(\beta R_2+\gamma''R_2^2 Re_c^{-1/2})\frac{\partial\langle U_i\rangle}{\partial R_1}\nonumber\\
	=&\beta\langle S_{12}\rangle\Big[ \delta_{i1}\frac{\partial\delta(\bm{R})}{\partial R_2}+\delta_{i2} \frac{\partial\delta(\bm{R})}{\partial R_1}\Big],\\
	\bm{\nabla}\cdot \langle\bm{U}\rangle=&0,
	\end{align} 
\end{subequations}
where $\langle\bm{U}\rangle$ satisfies:
\begin{subequations}
	\begin{align} 
	\langle U_i\rangle &\sim  \frac{3 \beta \langle S_{12}\rangle R_1 R_2 R_i}{4\pi R^5} \text{ for } \bm{R}\rightarrow 0,\\
	\langle\bm{U}\rangle&=0 \text{ at } R_2=-s\,Re_c^{1/2}, (1-s)\,Re_c^{1/2}.
	\end{align} \label{eq:Ch3NSHinchOuterTimeAvgBC}
\end{subequations}
Here, $\langle S_{12}\rangle(\kappa)$ corresponds to the stresslet averaged over the relevant stable Jeffery orbit, and has been given in \eqref{tumbling_prolatestresslet}-\eqref{spinning_oblatestresslet}. Due to the linearity of (\ref{eq:Ch3NSHinchOuterTimeAvg}a,b) and (\ref{eq:Ch3NSHinchOuterTimeAvgBC}a,b), and the fact that the Jeffery-averaged spheroid stresslet tensor differs from that for a sphere only by a scalar multiplicative factor, $\langle \bm{U} \rangle$ and $\langle P \rangle$ differ from their spherical analogs only by $-\frac{3\langle S_{12}\rangle(\kappa)}{10\pi}$, corresponding to the ratio of the aforementioned stresslets. This proportionality relation must hold for any linear functional of the disturbance fields, and in particular, for the lift velocity that is a linear functional of $\langle U_2 \rangle$; see (\ref{eq:Ch3VpHinchInverseFT}) below. Thus, similar to the case of $Re_c\ll1$, the Jeffery-averaged lift profiles at a given finite $Re_c$, for an arbitrary aspect ratio spheroid, have the same shape as those for a sphere at the same $Re_c$. It follows that the associated pair of equilibria are identical to those for a sphere regardless of $Re_c$, and as for a sphere\citep{schonberghinch1989}, must migrate wallward with increasing $Re_c$. It is worth reiterating that the requirement for a Jeffery-averaged analysis to remain valid becomes restrictive for extreme-aspect-ratio spheroids. The regime of validity was originally stated after equation \eqref{eq:Ch3Spheroidangles}, and expressed in terms of $Re_c$, is given by $Re_c\kappa/(\lambda^2\ln \kappa) \ll 1$ and $Re_c/(\lambda^2\kappa^2) \ll 1$ for $\kappa \gg 1$ and $\kappa \ll 1$, respectively; these point to the restriction becoming more severe with increasing $Re_c$. The implications of the finite-$Re_c$ Jeffery-averaged analysis above, for shape-sorting, remain the same as those discussed in section \ref{sec:smallRecresults}.

For purposes of completeness, we now follow along the lines of \cite{schonberghinch1989}, and briefly present the manner in which inertial lift is determined for $Re_c \gtrsim O(1)$. After implementing the partial Fourier transform defined in (\ref{eq:Ch3FTdefn}), followed by some algebraic manipulation, one obtains the following coupled ordinary differential equations for $\langle \hat{P} \rangle$ and $\langle \hat{U}_2 \rangle$:
\begin{subequations}
	\begin{align}
	\frac{d^2 \hat{\langle P\rangle}}{dR_2^2}-k_\perp^2\hat{\langle P\rangle}&=2\iota k_1 \hat{\langle U_2\rangle}(\beta+2\gamma''R_2 Re_c^{-1/2}),\\
	\frac{d^2\hat{\langle U_2\rangle}}{dR_2^2}-k_\perp^2\hat{\langle U_2\rangle}&=\frac{d\hat{\langle P\rangle}}{dR_2}-\iota k_1\hat{\langle U_2\rangle}(\beta R_2+\gamma''R_2^2 Re_c^{-1/2}),
	\end{align} \label{eq:Ch3HinchODEs}
\end{subequations}
with the conditions, 
\begin{subequations}
	\begin{align} 
	\hat{\langle U_2\rangle} &\sim\frac{\iota \beta\langle S_{12}\rangle k_1 |R_2|\,\,e^{-k_\perp |R_2|}}{2}  \text{ for } k_1,k_3\to\infty \text{ and } R_2\rightarrow 0,\\
	\hat{\langle U_2\rangle}=\frac{d\hat{\langle U_2\rangle}}{dR_2}&=0 \text{ at } R_2=-s\,Re_c^{1/2},R_2=(1-s)\,Re_c^{1/2},
	\end{align} \label{eq:Ch3HinchBC}
\end{subequations}
where $k_\perp^2=k_1^2+k_3^2$ as before. The delta-function forcing in (\ref{eq:Ch3NSHinchOuterTimeAvg}a) leads to the following jump conditions across the particle location\,($R_2=0$):
\begin{subequations}
	\begin{align}
	\hat{\langle P\rangle}^+(k_1,0^+,k_3)-\hat{\langle P\rangle}^-(k_1,0^-,k_3) &=2\iota k_1 \beta\langle S_{12}\rangle,\\
	\frac{d\hat{\langle P\rangle}^+}{dR_2}(k_1,0^+,k_3)-\frac{d\hat{\langle P\rangle}^-}{dR_2}(k_1,0^-,k_3) &=0,\\
	\hat{\langle U_2\rangle}^+(k_1,0^+,k_3)-\hat{\langle U_2\rangle}^-(k_1,0^-,k_3) &=0,\\
	\frac{d\hat{\langle U_2\rangle}^+}{dR_2}(k_1,0^+,k_3)-\frac{d\hat{\langle U_2\rangle}^-}{dR_2}(k_1,0^-,k_3) &=\iota k_1 \beta \langle S_{12}\rangle,
	\end{align} \label{eq:Ch3JumpCondns}
\end{subequations}
where the superscripts `+' and `$-$' denote the limiting values attained on approaching $R_2 = 0$ from the regions $0<R_2\leq(1-s)\,Re_c^{1/2}$ and $-s\,Re_c^{1/2}\leq R_2<0$, respectively; these conditions are derived in Appendix \ref{App:C}. The limiting form of $\langle \bm{U} \rangle$ in the matching region\,($\bm{R}\ll 1$) is the sum of the singular stresslet contribution given in (\ref{eq:Ch3NSHinchOuterTimeAvgBC}a), and a uniform flow along the gradient\,(cross-stream) direction that is a consequence of fluid inertia. The neutrally buoyant spheroid being force-free is convected by this uniform flow which therefore equals the inertial lift velocity, and may be determined from the limit of the inverse transform for $\bm{R}\to0$:
\begin{align}
\langle V_p\rangle&=\frac{Re_p}{4\pi^2}\,\,\Re\left\{\int_{-\infty}^\infty\int_{-\infty}^\infty \,\hat{\langle U_2\rangle}^\pm(k_1,0^\pm,k_3)\,dk_1\, dk_3\right\}.
\label{eq:Ch3VpHinchInverseFT}
\end{align}
Here, $\Re\{.\}$ denotes taking the real part which eliminates the purely imaginary stresslet contribution. As indicated, one may use either $\hat{\langle U_2\rangle}^-$ or $\hat{\langle U_2\rangle}^+$ on account of continuity; see (\ref{eq:Ch3JumpCondns}c). The governing ODEs (\ref{eq:Ch3HinchODEs}a,b), the boundary conditions (\ref{eq:Ch3HinchBC}a,b) and the jump conditions (\ref{eq:Ch3JumpCondns}a-d) can be solved using the shooting technique described in Appendix A of \cite{schmid2002stability}, a brief description of which is given in Appendix \ref{App:D}. 

The integral in \eqref{eq:Ch3VpHinchInverseFT} is evaluated numerically using a two-dimensional Gauss-Legendre quadrature over a circle of a sufficiently large radius $K_m$ in the $k_1\!-\!k_3$ plane. An analytical large-$k_\perp$ asymptote was calculated using the steps outlined in \cite{hogg1994}, and added to the numerical integral, to improve convergence. The analysis involves expanding $\hat{\langle U_2\rangle}$ and $\hat{\langle P\rangle}$ in inverse powers of $k_\perp$, an ansatz valid only in an $O(k_\perp^{-1})$ neighborhood of $R_2 = 0$. As a result, satisfaction of  (\ref{eq:Ch3HinchBC}a,b) is replaced by a farfield decay requirement for $R_2k_\perp \gg 1$. Including only the exponentially decaying solutions on either side of $R_2 = 0$, satisfying the jump conditions (\ref{eq:Ch3JumpCondns}a-d), and then performing the inverse Fourier transform, one obtains:
\begin{align}
\langle V_p\rangle^\text{far field}&\approx-\frac{9\beta\gamma''\langle S_{21}\rangle Re_p}{64\pi K_m Re_c^{1/2}},
\label{eq:Ch3Vplargek}
\end{align} 
 at leading order. As in $\S$\ref{sec:Ch3Recsmall}, the choice of $K_m$ is dictated by the distance of the particle from the walls. The relevant scale sufficiently close to either wall is still the spheroid-wall separation\,($s$ or $1-s$) - the near-wall lift remains the same as that for $Re_c\ll1$, being the farfield lift experienced by a particle in the presence of a single plane boundary. 
  While one needs to keep increasing $K_m$ with approach to either wall\,(that is, for sufficiently small $s$ or $1-s$), to obtain a converged result, this increase is only necessary once $s$ or $(1-s)$ becomes less than $O(Re_c^{-1/2})$. Thus, to capture the wall-induced repulsion for finite $Re_c$, one needs to ensure $K_m Re_c^{1/2} s_\text{min}, K_m Re_c^{1/2} (1-s)_\text{min}\gg1$. In our calculations, we chose $K_m=200$ for $0.1\leq Re_c\leq10$ to ensure accurate lift velocities down to $s, 1-s\approx0.03$. For $Re_c>10$, accurate lift profiles were obtained down to $s, 1-s\approx0.05$ for $K_m=40$. 
\begin{figure}
	\centering
	\includegraphics[width=\textwidth]{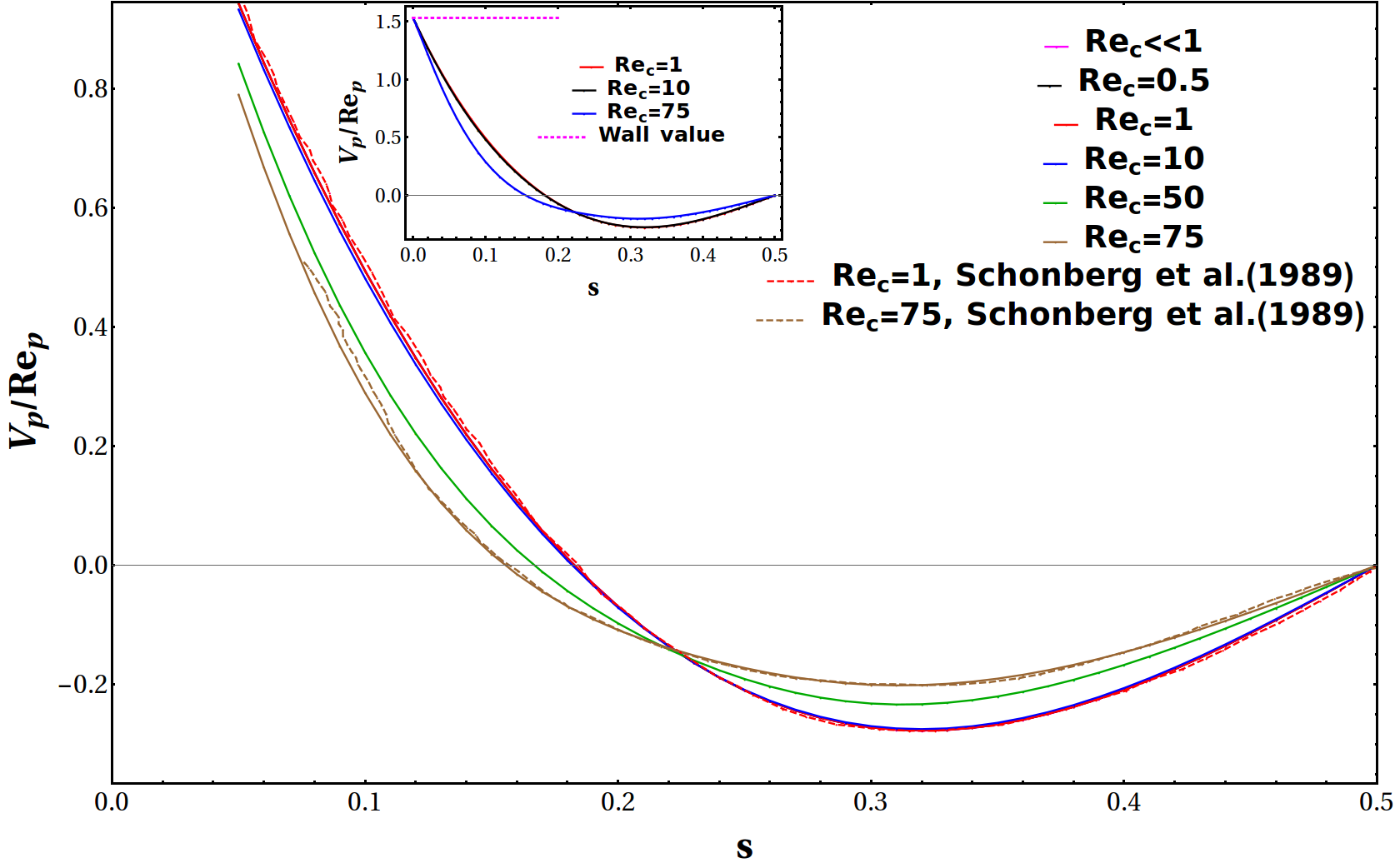}
	\caption{Comparison of finite-$Re_c$ lift velocity profiles for a sphere with the semi-analytical small-$Re_c$ profile, and the data from \cite{schonberghinch1989}. The inset shows the approach of the finite-$Re_c$ profiles towards the wall asymptote given by \eqref{eq:Ch3VpnearWall2}.}
	\label{fig:Ch3vsHoLealHigherRec}
\end{figure}

\subsection{Results and Discussion}\label{sec:finiteRecResults}

\begin{figure}
	\centering
	\includegraphics[width=\textwidth]{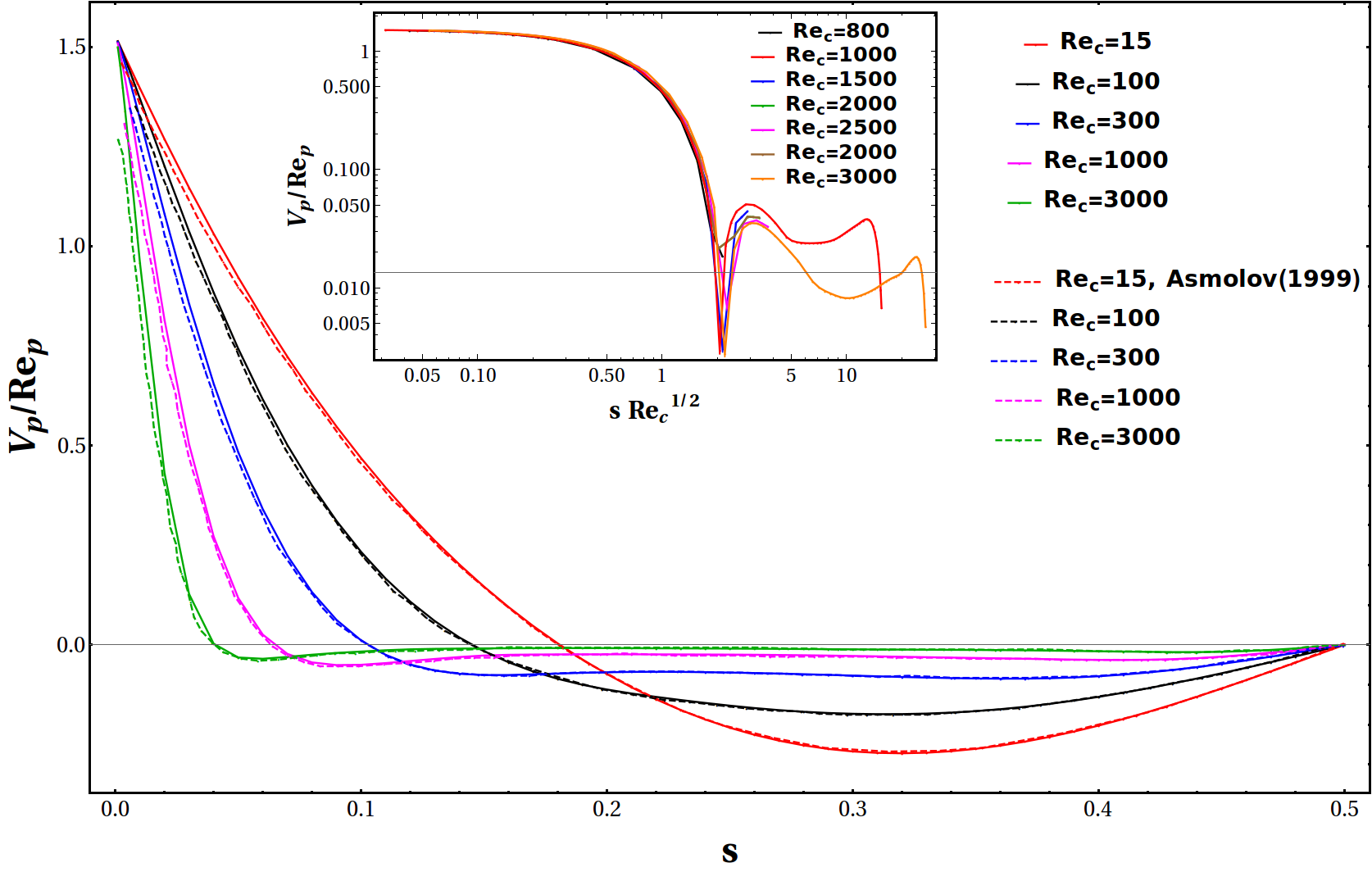}
	\caption{Comparison of lift velocity profiles for a sphere at higher $Re_c$ with \cite{asmolov1999}\,(dashed curves). The inset shows the drop in the magnitude of the lift velocity with increasing $Re_c$, except close to the wall\,(zero crossings appear as dips to negative infinity).}
	\label{fig:Ch3vsAsmolov}
\end{figure}

We begin with Figure \ref{fig:Ch3vsHoLealHigherRec} which shows the inertial lift profiles for a sphere for $0.5 \leq Re_c \leq 75$, along with the limiting small-$Re_c$ profile given by \eqref{eq:Ch3VpFinalExpressionsmallRec}; only profiles in the half-channel have been plotted owing to their anti-symmetry about the centerline. The profiles for $Re_c=1$ and $75$ exhibit good agreement with the data extracted from \cite{schonberghinch1989}. Interestingly, the profiles for $Re_c = 0.5, 1$ and $10$ compare closely with the small-$Re_c$ limiting form. The inset shows the approach of the finite-$Re_c$ profiles to the near-wall value given by \eqref{eq:Ch3VpnearWall2}. Figure \ref{fig:Ch3vsAsmolov} shows that the sphere lift profiles for higher $Re_c$ agree well with those extracted from Asmolov's data\citep{asmolov1999}, all the way upto $Re_c=3000$; note that our profiles have been continued to smaller $s$ to emphasize the approach to the common near-wall limiting value mentioned above. For $Re_c \gtrsim 300$, the lift profiles begin to exhibit an intermediate region of oppositely signed curvature. It has recently been shown that including finite-size effects pushes this intermediate region towards the zero-lift line, eventually leading to the emergence of new equilibria closer to the centerline for sufficiently large $Re_c$\citep{anandfinitesize2022}. The inset in the said figure, on a logarithmic ordinate scale, helps highlight the rapid decrease in the lift magnitude for $Re_c\gtrsim O(10)$, reflective of weakening particle-wall interactions. Inertia-induced faster decay of the velocity field, on scales larger than $O(HRe_c^{-\frac{1}{2}})$, is responsible for the reduced influence of the walls with increasing $Re_c$. The rescaled abscissa in the inset highlights the $O(Re_c^{-\frac{1}{2}})$ neighborhood of the wall where the lift profiles begin to rise towards to near-wall limit.

\begin{figure}
	\centering
	\includegraphics[width=0.9\textwidth]{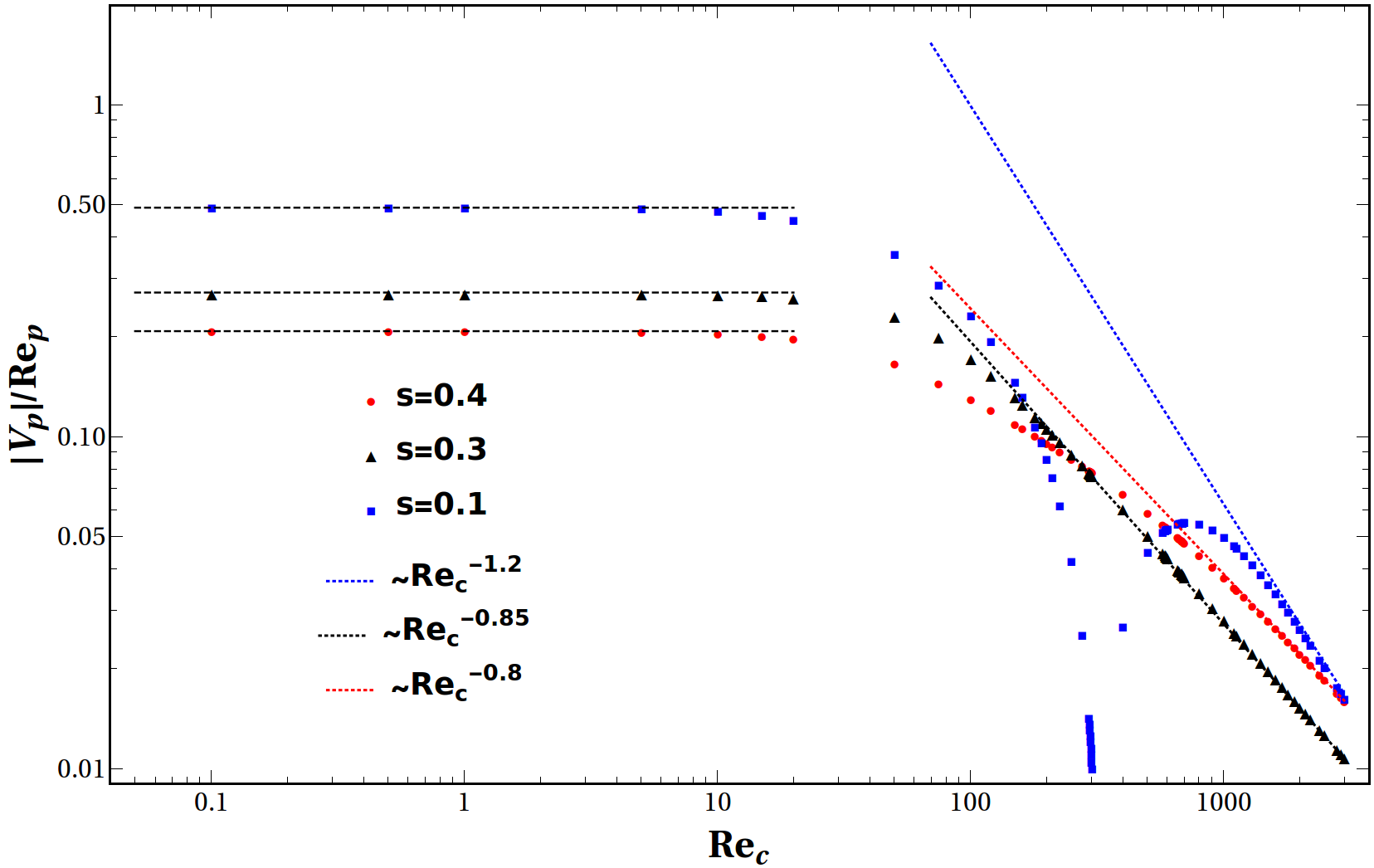}
	\caption{Sphere lift velocity\,(magnitude), as a function of $Re_c$, for different $s$. In all cases, $|V_p|/Re_p$ exhibits a plateau (dashed lines) until $Re_c \approx 10$, transitioning to an  algebraic decrease for sufficiently large $Re_c$. For $s < s_{eq}\,(=0.182)$, the transition is preceded by a zero-crossing that appears as a sharp dip to negative infinity. The dotted lines are empirical fits to the large-$Re_c$ behavior, and highlight the $s$-dependent decay exponent.}
	\label{fig:Ch3HoLealplateau}
\end{figure}

Figure \ref{fig:Ch3HoLealplateau} shows the magnitude of the sphere lift velocity, as a function of $Re_c$, at different locations on either side of the Segre-Silberberg equilibrium\,($s_\text{eq} = 0.182$). In accordance with the above discussion, for all $s$ values considered, the lift velocity starts off on a small-$Re_c$ plateau which extends until $Re_c \approx 10$. Thus, the small-$Re_c$ approximation remains a good approximation well beyond $Re_c$'s of order unity. For $s=0.3$ and $0.4$, with increasing $Re_c$, the lift velocity directly transitions from the plateau to an eventual algebraic decrease, this being typical for all $s>s_\text{eq}$. On the other hand, the lift velocity magnitude for $s=0.1$\,(and for all $s<s_\text{eq}$) exhibits a non-monotonic variation with an intermediate zero-crossing which, for $s=0.1$, is at $Re_c \approx 300$. This is due to the Segre-Silberberg equilibrium crossing the given $s$ in course of its wallward movement\,(with increasing $Re_c$). The lift velocity increases  again at larger $Re_c$, but to a value smaller than the small-$Re_c$ plateau, finally transitioning to a steeper algebraic decrease.

\begin{figure}
	\centering
    \begin{subfigure}[b]{0.49\textwidth}
		\includegraphics[width=\textwidth]{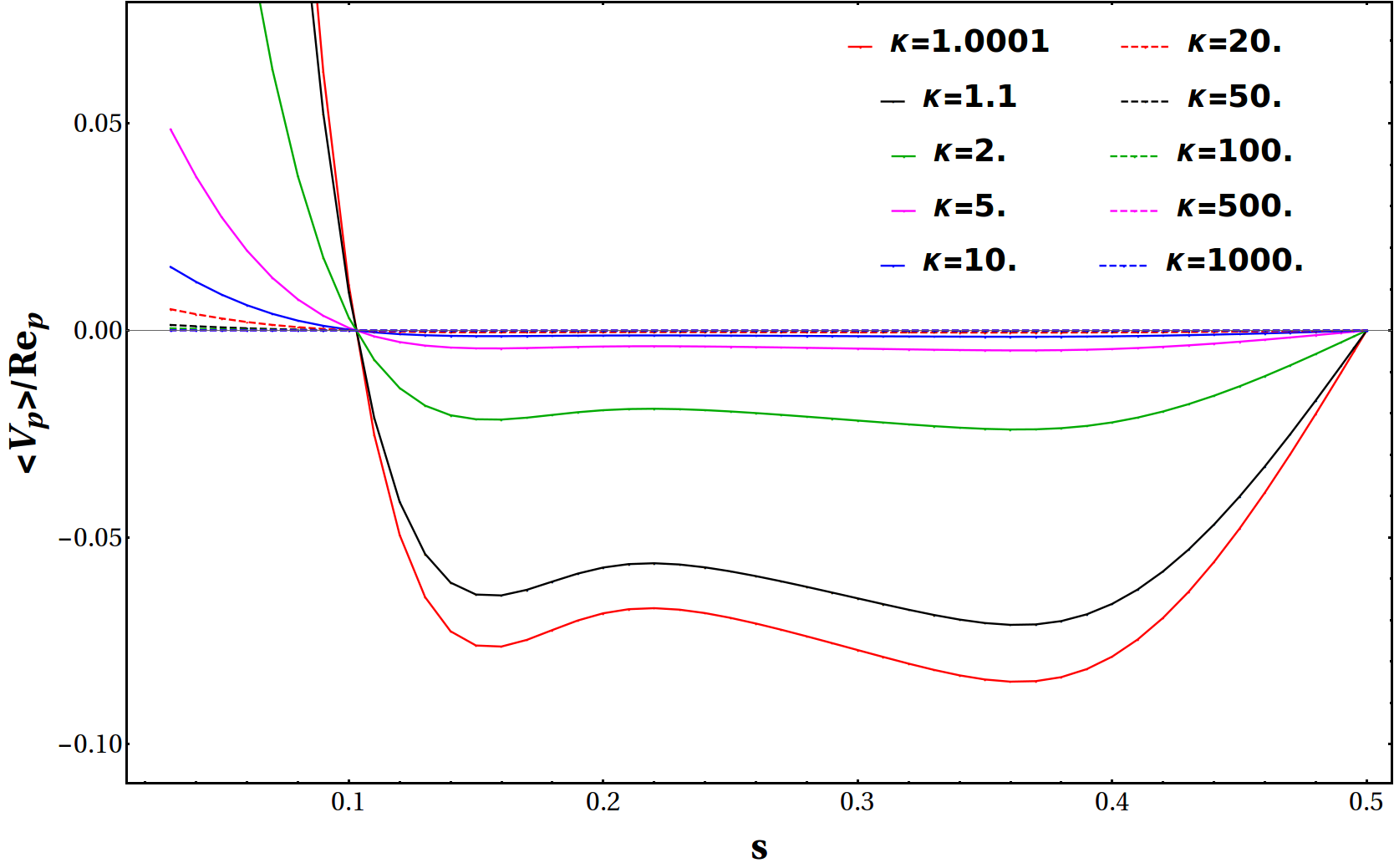}
        \caption{}
    \end{subfigure}
    \hfill
    \begin{subfigure}[b]{0.49\textwidth}
		\includegraphics[width=\textwidth]{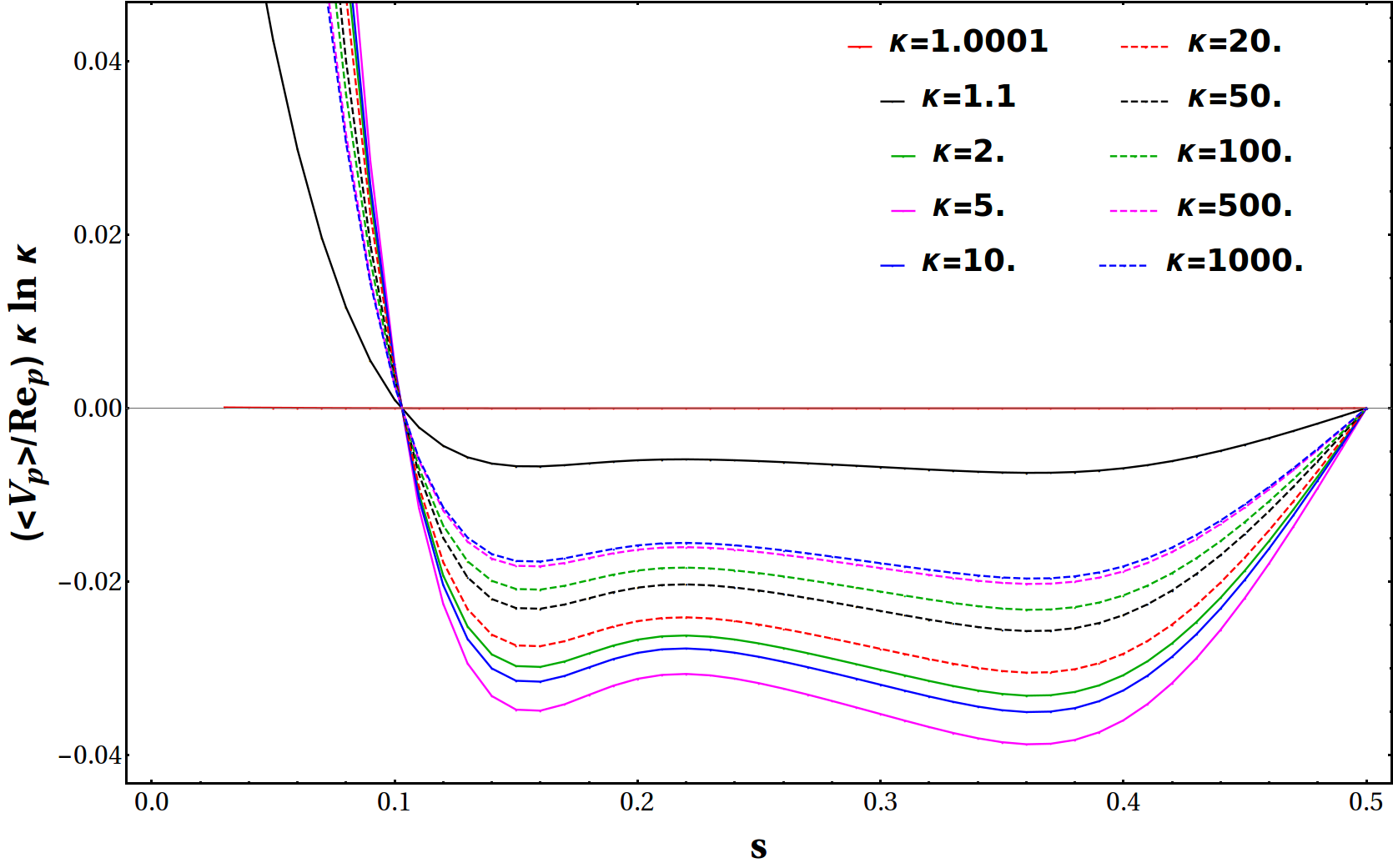}
        \caption{}
    \end{subfigure}
	\caption{(a) Lift velocity profiles for tumbling prolate spheroids of various aspect ratios for $Re_c=300$; (b) Lift profiles in (a) re-scaled using the $\kappa\to\infty$ limit of $\langle S_{12}\rangle$.}
\label{fig:Ch3VpTumblingProlatevskappaRec300}
\end{figure}

\begin{figure}
	\centering
    \begin{subfigure}[b]{0.49\textwidth}
		\includegraphics[width=\textwidth]{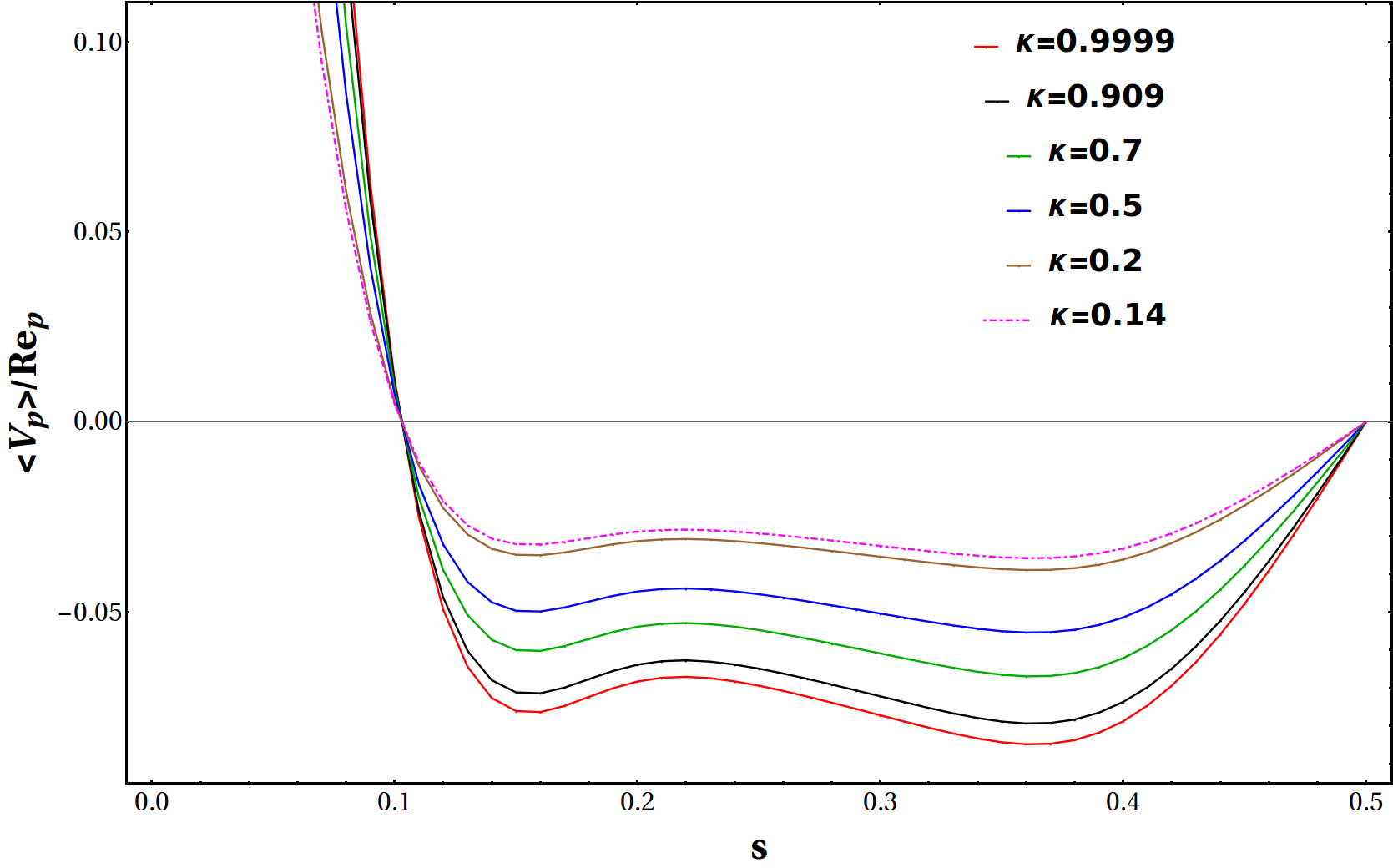}
        \caption{}
    \end{subfigure}
    \hfill
    \begin{subfigure}[b]{0.49\textwidth}
		\includegraphics[width=\textwidth]{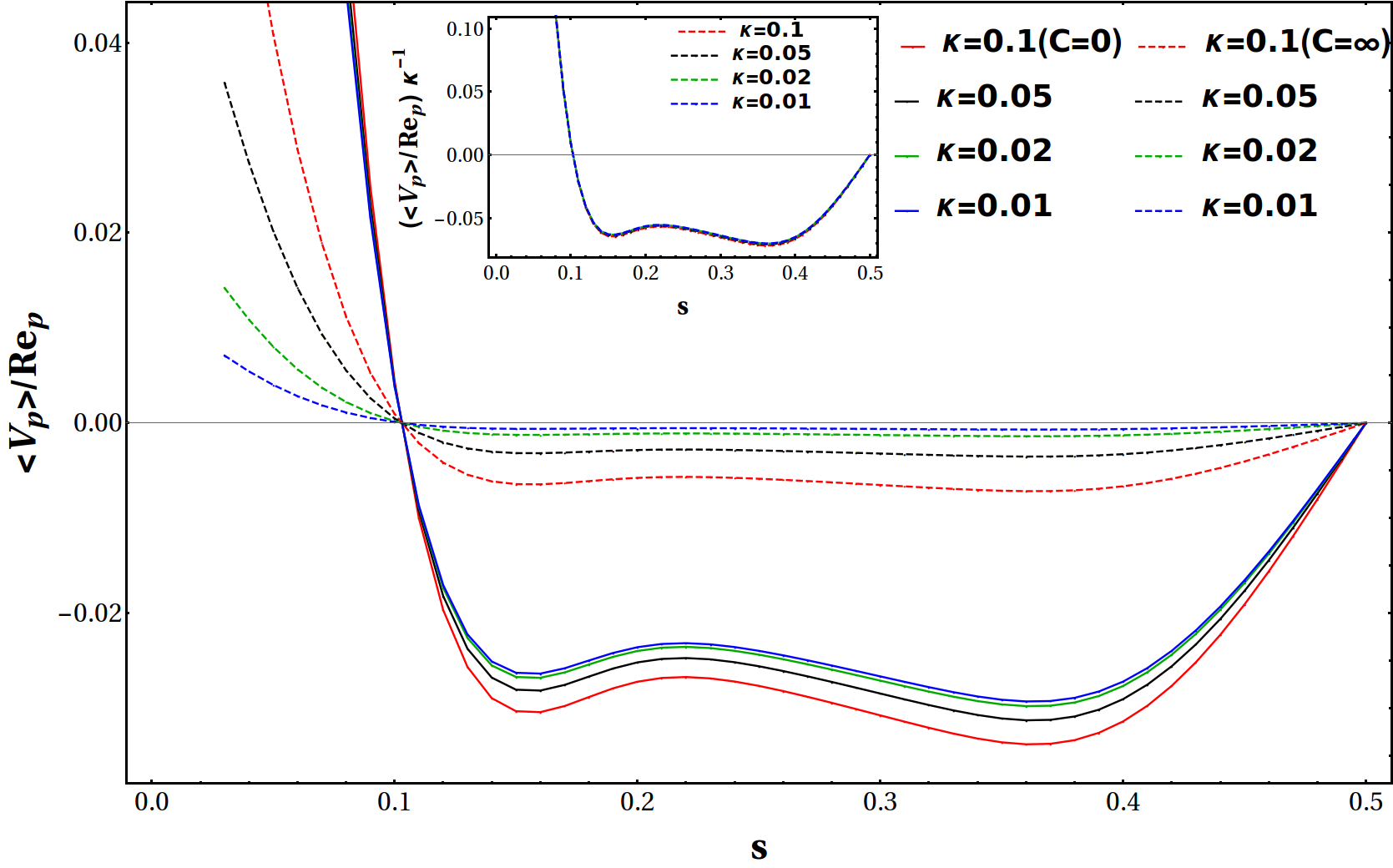}
        \caption{}
    \end{subfigure}
	\caption{(a) Lift velocity profiles for spinning oblate spheroids in the interval $0.14\leq\kappa<1$ for $Re_c=300$; (b) Lift profiles for spinning (solid lines) and tumbling (dashed lines) spheroids in the interval $0<\kappa<0.14$. The inset shows the profiles re-scaled using the $\kappa\to0$ limit of $\langle S_{12}\rangle$.}
\label{fig:Ch3VpTumblingOblatevskappaRec300}
\end{figure}

As mentioned earlier, the Jeffery-averaged lift profiles for a spheroid of an arbitrary aspect ratio may simply be obtained by multiplying the sphere lift profile at the same $Re_c$ by the ratio of the stresslets. Consequently, features pertaining to the $Re_c$-dependence mentioned above, including the range of validity of the small-$Re_c$ approximation, remain true for spheroids. Figures \ref{fig:Ch3VpTumblingProlatevskappaRec300}a and b show the inertial lift profiles for tumbling prolate spheroids over a range of $\kappa$, for $Re_c = 300$, with and without the large-$\kappa$ stresslet scaling. Figures \ref{fig:Ch3VpTumblingOblatevskappaRec300}a and b shows the inertial lift profiles for both spinning\,($0.14\leq\kappa<1$) and spinning/tumbling\,($0<\kappa<0.14)$ oblate spheroids, again for $Re_c = 300$, with and without the small-$\kappa$ stresslet scaling. Note that the latter scaling is only used for tumbling oblate spheroids in the inset of Figure \ref{fig:Ch3VpTumblingOblatevskappaRec300}b, since $\langle S_{12}\rangle$ remains of order unity for spinning spheroids even as $\kappa\to 0$. As in $\S$\ref{sec:Ch3Recsmall}, the scaled lift profiles in Figure \ref{fig:Ch3VpTumblingProlatevskappaRec300}b, and in the inset of \ref{fig:Ch3VpTumblingOblatevskappaRec300}b, approach $\kappa$-independent limiting forms for $\kappa\to\infty$ and $0$, respectively. The magnitudes of the lift velocity profiles reflect that of the disturbance field\,(via the stresslet), and therefore, decrease with increasing\,(decreasing) $\kappa$ for tumbling prolate\,(oblate) spheroids. Although not shown, the finite-$Re_c$ lift profiles may be evaluated for arbitrary $C$, and the $C$-dependence again correlates to the magnitude of the disturbance velocity field, being the largest for tumbling prolate and spinning oblate spheroids.


\section{Conclusions} \label{sec:Ch3conclusion}

The primary result of this manuscript is to show that the inertial lift velocity profiles for neutrally buoyant spheroids in plane Poiseuille flow, for sufficiently small $Re_p$, can be calculated within a Jeffery-averaged approximation. Within this framework, spheroid lift profiles differ from the ones for a sphere only by a multiplicative factor, regardless of $Re_c$, which leads to the original  Segre-Silberberg equilibria being dependent on $Re_c$ but not on the aspect ratio $\kappa$. The multiplicative factor above is a function of $C$ and $\kappa$ in the general case; for times relevant to cross-stream migration, and that are much longer than those characterizing the inertia-induced orientation drift, it is only a function of $\kappa$. Although the Jeffery-averaged approximation only requires $Re_p \ll 1$ for spheroids with $\kappa \sim O(1)$, it becomes increasingly restrictive for spheroids with asymptotically large and small $\kappa$, and with increasing $Re_c$. Deviations from this approximation owing to inertia-induced slow down of rotation and eventual arrest, that lead to $\kappa$-dependent equilibria for extreme-aspect-ratio spheroids, will be reported in a future investigation. Herein, we only note that accounting for the rotational slow down leads to the equilibria migrating back towards the centerline beyond a  certain $\kappa$-dependent $Re_c$ threshold, this being consistent with both experiments\citep{masaeli2012} and computations\citep{chen2012}.
    
En-route to deriving the Jeffery-averaged lift on a spheroid, we have presented, in some detail, the inertial lift profiles for a sphere in plane Poiseuille flow. While most of these results are known from earlier literature, and are spread across multiple efforts\citep{holeal1974,vasseur1976,schonberghinch1989,asmolov1999}, each of these only pertains to specific ranges of $Re_c$. Moreover, there are important features of the lift profile not reported in any of these efforts. We have expressed the small-$Re_c$ sphere lift velocity, obtained within the framework of a point-particle approximation, in terms of a one-dimensional Fourier integral, with detailed expressions for the integrands given in Appendix \ref{App:B}. These expressions allow one to analytically determine the near-wall lift value. The arguments leading to this limiting value also show that it must be independent of $Re_c$, numerical evidence for which is provided by the finite-$Re_c$ lift profiles obtained using a shooting technique. Further, the near-wall limiting lift was also identified with the farfield limit of the lift acting on a finite-sized sphere moving parallel to a single plane wall subject to a linear shearing flow. This connection allows one, in principle, to construct a uniformly valid lift profile across the entire channel. We have also established the surprisingly large range of validity\,(up to $Re_c \approx 10$) of the small-$Re_c$ approximation derived first by \cite{holeal1974} and \cite{vasseur1976}.

The scaling arguments in $\S$\ref{sec:Ch3Scaling} are a crucial element of the overall analysis, and clearly show that the dominant scales contributing to the inertial lift remain much greater than the particle size, regardless of $Re_c$. This leads to use of a  point-particle framework to obtain the leading order approximation for the inertial lift, with scaling arguments identifying both competing contributions due to profile curvature and wall-shear-induced repulsion. The first lift force calculation for a sphere, by \cite{holeal1974} for $Re_c\ll1$, used the same reciprocal theorem formulation as that given here. Although in a less transparent form, the said authors, via scaling arguments, did recognize the dominant contribution of the linearized inertial terms, on scales of $O(H)$, to the reciprocal theorem volume integral. The effect of confining plane boundaries was modeled differently. Rather than directly consider the relevant point singularity\,(Stokeslet or stresslet) between plane parallel walls, the authors used a partial Fourier representation of the full form of the unbounded domain velocity field, including both the finite-size terms and the quadrupolar disturbance induced by the quadratic component of the ambient flow, in order to derive the wall-induced contribution\,(the first reflection). For evaluation of the final volume integral, the physical space integration appears to have been done analytically, with the partial Fourier integral done numerically. Although more circuitous, carrying out the physical space integration should have led to a residual Fourier integral identical to the one obtained directly here from use of the convolution theorem. While the inaccuracy of the resulting profile, and the errors in the equilibrium locations, have been pointed out in the context of Figure \ref{fig:Ch3smallRecComparison}, the detailed calculational procedure in \cite{holeal1974} is correct, and their origin remains uncertain. In contrast to the intuitive scaling arguments here, the later small-$Re_c$ lift calculation of \cite{vasseur1976} made use of a velocity field that emerged from a formal matched asymptotics expansions approach, developed in earlier articles by Brenner and Cox \citep{cox1967,coxbrenner1968}.

The scaling arguments in \S\ref{sec:Ch3Scaling} can be generalized in several directions. First, the arguments apply virtually unchanged to pipe Poiseuille flow. Thus, for small pipe Reynolds numbers\,($Re$), the dominant contributions to the inertial lift must arise from scales of order the pipe radius, and a calculation of this lift would involve approximating the particle as a stresslet within a cylindrical domain. The required Stokesian velocity fields should either be obtainable from that known for a Stokeslet in this domain\citep{liron1978stokes}, or be derivable in a manner similar to that in Appendix \ref{App:A}\,(adapted to cylindrical coordinates). Such an analysis would yield the radius of the Segre-Silberberg annulus for $Re\to0$. Although finite-$Re$ lift force profiles for pipe Poiseuille flow have been computed earlier\citep{matas2009}, the profiles differ significantly for the two lowest $Re$'s examined\,($1$ and $30$), preventing one from inferring the range of validity of a small-$Re$ approximation. The small-$Re$ analysis in $\S$\ref{sec:Ch3Recsmall} can also be generalized to ducts of non-circular cross sections, provided one has the confined Stokeslet field for the relevant cross-sectional geometry. The latter may be derivable for rectangular or elliptical cross sections, using the procedure in Appendix \ref{App:A}, owing to availability of the Greens function the (2D)\,Laplacian. The broken symmetry for the duct case should then allow for the prediction of discrete inertial equilibria in the transverse plane, and their variation with cross-sectional aspect ratio. The requirement that the Stokeslet and stresslet fields be available for a given confined geometry, for a small-$Re$ inertial migration analysis in the same geometry to be possible, calls into question the effort of \cite{roper2015} who seem to have analyzed migration in rectangular ducts based on the approach of \cite{holeal1974} which was solely tailored to plane Poiseuille flow.

Importantly, the scaling arguments in $\S$\ref{sec:Ch3Scaling} allow for the systematic incorporation of finite-size effects that modify the leading order point-particle estimate of the inertial lift velocity for both spheres and anisotropic particles in plane Poiseuille flow. As mentioned in \S\ref{sec:Ch3Scaling}, and discussed in more detail in \cite{anandfinitesize2022}, the finite-size contributions involve an additional factor of $\lambda$, and pertain to the inner region\,(scales of order the particle size), implying that they arise independently of the outer-region point-particle contribution. It is shown in \cite{anandfinitesize2022} that, for spheres, these contributions invariably become important for sufficiently large $Re_c$, leading to the emergence of a pair of new equilibria closer to the channel centerline, consistent with the results of recent experiments. In contrast to the point-particle analysis presented here, calculation of the finite-size contributions involves consideration of the nonlinear terms, and also requires knowing the disturbance velocity field induced by a sphere in an ambient quadratic flow. The analogous calculation for a spheroid will proceed along similar lines. Although more complicated, the disturbance velocity field in an ambient quadratic flow, for a spheroid of an arbitrary aspect ratio and orientation, may be constructed using a superposition of the appropriate vector spheroidal harmonics\citep{navaneeth2015,navaneeth2016}. Crucially, the nonlinear inertial terms as well as the time dependence of the test velocity field, that arise in the inner region, imply that there is no longer a simple proportionality relationship between the finite-size contribution for a sphere and the Jeffery-averaged version of the same for a spheroid. As a result, one expects the incorporation of finite-size contributions to lead to $\kappa$-dependent equilibria, for neutrally buoyant spheroids, even within a Jeffery-averaged framework. This will be examined in a separate communication.
   
\appendix
\section{} \label{App:A}
Herein, we will solve for the disturbance field due to a Stokeslet\,(point force) confined between plane parallel boundaries. Recall that the partially Fourier transformed Stokeslet, $\hat{\bm{u}}_\text{St}$, appears in the final expression for the lift velocity integral viz. \eqref{eq:Ch3VpOuterTimeAvgFT}. In physical space, the disturbance field $\bm{u}_\text{St}$ satisfies the following equations:
\begin{subequations}  \label{eq:Ch3StokesletEqn}
	\begin{align}
	\nabla^2 \bm{u}_\text{St}-\bm{\nabla}p_\text{St}&=- \bm{1}_2\delta(\bm{r}),\\
	\bm{\nabla}\cdot\bm{u}_\text{St}&=0,
	\end{align} 
\end{subequations}
	with the boundary conditions: 
\begin{subequations} \label{eq:Ch3StokesletBC}
	\begin{align}
	\bm{u}_\text{St}&=0 \text{ at } r_2=-s\lambda^{-1}, (1-s)\lambda^{-1},\\
	\bm{u}_\text{St}&\rightarrow 0 \text{ for } r_1,r_3\rightarrow\infty.
	\end{align}
\end{subequations}
The problem of a Stokeslet in the vicinity of a single plane wall was solved for by \cite{blake1971}, using the method of images. \cite{liron1976} calculated the disturbance field due to a Stokeslet between two parallel walls, by taking repeated reflections of Blake's single wall solution and superposing these as an infinite but convergent series. This approach, however, turns out to be quite tedious, and one can instead derive $\bm{u}_\text{St}$ using another method described originally by \cite{vasseur1976}, and later used by \cite{brady2010}. In this procedure, instead of taking repeated reflections of the single wall solution, one can satisfy the no-slip condition on both channel walls at one go. One writes the velocity field as
\begin{align}
\bm{u}_\text{St}=\bm{u}^{\infty}_\text{St}+\bm{u}^{w}_\text{St}, 
\label{eq:Ch3Reflections}
\end{align}
with an analogous decomposition for the pressure field. Here, $\bm{u}^{\infty}_\text{St}$ is the Stokeslet velocity field in an unbounded domain, given by:
\begin{align}
\bm{u}^{\infty}_\text{St}=\bm{J}^\infty\cdot\bm{1}_2,
\label{eq:Ch3StokesletUnboundeddomain}
\end{align}
where $\bm{J}^\infty=\frac{1}{8\pi}\big(\frac{\bm{I}}{r}+\frac{\bm{rr}}{r^3}\big)$ is the Oseen-Burger's tensor; here $\bm{r}=\bm{x}-\bm{y}$ defines the position of any point $\bm{x}$ in the domain relative to the Stokeslet location $\bm{y}$. The second contribution in \eqref{eq:Ch3Reflections} is the one that accounts for the no-slip conditions at the channel walls, and satisfies,:
\begin{subequations}
	\begin{align}
	\nabla^2\bm{u}^{w}_\text{St}-\bm{\nabla} p_\text{St}^{w} &= 0, \\
	\bm{\nabla}\cdot\bm{u}^{w}_\text{St} &= 0,
	\end{align} 	\label{eq:Ch3Stokesletreflection} 
\end{subequations}
with no-slip boundary conditions written as:
\begin{align}
	\bm{u}^{w}_\text{St} = -\bm{u}^{\infty}_\text{St} \text{ at } r_2=-y_2, 1-y_2,
	\label{eq:Ch3wallBCFull}
\end{align}
where $y_2=s\lambda^{-1}$. The solution to (\ref{eq:Ch3Stokesletreflection}a,b) is easily obtained by Fourier transforming the flow and vorticity coordinates, with the partial Fourier transform being defined as:
\begin{align}
\hat{f}=\int\int dr_1 dr_3\,\,e^{\iota(k_1 r_1+k_3 r_3)}\,\,f.
\label{eq:Ch3FT}
\end{align}
Fourier-transforming (\ref{eq:Ch3Stokesletreflection}a,b) in accordance with \eqref{eq:Ch3FT}, one obtains:
\begin{subequations}
	\begin{align}
	&\frac{d^2 \hat{u}^{w}_\text{St,i}}{d r_2^2}\,-k_\perp^2\hat{u}^{w}_\text{St,i}+\iota (k_1\delta_{i1}+k_3\delta_{i3})\,\hat{p}_\text{St}^{w}-\delta_{i2}\frac{d\hat{p}_\text{St}^{w}}{d r_2} = 0, \\
	&\frac{d\hat{u}^{w}_\text{St,2}}{d r_2}-\iota (k_1\,\, \hat{u}^{w}_\text{St,1}+k_3\,\hat{u}^{w}_\text{St,3})=0,
	\end{align} \label{eq:Ch3stokesletFT}
\end{subequations}
where $i=1,2, 3$, and $k_\perp^2=k_1^2+k_3^2$. The partial Fourier transform, $\bm{\hat{u}}^{w}_\text{St}$, satisfies the same boundary conditions as the physical space velocity field above except that one uses $\hat{\bm{u}}^{\infty}_\text{St}$ instead of $\bm{u}^{\infty}_\text{St}$ on the RHS.
Here, $\hat{\bm{u}}^{\infty}_\text{St} =\hat{\bm{J}}^\infty\cdot\bm{1}_2$ with the partial Fourier-transform of the Oseen-Burger's tensor given by:
\begin{align}
\hat{\bm{J}}^\infty=\frac{1}{8\pi}\left(
\begin{array}{ccc}
\frac{2 e^{-k_\perp \left| r_2\right| } \pi  \left(k_3^2+k_\perp^2+k_\perp \left(k_3^2-k_\perp^2\right) \left| r_2\right| \right)}{k_\perp^3} & \frac{2 \iota e^{-k_\perp \left| r_2\right| } k_1 \pi  r_2}{k_\perp} & -\frac{2 e^{-k_\perp \left| r_2\right| } k_1 k_3 \pi (k_\perp \left| r_2\right| +1)}{k_\perp^3} \\
\frac{2 \iota e^{-k_\perp \left| r_2\right| } k_1 \pi  r_2}{k_\perp} & \frac{2 e^{-k_\perp \left| r_2\right| } \pi  (k_\perp \left| r_2\right| +1)}{k_\perp} & \frac{2 \iota e^{-k_\perp \left| r_2\right| } k_3 \pi  r_2}{k_\perp} \\
-\frac{2 e^{-k_\perp \left| r_2\right| } k_1 k_3 \pi  (k_\perp \left| r_2\right| +1)}{k_\perp^3} & \frac{2 \iota e^{-k_\perp \left| r_2\right| } k_3 \pi r_2}{k_\perp} & -\frac{2 e^{-k_\perp \left| r_2\right| } \pi  \left(k_\perp \left| r_2\right|  k_3^2+k_3^2-2 k_\perp^2\right)}{k_\perp^3} \\
\end{array}
\right).
\label{eq:Ch3Jhatinfinity}
\end{align}
It is worth mentioning that, in contrast to the velocity field due to a Stokeslet in an unbounded domain \eqref{eq:Ch3StokesletUnboundeddomain} which only depends on $\bm{y}$ via the position vector $\bm{r}=\bm{x}-\bm{y}$, the bounded domain contribution, apart from its dependence on $\bm{r}$, also depends \textit{explicitly} on the location of the singularity via the wall boundary conditions \eqref{eq:Ch3wallBCFull}. Thus, $\bm{u}^{w}_\text{St} \equiv\bm{u}^{w}_\text{St}(\bm{r};\bm{y})$. 

To begin with, one derives the equation governing the pressure field by taking the divergence of both sides in (\ref{eq:Ch3stokesletFT}a) and using the incompressibility condition (\ref{eq:Ch3stokesletFT}b), leading to,
\begin{align}
	\frac{d^2}{d r_2^2}\,\,\hat{p}^{w}_\text{St}-k_\perp^2\,\, \hat{p}^{w}_\text{St}=0.
\end{align} \label{eq:Ch3PressureFT}
This ODE can be solved to give: $\hat{p}^w_\text{St}=[A_m(\bm{k}_\perp;y_2) e^{-k_\perp r_2}+B_m(\bm{k}_\perp;y2)e^{-k_\perp r_2}]\delta_{m2}$, where $A_m$ and $B_m$ are unknown vectors. After substituting this solution in (\ref{eq:Ch3stokesletFT}a), one may use variation of parameters \citep{arfkenweber} to solve the resulting inhomogeneous ODE to obtain,
\begin{align}
\hat{\bm{u}}^{w}_\text{St}=\hat{\bm{J}}^w\cdot\bm{1}_2,
\end{align}
where $\hat{\bm{J}}^w$, the Fourier transform of the second order tensor $\bm{J}^w$\,($\bm{u}^{w}_\text{St} = \bm{J}^w \cdot \bm{1}_s$), is defined as: 
\begin{align}
\hat{J}_{im}^w=C_{im}(\bm{k}_\perp;y_2)e^{k_\perp r_2}+&D_{im}(\bm{k}_\perp;y_2)e^{-k_\perp r_2}+\frac{1}{4k_\perp^2}\big[A_m(\bm{k}_\perp;y_2) d_i e^{-k_\perp r_2}(2k_\perp r_2+1)\nonumber\\
&+B_m(\bm{k}_\perp;y_2)\bar{d}_i e^{k_\perp r_2}(2k_\perp r_2-1)\big].
\label{eq:Ch3Jhatw}
\end{align}
Here, $d_i=k_\perp\delta_{i2}+\iota (k_1\delta_{i1}+k_3 \delta_{i3})$ and
$\bar{d}_i=k_\perp\delta_{i2}-\iota (k_1\delta_{i1}+k_3 \delta_{i3})$. We will now determine the unknown second order tensors $C_{im}$ and $D_{im}$ and the vectors $A_m$ and $B_m$, using the no-slip conditions on the walls, which can be written as:
\begin{align}
\hat{\bm{u}}^{w}_\text{St}& = -\hat{\bm{J}}^\infty|^L\cdot\bm{1}_2 \text{ at } r_2=-y_2,\label{eq:Ch3wallBCReflection1}\\
\hat{\bm{u}}^{w}_\text{St}& = -\hat{\bm{J}}^\infty|^U\cdot\bm{1}_2 \text{ at } r_2=1-y_2\label{eq:Ch3wallBCReflection2},
\end{align}
where the superscripts `$L$' and `$U$' denote the value of the Fourier-transformed Oseen-Burger's tensor calculated on the lower wall and upper wall, respectively. Using the incompressibility condition (\ref{eq:Ch3stokesletFT}b), 
\begin{align}
A_m&=2 D_{im}\,d_i, \label{eq:Ch3IncomprReflection1}\\
B_m&=-2 C_{im}\,\bar{d}_i. \label{eq:Ch3IncomprReflection2}
\end{align}
The wall boundary conditions \eqref{eq:Ch3wallBCReflection1} and \eqref{eq:Ch3wallBCReflection2} along with the relations \eqref{eq:Ch3IncomprReflection1} and \eqref{eq:Ch3IncomprReflection2} can be solved simultaneously to obtain the following:

\begin{align}
A_m&=\frac{Y_m \sinh (k_\perp\lambda^{-1})+Z_m k_\perp\lambda^{-1}e^{k_\perp(\lambda^{-1}-2 y_2)}}{\sinh^2 (k_\perp\lambda^{-1})-(k_\perp\lambda^{-1})^2},\\
B_m&=\frac{Y_m k_\perp\lambda^{-1} e^{-k_\perp(\lambda^{-1}-2y_2)}+ Z_m \sinh (k_\perp\lambda^{-1})}{\sinh^2 (k_\perp\lambda^{-1})-(k_\perp\lambda^{-1})^2},\\
Y_m&=-d_j(\hat{J}^\infty_{jm}|^L e^{k_\perp(\lambda^{-1}-y_2)}-\hat{J}^\infty_{jm}|^U e^{-k_\perp y_2}),\\
Z_m&=-\bar{d}_j(\hat{J}^\infty_{jm}|^L e^{-k_\perp (\lambda^{-1}-y_2)}-\hat{J}^\infty_{jm}|^U e^{k_\perp y_2}),\\
C_{im}&=\frac{F_{im}e^{-k_\perp(\lambda^{-1}-y_2)}-G_{im}e^{k_\perp y_2}}{e^{-k_\perp\lambda^{-1}}-e^{k_\perp\lambda^{-1}}},\\
D_{im}&=\frac{G_{im}e^{-k_\perp y_2}-F_{im}e^{k_\perp(\lambda^{-1}-y_2)}}{e^{-k_\perp\lambda^{-1}}-e^{k_\perp\lambda^{-1}}},\\
F_{im}&=-\hat{J}^\infty_{im}|^L-\frac{1}{4k_\perp^2}[A_m d_i e^{k_\perp y_2}(1-2k_\perp y_2)-B_m\bar{d}_i e^{-k_\perp y_2}(1+2k_\perp y_2)],\\
G_{im}&=-\hat{J}^\infty_{im}|^U-\frac{1}{4k_\perp^2}[A_m d_i e^{-k_\perp(\lambda^{-1}-y_2)}(1+2k_\perp(\lambda^{-1}-y_2)),\nonumber\\
&+B_m\bar{d}_i e^{k_\perp(\lambda^{-1}-y_2)}(2k_\perp(\lambda^{-1}-y_2)-1)].
\end{align}
Finally, the partial Fourier transforme of the velocity field due to the Stokeslet confined between plane parallel walls is written as:
\begin{align}
\hat{\bm{u}}_{\text{St}}=\hat{\bm{J}}\cdot\bm{1}_2,
\label{eq:Ch3BoundedStokesletfield}
\end{align}
where $\hat{\bm{J}}(k_1,r_2,k_3;y_2)=\hat{\bm{J}}^\infty(k_1,r_2,k_3) +\hat{\bm{J}}^w(k_1,r_2,k_3;y_2)$, with $\hat{\bm{J}}^\infty$ and $\hat{\bm{J}}^w$ being defined in \eqref{eq:Ch3Jhatinfinity} and \eqref{eq:Ch3Jhatw}, respectively.

\section{} \label{App:B}
The functions $I(k_\perp'',s)$ and $J(k_\perp'',s)$ that appear in the integrands, in the expressions for $F(s)$ and $G(s)$ given by (\ref{eq:Ch3Fs}) and (\ref{eq:Ch3Gs}) in the main paper, are defined below:
\begin{align}
I(k_\perp'',s)&=-e^{k_\perp'' (25 s+18)}(s-1)^2\left[3 k_\perp''^2 (s-1)^2-2 k_\perp'' (s-1)+3\right]+e^{k_\perp'' (29 s+24)}(s-1)^2\nonumber\\
&\big[3 k_\perp''^2 (s-1)^2+2 k_\perp'' (s-1)+3\big]-2 (2 s-1) e^{3 k_\perp'' (9 s+8)} \big[6 k_\perp''^3 (s-1) s\nonumber\\
&-4 k_\perp''^2 (s-1) s-3\big]-2 (2 s-1) e^{9 k_\perp'' (3 s+2)} \left[6 k_\perp''^3 (s-1) s+4 k_\perp''^2 (s-1) s+3\right]\nonumber\\
&-s^2 e^{k_\perp'' (25 s+26)} \left(3 k_\perp''^2 s^2-2 k_\perp'' s+3\right)+s^2 e^{k_\perp'' (29 s+16)} \left(3 k_\perp''^2 s^2+2 k_\perp'' s+3\right)\nonumber\\
&-2 e^{k_\perp'' (27 s+20)} \big[8 k_\perp''^4 s \big(2 s^2-3 s+1\big)-6 k_\perp''^3 s \left(2 s^2-3 s+1\right)\nonumber\\
&-12 k_\perp''^2 \left(2 s^3-3 s^2+3 s-1\right)-18 s+9\big]+2 e^{k_\perp'' (27 s+22)} \big[8 k_\perp''^4 s \left(2 s^2-3 s+1\right)\nonumber\\
&+6 k_\perp''^3 s \left(2 s^2-3 s+1\right)-12 k_\perp''^2 \left(2 s^3-3 s^2+3 s-1\right)-18 s+9\big]\nonumber\\
&+e^{5 k_\perp'' (5 s+4)} \big[12 k_\perp''^4 (s-1)^2 s^2+4 k_\perp''^3 (s-1)^2 (4 s-1)+3 k_\perp''^2 \big(4 s^4-12 s^3+14 s^2\nonumber\\
&-12 s+5\big)-2 k_\perp'' \left(4 s^3-9 s^2+9 s-3\right)+3 \left(4 s^2-6 s+3\right)\big]\nonumber\\
&-e^{k_\perp'' (29 s+22)} \big[12 k_\perp''^4 (s-1)^2 s^2-4 k_\perp''^3 (s-1)^2 (4 s-1)+3 k_\perp''^2 \big(4 s^4-12 s^3+14 s^2\nonumber\\
&-12 s+5\big)+2 k_\perp'' \left(4 s^3-9 s^2+9 s-3\right)+3 \left(4 s^2-6 s+3\right)\big]\nonumber\\
&+e^{k_\perp'' (25 s+24)} \big[12 k_\perp''^4 (s-1)^2 s^2+4 k_\perp''^3 s^2 (4 s-3)+3 k_\perp''^2 \left(4 s^4-4 s^3+2 s^2+4 s-1\right)\nonumber\\
&+k_\perp'' \left(-8 s^3+6 s^2-6 s+2\right)+12 s^2-6 s+3\big]-e^{k_\perp'' (29 s+18)} \big[12 k_\perp''^4 (s-1)^2 s^2\nonumber\\
&-4 k_\perp''^3 s^2 (4 s-3)+3 k_\perp''^2 \left(4 s^4-4 s^3+2 s^2+4 s-1\right)+k_\perp'' \left(8 s^3-6 s^2+6 s-2\right)\nonumber\\
&+12 s^2-6 s+3\big]-e^{k_\perp'' (25 s+22)} \big[24 k_\perp''^4 (s-1)^2 s^2+4 k_\perp''^3 (2 s-1)^3\nonumber\\
&+3 k_\perp''^2 \big(6 s^4-12 s^3+10 s^2-4 s+3\big)-6 k_\perp'' \left(2 s^3-3 s^2+3 s-1\right)+9 \left(2 s^2-2 s+1\right)\big]\nonumber\\
&+e^{k_\perp'' (29 s+20)} \big[24 k_\perp''^4 (s-1)^2 s^2-4 k_\perp''^3 (2 s-1)^3+3 k_\perp''^2 \left(6 s^4-12 s^3+10 s^2-4 s+3\right)\nonumber\\
&+6 k_\perp'' \left(2 s^3-3 s^2+3 s-1\right)+9 \left(2 s^2-2 s+1\right)\big],
\end{align}

\begin{align}
J(k_\perp'',s)&=e^{k_\perp'' (25 s+18)} \big[8 k_\perp''^5 (s-1)^5-6 k_\perp''^4 (s-1)^4+60 k_\perp''^3 (s-1)^3\nonumber\\
&+24 k_\perp''^2 (s-1)^2+54 k_\perp'' (s-1)+27\big]-27 e^{k_\perp'' (27 s+16)}-27 e^{k_\perp'' (27 s+26)}\nonumber\\
&+e^{k_\perp'' (29 s+24)} \big[-8 k_\perp''^5 (s-1)^5-6 k_\perp''^4 (s-1)^4-60 k_\perp''^3 (s-1)^3+24 k_\perp''^2 (s-1)^2\nonumber\\
&-54 k_\perp'' (s-1)+27\big]-e^{k_\perp'' (29 s+16)}\big[8 k_\perp''^5 s^5+6 k_\perp''^4 s^4+60 k_\perp''^3 s^3-24 k_\perp''^2 s^2+54 k_\perp'' s-27\big]\nonumber\\
&+e^{k_\perp'' (25 s+26)} \big[8 k_\perp''^5 s^5-6 k_\perp''^4 s^4+60 k_\perp''^3 s^3+24 k_\perp''^2 s^2+54 k_\perp'' s+27\big]\nonumber\\
&+e^{9 k_\perp'' (3 s+2)} \big[32 s (3 s^3-6 s^2+4 s-1) k_\perp''^6+8 s (7 s^3-14 s^2+10 s-3) k_\perp''^5\nonumber\\
&+240 (s-1) s k_\perp''^4+24 (s-1) s k_\perp''^3+108 k_\perp''^2+108 k_\perp''+81\big]\nonumber\\
&+e^{3 k_\perp'' (9 s+8)} \big[32 s (3 s^3-6 s^2+4 s-1) k_\perp''^6-8 s (7 s^3-14 s^2+10 s-3) k_\perp''^5\nonumber\\
&+240 (s-1) s k_\perp''^4-24 (s-1) s k_\perp''^3+108 k_\perp''^2-108 k_\perp''+81\big]\nonumber\\
&-2 e^{k_\perp'' (27 s+22)} \big[16 s (5 s^3-10 s^2+6 s-1) k_\perp''^7+16 s (3 s^3-6 s^2+4 s-1) k_\perp''^6\nonumber\\
&-12 s (7 s^3-14 s^2-26 s+33) k_\perp''^5+120 (s-1) s k_\perp''^4-36 (s^2-s+6) k_\perp''^3+54 k_\perp''^2\nonumber\\
&-162 k_\perp''+27\big]+2 e^{k_\perp'' (27 s+20)} \big[16 s \left(5 s^3-10 s^2+6 s-1\right) k_\perp''^7\nonumber\\
&-16 s (3 s^3-6 s^2+4 s-1) k_\perp''^6-12 s (7 s^3-14 s^2-26 s+33) k_\perp''^5\nonumber\\
&-120 (s-1) s k_\perp''^4-36 (s^2-s+6) k_\perp''^3-54 k_\perp''^2-162 k_\perp''-27\big]\nonumber\\
&+2 e^{k_\perp'' (29 s+18)} \big[16 (s-1)^3 s^2 k_\perp''^7-4 (2-3 s)^2 s^2 k_\perp''^6+4 (4 s^5-5 s^4+28 s^3\nonumber\\
&-22 s^2-5 s+1) k_\perp''^5+3 (4 s^4-4 s^3-50 s^2-4 s+1) k_\perp''^4+6 (20 s^3-15 s^2+13 s+5) k_\perp''^3\nonumber\\
&-6 (8 s^2-4 s+11) k_\perp''^2+27 (4 s-1) k_\perp''-54\big]-2 e^{k_\perp'' (25 s+24)} \big[16 (s-1)^3 s^2 k_\perp''^7\nonumber\\
&+4 s^2 (2-3 s)^2 k_\perp''^6+4 (4 s^5-5 s^4+28 s^3-22 s^2-5 s+1) k_\perp''^5-3 (4 s^4-4 s^3-50 s^2\nonumber\\
&-4 s+1) k_\perp''^4+6 (20 s^3-15 s^2+13 s+5) k_\perp''^3+6 (8 s^2-4 s+11) k_\perp''^2+27 (4 s-1) k_\perp''\nonumber\\
&+54\big]+2 e^{k_\perp'' (29 s+22)} \big[16 (s-1)^2 s^3 k_\perp''^7-4 (3 s^2-4 s+1)^2 k_\perp''^6+4 (4 s^5-15 s^4\nonumber\\
&+48 s^3-72 s^2+35 s-1) k_\perp''^5+3 (4 s^4-12 s^3-38 s^2+100 s-53) k_\perp''^4\nonumber\\
&+6 (20 s^3-45 s^2+43 s-23) k_\perp''^3-6 (8 s^2-12 s+15) k_\perp''^2+27 (4 s-3) k_\perp''-54\big]\nonumber\\
&-2 e^{5 k_\perp'' (5 s+4)} \big[16 (s-1)^2 s^3 k_\perp''^7+4 (3 s^2-4 s+1)^2 k_\perp''^6+4 (4 s^5-15 s^4+48 s^3-72 s^2\nonumber\\
&+35 s-1) k_\perp''^5-3 (4 s^4-12 s^3-38 s^2+100 s-53) k_\perp''^4+6 (20 s^3-45 s^2+43 s-23) k_\perp''^3\nonumber\\
&+6 (8 s^2-12 s+15) k_\perp''^2+27 (4 s-3) k_\perp''+54\big]+2 e^{k_\perp'' (25 s+22)} \big[16 (s-1)^2 s^2 (2 s-1) k_\perp''^7\nonumber\\
&+4 (18 s^4-36 s^3+26 s^2-8 s+1) k_\perp''^6+4 (6 s^5-15 s^4+66 s^3-84 s^2+25 s+1) k_\perp''^5\nonumber\\
&-3 (6 s^4-12 s^3-94 s^2+100 s-53) k_\perp''^4+6 (30 s^3-45 s^2+41 s-13) k_\perp''^3\nonumber\\
&+72 (s^2-s+2) k_\perp''^2+81 (2 s-1) k_\perp''+81\big]-2 e^{k_\perp'' (29 s+20)} \big[16 (s-1)^2 s^2 (2 s-1) k_\perp''^7\nonumber\\
&-4 (18 s^4-36 s^3+26 s^2-8 s+1) k_\perp''^6+4 (6 s^5-15 s^4+66 s^3-84 s^2+25 s+1) k_\perp''^5\nonumber\\
&+3 (6 s^4-12 s^3-94 s^2+100 s-53) k_\perp''^4+6 (30 s^3-45 s^2+41 s-13) k_\perp''^3\nonumber\\
&-72 (s^2-s+2) k_\perp''^2+81 (2 s-1) k_\perp''-81\big].
\end{align}

\section{} \label{App:C}
Herein, we derive the jump conditions (\ref{eq:Ch3JumpCondns}a-d). Starting off with the governing equations (\ref{eq:Ch3HinchODEs}ab),
\begin{subequations}
	\begin{align}
	\frac{d^2 \hat{\langle P\rangle}}{dR_2^2}-k_\perp^2\hat{\langle P\rangle}&=2\iota k_1 \hat{\langle U_2\rangle}(\beta+2\gamma''R_2 Re_c^{-1/2})+2\beta\langle S_{21}\rangle\iota k_1 \delta'(R_2)
	\label{eq:Ch3PprimeprimeODEwithdeltaForcings},\\
	\frac{d^2\hat{\langle U_2\rangle}}{dR_2^2}-k_\perp^2\hat{\langle U_2\rangle}&=\frac{d\hat{\langle P\rangle}}{dR_2}-\iota k_1\hat{\langle U_2\rangle}(\beta R_2+\gamma''R_2^2 Re_c^{-1/2})-\beta\langle S_{21}\rangle\iota k_1 \delta(R_2),
	\label{eq:Ch3UprimeprimeODEwithdeltaForcings}
	\end{align} 
\end{subequations}
where the prime ($'$) denotes differentiation wrt $R_2$. Since the singular forcings on the RHS arise from the highest order derivative, one can postulate the following forms:
\begin{align}
\hat{\langle P\rangle}&=\hat{\langle P\rangle}^-\!\!(R_2)+[\hat{\langle P\rangle}^+(R_2)-\hat{\langle P\rangle}^-(R_2)]\mathcal{H}(R_2), \label{eq:Ch3PPostulate1}\\
\hat{\langle U_2\rangle}'&=\hat{\langle U_2\rangle}^{-'}(R_2)+[\hat{\langle U_2\rangle}^{+'}(R_2)-\hat{\langle U_2\rangle}^{-'}(R_2)]\mathcal{H}(R_2),
\label{eq:Ch3UprimePostulate1}
\end{align}
where the superscripts `$+$' and `$-$' denote the function definitions for $R_2>0$ and $R_2<0$, respectively, and $\mathcal{H}(R_2)$ is the Heaviside function. Differentiating \eqref{eq:Ch3PPostulate1} twice gives, 
\begin{align}
\hat{\langle P\rangle}'&=\hat{\langle P\rangle}^{-'}(R_2)+[\hat{\langle P\rangle}^{+'}(R_2)-\hat{\langle P\rangle}^{-'}(R_2)]\mathcal{H}(R_2)+[\hat{\langle P\rangle}^+(R_2)-\hat{\langle P\rangle}^-(R_2)]\delta(R_2), \label{eq:Ch3PprimePostulate1}\\
\hat{\langle P\rangle}''&=\hat{\langle P\rangle}^{-''}(R_2)+[\hat{\langle P\rangle}^{+''}(R_2)-\hat{\langle P\rangle}^{-''}(R_2)]\mathcal{H}(R_2)+2[\hat{\langle P\rangle}^{+'}(R_2)-\hat{\langle P\rangle}^{-'}(R_2)]\delta(R_2)\nonumber\\
&+[\hat{\langle P\rangle}^+(R_2)-\hat{\langle P\rangle}^-(R_2)]\delta'(R_2). \label{eq:Ch3PprimeprimePostulate1}
\end{align}
Differentiating \eqref{eq:Ch3UprimePostulate1} once gives,
\begin{align}
\hat{\langle U_2\rangle}''&=\hat{\langle U_2\rangle}^{-''}(R_2)+[\hat{\langle U_2\rangle}^{+''}(R_2)-\hat{\langle U_2\rangle}^{-''}(R_2)]\mathcal{H}(R_2)+[\hat{\langle U_2\rangle}^{+'}(R_2)-\hat{\langle U_2\rangle}^{-'}(R_2)]\delta(R_2).
\label{eq:Ch3UprimeprimePostulate1}
\end{align}
Integrating \eqref{eq:Ch3UprimePostulate1} gives,
\begin{align}
\hat{\langle U_2\rangle}&=\hat{\langle U_2\rangle}^{-}(R_2)+[\hat{\langle U_2\rangle}^{+}(R_2)-\hat{\langle U_2\rangle}^{-}(R_2)]\mathcal{H}(R_2)-[\hat{\langle U_2\rangle}^{+}(0^+)-\hat{\langle U_2\rangle}^{-}(0^-)].
\label{eq:Ch3UPostulate1}
\end{align}

To obtain the jump in the aforementioned fields across $R_2 = 0$, one can multiply \eqref{eq:Ch3PprimeprimeODEwithdeltaForcings} and \eqref{eq:Ch3UprimeprimeODEwithdeltaForcings} with $R_2^n$, where $n=0$ and $1$. The resulting expressions can then be integrated with the help of the expressions \eqref{eq:Ch3PPostulate1}-\eqref{eq:Ch3UPostulate1}. Multiplying \eqref{eq:Ch3PprimeprimeODEwithdeltaForcings} with $R_2$ and integrating the resulting expression from $R_2=0^-$ to $0^+$ gives,
\begin{align}
\hat{\langle P\rangle}^+(0^+)-\hat{\langle P\rangle}^-(0^-)=2\iota \beta k_1 \langle S_{12}\rangle.
\label{eq:Ch3JumpP}
\end{align}
Next, \eqref{eq:Ch3PprimeprimeODEwithdeltaForcings} can be integrated across $R_2 =0$ to yield,
\begin{align}
\hat{\langle P\rangle}^{+'}(0^+)=\hat{\langle P\rangle}^{-'}(0^-).
\label{eq:Ch3JumpPprime}
\end{align}
Equation \eqref{eq:Ch3UprimeprimeODEwithdeltaForcings} can be integrated across $R_2 =0$ to yield,
\begin{align}
\hat{\langle U_2\rangle}^{+'}(0^+)-\hat{\langle U_2\rangle}^{-'}(0^-)=\iota \beta k_1 \langle S_{12}\rangle.
\label{eq:Ch3JumpUprime}	
\end{align}
Finally, one can multiply \eqref{eq:Ch3UprimeprimeODEwithdeltaForcings} with $R_2$ and integrate from $R_2=0^-$ to $0^+$ to obtain:
\begin{align}
\hat{\langle U_2\rangle}^{+}(0^+)=\hat{\langle U_2\rangle}^{-}(0^-).
\label{eq:Ch3JumpU}	
\end{align}
This concludes the derivation of all the jump conditions.

\section{} \label{App:D}
Equations (\ref{eq:Ch3HinchODEs}a,b), written as a set of four first order ODEs, take the following form: 
\begin{align} \label{eq:Ch3HinchForm1}
\bm{\Phi}'=\bm{B}\cdot\bm{\Phi},
\end{align} 
where 
\begin{subequations} \label{eq:Ch3HinchForm2}
	\begin{align}
	&\bm{\Phi}' =\begin{bmatrix}
	d\hat{\langle U_2\rangle}/dR_2 \\
	d\hat{\langle U_2\rangle}'/dR_2 \\
	d \hat{\langle P\rangle}/dR_2 \\
	d \hat{\langle P\rangle}'/dR_2
	\end{bmatrix}, \,\,\,
	\bm{\Phi} =\begin{bmatrix}
	\hat{\langle U_2\rangle} \\
	\hat{\langle U_2\rangle}' \\
	\hat{\langle P\rangle} \\
	\hat{\langle P\rangle}'
	\end{bmatrix}, \\
	&\bm{B}=\begin{bmatrix}
	0 & 1 & 0 & 0\\
	k_\perp^2-\iota k_1(\beta R_2+\gamma''R_2^2 Re_c^{-1/2}) & 0 & 0 & 1 \\
	0 & 0 & 0 & 1 \\
    2\iota k_1 (\beta+2\gamma''R_2 Re_c^{-1/2})  & 0 & k_\perp^2 & 0
	\end{bmatrix}.
	\end{align}
\end{subequations}
The general solution to equation \eqref{eq:Ch3HinchForm1} can be written as:
\begin{align} 
\bm{\Phi}^-(R_2)&=c_1^- \bm{\Phi}_1^- + c_2^- \bm{\Phi}_2^- + c_3^- \bm{\Phi}_3^- + c_4^- \bm{\Phi}_4^-\,\,\,\text{for}\,\,-sRe_c^{1/2}\leq R_2<0,\label{eq:Ch3HinchForm3}\\
\bm{\Phi}^+(R_2)&=c_1^+ \bm{\Phi}_1^+ + c_2^+ \bm{\Phi}_2^+ + c_3^+ \bm{\Phi}_3^+ + c_4^+ \bm{\Phi}_4^+ \,\,\,\text{for}\,\, 0<R_2\leq(1-s)Re_c^{1/2},  \label{eq:Ch3HinchForm4}
\end{align} 
where each of the $\bm{\Phi}_i^-$'s and $\bm{\Phi}_i^+$'s constitute a set of four linearly independent solution vectors, with the $c_i^-$'s and $c_i^+$'s being the corresponding integration constants. We choose the following $\bm{\Phi}_i^-$'s and $\bm{\Phi}_i^+$'s on the walls:
 
 \begin{align}
 \big[\bm{\Phi}_1^-(-s\,Re_c^{1/2}),\bm{\Phi}_2^-(-s\,Re_c^{1/2}),\bm{\Phi}_3^-(-s\,Re_c^{1/2}),\bm{\Phi}_4^-(-s\,Re_c^{1/2})\big] =\begin{bmatrix}
 0 & 0 & 0 & 1\\
 0 & 0 & 1 & 0\\
 0 & 1 & 0 & 0\\
 1 & 0 & 0 & 0
 \end{bmatrix},
 \label{eq:Ch3HinchWallCondns}
 \end{align}
 
  \begin{align}
 \big[\bm{\Phi}_1^+((1-s)\,Re_c^{1/2}),\bm{\Phi}_2^+((1-s)\,Re_c^{1/2}),\bm{\Phi}_3^+((1-s)\,Re_c^{1/2}),\bm{\Phi}_4^+((1-s)\,Re_c^{1/2})\big] =\begin{bmatrix}
 0 & 0 & 0 & 1\\
 0 & 0 & 1 & 0\\
 0 & 1 & 0 & 0\\
 1 & 0 & 0 & 0
 \end{bmatrix}.
 \label{eq:Ch3HinchWallCondns2}
 \end{align}
Using the boundary coundition at the lower wall \eqref{eq:Ch3HinchWallCondns} in \eqref{eq:Ch3HinchForm3}, and the boundary condition on the upper wall \eqref{eq:Ch3HinchWallCondns2} in \eqref{eq:Ch3HinchForm4}, one obtains: $c_3^-=c_4^-=c_3^+=c_4^+=0$. Therefore,
\begin{align} 
\bm{\Phi}^-(R_2)=c_1^- \bm{\Phi}_1^- + c_2^- \bm{\Phi}_2^-,\label{eq:Ch3HinchForm5}\\
\bm{\Phi}^+(R_2)=c_1^+ \bm{\Phi}_1^+ + c_2^+ \bm{\Phi}_2^+,\label{eq:Ch3HinchForm6}
\end{align} 
The \textit{NDSOLVE} subroutine in the symbolic computation software \textit{Mathematica} may be employed to solve for the solutions that appear in \eqref{eq:Ch3HinchForm5} and \eqref{eq:Ch3HinchForm6}.  To accomplish this, the entire integration interval in  $R_2$ can be divided into sub-intervals bounded by `orthonormalization' points $y_i$. One may then use \textit{NDSOLVE} to integrate the relevant solution from $y_i$ to $y_{i+1}$, starting from either the upper or lower boundary. At $y_{i+1}$, one may use Gram-Schmidt orthogonalization to orthonormalize the solution vectors $\bm{\Phi}_1^-$ and $\bm{\Phi}_2^-$, and $\bm{\Phi}_1^+$ and $\bm{\Phi}_2^+$. The orthonormalization at regular intervals is necessary since the initially linearly independent vectors $\bm{\Phi}_1^-$ and $\bm{\Phi}_2^-$ (for $R_2<0$), and $\bm{\Phi}_1^+$ and $\bm{\Phi}_2^+$ (for $R_2>0$), become increasingly collinear as the numerical integration progresses, leading to a loss of accuracy.

Following the above procedure, one shoots all the way upto the particle location, $R_2=0$, to obtain: 
\begin{align} \label{eq:Ch3HinchForm7}
\bm{\Phi}^-(0^-)&=c_1^- \bm{\Phi}_1^-(0^-) + c_2^- \bm{\Phi}_2^-(0^-),\\
\bm{\Phi}^+(0^+)&=c_1^+ \bm{\Phi}_1^+(0^+) + c_2^+ \bm{\Phi}_2^+(0^+).
\end{align}
One may now use the jump conditions (\ref{eq:Ch3JumpCondns}a-d), which yields:
\begin{align}
\bm{\Phi}^+(0^+)-\bm{\Phi}^-(0^-)=\bm{C}\cdot
								  \begin{bmatrix}
								  c_1^+ \\
								  c_2^+ \\
								  c_1^- \\
								  c_2^-
								  \end{bmatrix}
								  =\begin{bmatrix}
								  0 \\
								  \iota k_1 \beta \langle S_{12}\rangle \\
								  2\iota k_1 \beta \langle S_{12}\rangle \\
								  0
								  \end{bmatrix},
\end{align}
where 
\begin{align}
\bm{C}=\begin{bmatrix}
\Phi_{11}^+(0^+) & \Phi_{21}^+(0^+) & -\Phi_{11}^-(0^-) & -\Phi_{21}^-(0^-)\\
\Phi_{12}^+(0^+) & \Phi_{22}^+(0^+) & -\Phi_{12}^-(0^-) & -\Phi_{22}^-(0^-) \\
\Phi_{13}^+(0^+) & \Phi_{23}^+(0^+) & -\Phi_{13}^-(0^-) & -\Phi_{23}^-(0^-) \\
\Phi_{14}^+(0^+) & \Phi_{24}^+(0^+) & -\Phi_{14}^-(0^-) & -\Phi_{24}^-(0^-)
\end{bmatrix}.
\end{align}
The matrix $\bm{C}$ can be inverted to obtain the constants $c_1^+$, $c_2^+$, $c_1^-$ and $c_2^-$, which can then be used to write:
\begin{align}
\hat{\langle U_2\rangle}^{-}(k_1,0^-,k_3)&= c_1^-\Phi_{11}^-(0^-)+ c_2^-\Phi_{12}^-(0^-),\\
\hat{\langle U_2\rangle}^{+}(k_1,0^+,k_3)&= c_1^+\Phi_{11}^+(0^+)+ c_2^+\Phi_{12}^+(0^+).
\end{align}
Either $\hat{\langle U_2\rangle}^{+}(k_1,0^+,k_3)$ or $\hat{\langle U_2\rangle}^{-}(k_1,0^-,k_3)$ can be used in \eqref{eq:Ch3VpHinchInverseFT} to obtain the spheroid migration velocity.

\bibliographystyle{jfm}
\bibliography{references}

\end{document}